\documentclass{pasj01}

\newcommand{\Ka}{Fe\emissiontype{I}-K$\alpha$}
\newcommand{\Kb}{Fe\emissiontype{I}-K$\beta$}

\newcommand{\Hea}{Fe\emissiontype{XXV}-He$\alpha$}
\newcommand{\Heb}{Fe\emissiontype{XXV}-He$\beta$}
\newcommand{\Hec}{Fe\emissiontype{XXV}-He$\gamma$}
\newcommand{\FeK}{Fe-K$\alpha$}

\newcommand{\Lya}{Fe\emissiontype{XXVI}-Ly$\alpha$}
\newcommand{\Lyb}{Fe\emissiontype{XXVI}-Ly$\beta$}

\newcommand{\NiHea}{Ni\emissiontype{XXVII}-He$\alpha$}

\newcommand{\SiHea}{Si\emissiontype{XVIII}-He$\alpha$}
\newcommand{\SiLya}{Si\emissiontype{XVIV}-Ly$\alpha$}

\newcommand{\SHea}{S\emissiontype{XV}-He$\alpha$}
\newcommand{\SLya}{S\emissiontype{XVI}-Ly$\alpha$}

\newcommand{\ArHea}{Ar\emissiontype{XVII}-He$\alpha$}
\newcommand{\ArLya}{Ar\emissiontype{XVIII}-Ly$\alpha$}

\newcommand{\CaHea}{Ca\emissiontype{XIX}-He$\alpha$}

\newcommand{\EWKa}{EW$_{6.4}$}
\newcommand{\EWHea}{EW$_{6.7}$}
\newcommand{\EWLya}{EW$_{6.97}$}
\newcommand{\EWFeK}{EW$_{\rm Fe-K}$}

\newcommand{\SHKa}{SH$_{6.4}$}
\newcommand{\SHHea}{SH$_{6.7}$}
\newcommand{\SHLya}{SH$_{6.97}$}
\newcommand{\SHFeK}{SH$_{\rm Fe-K}$}

\newcommand{\Fu}{\,erg\,cm$^{-2}$\,s$^{-1}$}
\newcommand{\Lu}{\,erg\,s$^{-1}$}
\newcommand{\NH}{$N_{\rm H}$}

\newcommand{\SGRA}{Sgr\,A$^*$}

\begin{document}
\SetRunningHead{K. Koyama}{Galactic Center X-rays} 
\setcounter{tocdepth}{3}
\Received{2017/01/25}
\Accepted{2017/06/09}

\title{Diffuse X-Ray Sky in the Galactic Center}

\author{Katsuji \textsc{Koyama}}
\altaffiltext{}{Department of Physics, Graduate School of Science, Kyoto University, Sakyo-ku, Kyoto, 606-8502, Japan}
\email{koyama@cr.scphys.kyoto-u.ac.jp}

\maketitle
\begin{abstract}
The Galactic Diffuse X-ray Emission (GDXE) in the Milky Way Galaxy is spatially and spectrally decomposed into the Galactic Center X-ray Emission (GCXE), 
the Galactic Ridge X-ray Emission (GRXE) and the Galactic Bulge X-ray Emission (GBXE).  The X-ray spectra of the GDXE are  characterized by the strong K-shell lines of the highly ionized atoms, the brightest are the K-shell transition (principal quantum number transition of $n=2\rightarrow1$) of neutral iron  (\Ka), He-like iron (\Hea) and He-like sulfur (\SHea)~lines.
Accordingly, the GDXE is composed of a high-temperature plasma of $\sim$7\,keV (HTP) and a low-temperature plasma of $\sim$1\,keV (LTP), which emit the \Hea~and \SHea~lines, respectively.
The \Ka~line is emitted from nearly neutral irons, and hence the third component of the GDXE is a Cool Gas (CG). The \Ka~distribution in the GCXE 
region is clumpy (\Ka~clump), associated with giant molecular cloud (MC) complexes, Sagittarius A, B, C, D and E, in the central molecular zone. The origin of the \Ka~clumps is the fluorescence and Thomson scattering from the MCs irradiated by past big flares of the super massive  black hole Sagittarius A$^*$.
The scale heights and equivalent widths of the \Ka, \Hea~and \Lya~( $n=2\rightarrow1$ transition of H-like iron) lines are different among the GCXE, GBXE and GRXE. Therefore the structure and origin are separately examined. This paper overviews the research history and the present understandings of the GDXE, in particular focus on the origin of the HTP and CG in the GCXE.
\end{abstract} 

\tableofcontents 

\section{Introduction} %section 1

The center of our Milky Way Galaxy (the Galactic Center: GC) and Sagittarius A$^*$ (\SGRA) are the nearest galactic center and the Super Massive Black hole (SMBH) from the Earth, and hence are unique and ideal laboratories for the study of various astrophysical processes. Therefore,  many observations and theoretical works have been made in the wide band of electromagnetic radiations.
The recent reviews of the GC and \SGRA~are found in \citet{Ge10} and 
\citet{Mo12}, which focused mainly on the infrared and radio bands, respectively.  This paper, therefore, focuses on the review of the X-ray sky near the GC and related activities of \SGRA.

The X-ray astronomy was opened by the discovery of a bright extraterrestrial X-ray source with the sounding 
rocket \citep{Gi62}. This source is now known as Sco X-1, the brightest Galactic X-ray star.  Diffuse X-ray emissions, later called the Cosmic X-ray Background (CXB),  were also found. After many studies and debates, the origin of the CXB, at least in the energy range below $\sim$10\,keV, has come to a common consensus that the CXB is an integrated emission of the extragalactic sources such as Active Galactic Nuclei (AGN) and active galaxies. 

The first X-ray satellite,  Uhuru discovered many point-like sources, majority of the bright sources are 
concentrated in the Galactic plane (e.g., see the fourth Uhuru catalog; \cite{Fo78}). Most of them  are close binaries of  neutron star (NS) or black hole (BH) with normal stars, and are named the X-ray Binary (XB). 
Sco X-1 is the brightest XB observed from the Earth. In addition to the CXB and XBs, diffuse X-rays of the Galactic origin were found with Uhuru, Ariel\,5 and HEAO-1. 
These emissions extended to a high Galactic latitude of the scale height (SH) of $\gtrsim$500\,pc\,--\,1.5\,kpc.  Since the surface brightness is very faint, less than $\sim$10\,\% of the CXB, any quantitative study has been limited, which places this emission out of the scope of this review.
Soon after, the Galactic diffuse X-ray emission with the surface brightness nearly or larger than the CXB, has been discovered.  This emission is  more concentrated toward the Galactic plane, with the  SH of less than  a few 100\,pc (see section 2).  This review specifies this X-ray emission as the Galactic Diffuse X-ray Emission (GDXE).

In this review, the transition lines from the first excited to the ground states (the principle quantum 
number $n=2\rightarrow1$) in the neutral or low ionization atoms, He-like (ions with two electrons), and H-like (ions
with one electron) atoms are designated as the K$\alpha$, He$\alpha$ and  Ly$\alpha$~lines, respectively. 
Likewise, the transition lines from the second excited to the ground states 
(the principle quantum number $n=3\rightarrow1$) are the K$\beta$,  He$\beta$~and Ly$\beta$~lines, while the transitions from the third  excited to the ground states (the principle quantum number $n=4\rightarrow1$) are the He$\gamma$ and Ly$\gamma$~lines \footnote {In some of the reference papers, notation of K$\alpha, \beta, \gamma$~is used instead of He$\alpha, \beta, \gamma$ or Ly$\alpha, \beta, \gamma$.}.
The Equivalent Width (EW) and Scale Height (SH) of these iron lines are expressed as  \EWKa, \EWHea, \EWLya, \SHKa,  \SHHea, and \SHLya, where the subscript is the energy of the lines. For brevity and\,/\,or in the case that the \Ka, \Hea~and \Lya~lines are not resolved,
the notation of \FeK~with the EW of  \EWFeK~(\EWKa, \EWHea~and \EWLya) and the SH of \SHFeK~(\SHKa, \SHHea~and \SHLya) are used.  
These and the other abbreviations and symbols frequently used in this review are summarized in table 1.

After long and extensive studies, the GDXE is now decomposed to three spatial components, the Galactic Center X-ray Emission (GCXE), the Galactic Bulge X-ray Emission (GBXE) and the Galactic Ridge X-ray Emission (GRXE). The GDXE exhibits various  atomic  lines, the brightest are the K-shell transition lines from  the He-like Fe (\Hea),  He-like S (\SHea)~ and from the neutral or low ionized Fe (\Ka). These atomic lines are  emitted from a  High-temperature Plasma (HTP) and a Low-temperature Plasma (LTP) and an X-ray re-emitting Cool Gas (CG), respectively.
%%% 
\begin{table*}[!ht] %table 1
\caption{Abbreviations and Symbols frequently used in this text.}
\begin{center} 
\footnotesize
\begin{tabular}{ll}
\hline 
Symbol, Acronym 	&Explanation \\
\hline
GC 	  	& Galactic Center, The center of our Milky Way Galaxy \\
GDXE 	  	& Galactic Diffuse X-ray Emission which is composed of GCXE, GBXE and GRXE \\
GCXE		& Galactic Center X-ray Emission  \\
GBXE 		& Galactic Bulge X-ray Emission \\
GRXE		& Galactic Ridge X-ray Emission \\
CXB 		& Cosmic X-ray Background \\
NXB 		&  Non X-ray Background, the cosmic ray induced background \\
\Ka 	& $n=2\rightarrow1$ transition of the neutral or low ionization iron at the energy of $\sim$6.40 keV\\
\Kb	& $n=3\rightarrow1$ transition of the neutral or low ionization iron at the energy of $\sim$7.06 keV\\

\Hea 		& $n=2\rightarrow1$ transition of the He-like (two electrons are left) iron at the energy of $\sim$6.68 keV\\
\Heb	& $n=3\rightarrow1$ transition of the He-like iron at the energy of $\sim$7.88 keV \\

\Lya		& $n=2\rightarrow1$ transition of the H-like (one electron is left) iron at the energy of $\sim$6.97 keV\\

\FeK	& $n=2\rightarrow1$ transition  of the neutral, He-like and H-like iron\\

\NiHea	& $n=2\rightarrow1$ transition of the He-like nickel at the energy of $\sim$7.80 keV \\

\SiHea	& $n=2\rightarrow1$ transition of the He-like silicon at the energy of $\sim$1.86 keV\\ 

\SiLya	& $n=2\rightarrow1$ transition of the H-like silicon at the energy of  $\sim$2.00 keV\\

\SHea & $n=2\rightarrow1$ transition of the He-like sulfur at the energy of $\sim$2.46 keV \\

\SLya & $n=2\rightarrow1$ transition of the H-like sulfur at the energy of $\sim$2.62 keV \\

EW		& Equivalent Width, the flux ratio of the line to the continuum emission. \\
EW$_{6.4}$	& Equivalent width of the 6.4 keV line, \Ka~line \\   
EW$_{6.7}$	& Equivalent width of the 6.7 keV line, \Hea~line \\
EW$_{6.97}$	& Equivalent width of the 6.97 keV line,  \Lya~line \\
EW$_{\rm Fe-K}$	& Equivalent width of the iron K-shell line, the sum of EW$_{6.4}$, EW$_{6.7}$ and EW$_{6.97}$ \\
SH		& Scale Height, the longitude distance from the Galactic plane where the flux falls by a factor of $1/e$.\\
SH$_{6.4}$	& Scale Height  of  the 6.4 keV line,  \Ka~line \\ 
SH$_{6.7}$	& Scale Height  of the 6.7 keV line, \Hea~line \\
SH$_{6.97}$	& Scale Height  of the  6.97 keV line,  \Lya~line \\
SH$_{\rm Fe-K}$	& Scale height of the iron K-shell line, the mean of SH$_{6.4}$, SH$_{6.7}$ and SH$_{6.97}$ \\
HTP		& High temperature Plasma ($kT\sim$6--7 keV) in the GDXE \\
LTP		& Low temperature Plasma ($kT\sim$0.8--1 keV) in the GDXE\\
CG		& Cool Gas in the GDXE which emits the K-shell lines of neutral atoms\\
MC		& Molecular Cloud \\
CMZ		&Central Molecular Zone\\
XRN		& X-ray Reflection Nenula:  Fluorescence/Thomson scattered X-ray nebula by the flare of \SGRA.\\
NTF		& Non Thermal Filaments, mostly radio source \\
PWN		& Pulsar Wind Nebula as a non thermal X-ray source \\  
CV 		& Cataclysmic Variable, Binary of normal star and white dwarf \\   
mCV 		& Magnetized CV which are intermediate polar, polar and symbiotic stars\\   
non-mCV 	& Non magnetized CV, or dwarf nova\\   
AB		& Coronal Active close Binary of low mass star like RS CVn and Algol types.\\
XAS		& X-ray Active Stars. The main components are mCV, non-mCV and  AB \\
SMD		& Stellar Mass Distribution made from the infrared flux \\
SMBH		& Super Massive Black Hole \\
SN		& Supernova \\
SNR		& Supernova Remnant　\\
LECR		& Low Energy Cosmic Ray \\
LECRe		& Low Energy Cosmic Ray electrons, typical energy is $\sim$ a few 10~keV \\
LECRp		& Low Energy Cosmic Ray protons, typical energy is $\sim$ a few 10~MeV \\
CIE		& Collisional Ionization Equilibrium \\
IP		& Ionizing Plasma (Non Equilibrium Ionization: NEI) \\
RP		& Recombining Plasma (Non Equilibrium Ionization: NEI) \\
XLF		& Cumulative X-ray Luminosity as a Function of point source luminosity\\
FIM		& Flux Integration Method of the point sources \\
SAM		& Spectrum Accumulation Method of the point sources \\
\hline	
\end{tabular}

 %table 1
\normalsize
\end{center} 
\end{table*} 

This paper reviews  the early studies of the GDXE, then moves on the reviews of the separate study of the GCXE, GBXE and GRXE. The reviews gradually focus on the origins and structures of the HTP and CG in the GCXE and its implications. The GBXE and GRXE are also reviewed, because the origin and structure of the GCXE are closely related to those of the GBXE and GRXE.  Since the EW and SH of the \FeK~lines are significantly different among the GCXE, GBXE and GRXE, the long-standing debate, the origin and structure of the GDXE, are re-examined by the separate, but coordinated studies on the GCXE, GBXE and GRXE. 

The contents are organized as follows. The early results taken before Chandra, XMM-Newton and Suzaku on the GRXE are reviewed in section 2. The structure of the GRXE and its possible origin are discussed in section 2.1. Discoveries of new components, the GCXE and GBXE, and their characteristics are  given in sections 2.2 and  2.3, respectively. 

Section 3 overviews the recent observational results of the GDXE made with Chandra, XMM-Newton and Suzaku.  
Section 3.1 gives the global spatial structure of the \FeK, \SHea~and \SLya~lines in the GDXE, along and perpendicular to the Galactic plane, which leads 
to the decomposition of the GDXE into the GCXE, GBXE and GRXE.
Section 3.2 reports the X-ray spectra and luminosity of the GCXE, GBXE and GRXE.
The spectra are significantly different among these components, and hence verify
the decomposition of the GDXE into these components.
Section 3.3 discusses the characteristics of the HTP and CG in the central region of the GCXE, based mainly on the observed flux of \FeK~and \EWFeK.

Section 4 reviews the local enhancements of the HTP and LTP in the GCXE obtained mainly with Suzaku.  
Sections 4.1 is devoted to the description of young Supernova Remnants (SNR) or candidates, which emit strong  \Hea~lines, 
while section 4.2 reviews the soft X-ray spots with strong \SHea~lines, which are intermediate aged  SNRs or candidates. 

Section 5 reviews the \Ka~line emitting component, the CG components observed mainly with Suzaku, Chandra and XMM-Newton. The emission mechanisms of the \Ka~line and resultant \EWKa~are given in section 5.1.
Section 5.2 is devoted to the X-ray Reflection Nebula (XRN), which is a source of fluorescence and scattered X-ray by past activities 
(flares) of \SGRA.  Section 5.3 concerns the other \Ka~ clumps, which may be unrelated to the flares of \SGRA. 

Section 6 summarizes small size diffuse X-ray emissions, found mainly with Chandra and XMM-Newton in the central GCXE region (section 6.1) and outer GCXE region (section 6.2). These are mostly power-law (non-thermal) X-ray filaments of length $\lesssim$ a few $10\arcsec$. 
 
Section 7 presents  activity history of \SGRA. The past activities are suggested by the XRNe, a Recombining Plasma (RP) and outflows or jet-like structures pointing to \SGRA, which are given in sections 7.1, 7.2 and 7.3, respectively.  

Sections 8 specified the methodology to the origin of the GDXE, together with the summary of  the \EWFeK~and \SHFeK~of the  magnetic Cataclysmic Variables (mCVs), non magnetic Cataclysmic Variables (non-mCVs) or dwarf nova and coronal Active Binaries (ABs). The spectral fit of the GCXE, GBXE and GRXE by a combination of the mCVs, non-mCVs (DN) and ABs spectra are presented.  

Section 9 discuss on the origin of the GBXE, GRXE and GCXE based on the results given in section 8. The origin of the GDXE are separately discussed in section 9.1
(GBXE) section 9.2 (GRXE) and section 9.3 (GCXE).

In the reference papers, the physical parameters, luminosity, plasma size and the other physical parameters, have been derived under the assumption of the GC distance of 8.5 or 8.0\,kpc.  This review, therefore unifies the physical parameters assuming the GC distance to be 8.0\,kpc. Then, the angular size and the X-ray flux of $1\arcmin$~and $10^{-12}$\Fu~ correspond to the physical size of 2.33\,pc and X-ray luminosity of $7.63\times10^{33}$\Lu, respectively. 
The cited errors in the reference papers were either 90\,\%, or 1\,$\sigma$ confidence levels depending on the physical parameters and\,/\,or authors.  This paper unifies the error to be 90\,\% confidence level, unless otherwise stated.  The metallic abundances of the solar photosphere are those in Anders and Grevesse (1989).
%Note that some figures are missing  from the original PASJ paper, because of the limit of the memory size.
%%%

\section{Early Studies of the Galactic Diffuse X-Ray Emission (GDXE)} % section 2
This section reports the start lines  in the studies of the GDXE, the early results of the GRXE (section 2.1), GCXE (section 2.2) and GBXE (section 2.3), using the results taken  before the era of Chandra, XMM-Newton and Suzaku. In these sub-sections, the contents are not exactly in the chronological orders, but are organized in subject-oriented styles.

\subsection{The Galactic Ridge X-Ray Emission (GRXE)} % section 2.1

In this section, the history of the GRXE survey is reported.  Section 2.1.1 is that oriented to the point source fraction of the GRXE, while section 2.1.2 is oriented to the GRXE spectrum. The discovery history of non-thermal emissions is given in section 2.1.3.

 \subsubsection{The GRXE and Point Sources} % section 2.1.1

The global structure of the GRXE is first reported with HEAO-1 by \citet{Wo82}. It is a diffuse X-ray emission 
in the 2\,--\,10\,keV band along the Galactic plane. Due to the large beam size of $\timeform{3D}\times\timeform{1.5D}$ (FWHM), the regions free from contamination of bright XBs, are limited to be $l\gtrsim\timeform{50D}$ ($\gtrsim7$\,kpc from \SGRA). Nevertheless, the overall  profile is estimated to be an exponential function of the e-folding radius of $\sim3.5$\,kpc with the half thickness (SH) of $\sim240$\,pc. Extrapolating the flux distribution to a radius of $\lesssim7$\,kpc, the total luminosity of the GRXE is 
estimated to be $\sim 10^{38}$\Lu~(2\,--\,10 keV).
They proposed that most probable origin of the GRXE is an integrated emission of many unresolved faint discrete sources.

\citet{Wo83} compared the results of \citet{Wo82} to the number density of serendipitous  sources in the Galactic 
plane discovered with the Imaging Proportional Counter (IPC) on board the Einstein Observatory. They
concluded that X-ray point sources with the 2\,--\,10\,keV band luminosity of $8\times10^{32}$\,--\,$3\times10^{34}$\Lu~are 
not dominant contributors to the GRXE.  In particular, contributions of Be/neutron star systems such as X Persei would 
be minor, because these systems have the 2\,--\,10\,keV band luminosity of $\sim10^{33}$~\Lu, 
and have smaller SH than the GRXE.  Lower luminosity stellar systems of $\le4\times10^{32}$\Lu~are likely major 
contributors to the GRXE. They predicted that coronal Active Binaries (ABs) and Cataclysmic Variables (CVs) with 
the 2\,--\,10\,keV band luminosity of $2\times10^{30}$\,--\, 4$\times10^{32}$\Lu~may 
contribute $43\pm{18}$\,\% of the GRXE. 

Hertz \& Grindlay (1984) found 71 point-like sources with the IPC in the regions of the Galactic latitude of $|b|\le 15^\circ$.   
In the sample, $\sim$46\,\%, $\sim$31\,\% and $\sim$23\,\% are coronal 
emission from non-degenerate stars, extragalactic sources and unidentified Galactic sources, respectively.
The approximated number density of the Galactic sources is consistent with Cataclysmic Variables (CVs) and other accreting white dwarfs.
Faint Galactic plane sources are concentrated toward the Galactic bulge, and have a flatter number-flux 
relation than that at higher Galactic latitude and longitude. 

\citet{Wa85} observed the inner GRXE with EXOSAT having a small beam size of $\timeform{0.75D}\times\timeform{0.75D}$ (FWHM).
The flux distribution (2\,--\,6 keV) of the unresolved emissions 
extends to the inner Galactic plane in the longitude of $|l| \lesssim40^{\circ}$. They found very small SH of $|b|\lesssim1^\circ$. This small SH excludes old population stars as the origin of the GRXE.  The overall profile of the GRXE is exponential shape with the e-folding $l$ and $b$ of $\sim3.5$~kpc and of $\sim100$~pc, respectively.  
The total luminosity is $\sim10^{38}$\Lu, consistent with the results of \citet{Wo82}.

With RXTE, \citet{Re06a} made the GRXE profile in the 3\,--\,20 keV band along the Galactic plane of $|l|\lesssim100\arcdeg$, and perpendicular to the plane of $|b|\lesssim6\arcdeg$ at $|l|\lesssim4\arcdeg$. The SH at $l$\,=\,$20\arcdeg$ is
$\sim$130\,eV.  They found the longitude profiles are similar to the infrared surface brightness distribution. 
\citet{Re06b} further investigated the RXTE data of the inner Galaxy  ($|l|\lesssim$25$\arcdeg$) and  the Galactic ridge emission up to $|l|\lesssim$120$\arcdeg$.  In order to reduce possible contamination of bright point-sources (XBs) to the GRXE, they used the \FeK~(6.7 keV) line, instead of the continuum (3\,--\,20\,keV) band following the Ginga results of \citet{Ko86a, Koy89} (see section 2.1.2). The SH of the \FeK~line is similar to that of the continuum band by \citet{Re06a}.  
They found  that the  surface brightness distributions along the Galactic plane of the \FeK~lines are similar to the infrared surface brightness distribution.

\citet{Re06a, Re06b} assumed that the infrared distribution represents the Galactic stellar mass  distribution (SMD), then proposed that the origin of the GRXE is discrete stellar sources.  The ratio of the X-ray luminosity in the 3\,--\,20\,keV band to the  
near-infrared luminosity is $L_{3-20\,{\rm keV}}/L_{3-4\,\mu{\rm m}} \sim 4\times10^{-5}$, which corresponds to $\sim3.5\times 10^{27}$\Lu\,$M_\odot^{-1}$.
This luminosity per stellar mass agrees within an uncertainty of $\sim$50\,\% to that of the solar neighborhood \citep{Saz06}. Then, they suggested that observations with  the sensitivity limit of $\sim10^{-16}$\Fu~(2\,--\,10\,keV)  may resolve
$\sim$90\,\% of the GRXE into discrete stellar sources. 

\citet {Re12} performed deep scans with RXTE  across the Galactic plane
in the energy band of 4.3\,--\,10.5\,keV from $b =\timeform{0D}$ to  
$\timeform{-30D}$  at $l =\timeform{18. 5D}$. The SH of the GRXE is estimated to be $\sim$110\, pc.  In the point source origin, they argued that the candidate stars with SH$\sim 260$\,pc contribute less than $\sim$0.3 of the total cumulative fractional emissivity of point sources in the Galactic plane.
The cumulative fractional emissivity of the GRXE in the energy band of 2\,--\,10\,keV is $\sim3\times10^{27}$erg s$^{-1}M_\odot^{-1}$, consistent with \citet{Re06a} in the energy band of 3\,--\,20\,keV.

One note is that the spatial resolution of both the GDXE and SMD are sub arc-degree.
Therefore, the comparison of the GDXE distribution to the SMD did not go into detailed spatial structure of the GDXE, but is limited in the GRXE. The scale heights (SHs) show large variations from author to author, possibly due the large and different beam sizes with each author, or due to the contribution of the GBXE (section 2.2).  These prevents to judge the point source populations, e.g. whether high mass stars (SH$\lesssim100$ pc) or low mass stars (SH$\gtrsim$100 pc).

\subsubsection{K-Shell Lines in the GRXE and Thermal Plasma Origin} % section 2.1.2

The Gas Scintillation Proportional Counter (GSPC) on board the Tenma 
satellite had higher spectral resolution than the ordinary proportional counter. With the GSPC, \citet{Ko86a} discovered  an intense emission line of \FeK~at $\sim$6.7\,keV. Since the field of view of the GSPC is $\timeform{3.1D}$ (FWHW), the observed sky, which is free from bright XBs are limited to be eight fields in the Galactic inner disk of $\timeform{280D}<l<\timeform{340D}$ (GRXE). 
The \EWFeK~is in the range of $\sim$500\,--\,700 eV. 
They interpreted that the \FeK~line  is due to
an optically thin plasma, because the line center energy of $\sim$6.7\,keV is consistent with \Hea \footnote{Soon after, this 6.7 keV line was found to be a complex of the \Ka, \Hea~ and \Lya~lines (section 2.2).}.  The plasma temperatures are variable from region to region in the range of $\sim$5\,--\,10\,keV. They claimed that the temperature variations do not favor the  origin of many faint point sources. Even from the limited sample of the eight fields, they determined  that the intensity distribution in the 2\,--\,10 keV band is a disk-shape with the SH of $\sim$100\,--\,300\,pc and the radius of $\sim$8\,kpc. The total luminosity of the GRXE is estimated to be $\sim10^{38}$\Lu.  

\citet{Ko86b} estimated a possible contribution of unidentified SNRs to explain the \FeK~line, and argued that if the Supernova (SN) rate is $\sim$10\,/\,century, the observed GRXE flux and the value of \EWFeK~would be explained. 
However, this SN rate is $\sim$3\,--\,10 times larger than  the canonical value of $\sim$1\,--\,3\,/\,century.
The allowed region of $n_{\rm e}$ (cm$^{-3}$) and $t$ (s), where $n_{\rm e}$ (cm$^{-3}$) and $t$ (s) are the electron density and time after the shock heating, respectively,  are $\sim10^{-3}$ \,--\, $4\times10^{-1}$ cm$^{-3}$ and $\lesssim3 \times10^{12}$\,s. Then, the ionization parameter $n_{\rm e}t$ is $\lesssim 10^{12}$cm$^{-3}$\,s. 
Therefore, the candidate sources for the origin of the GRXE are young-intermediated aged SNRs in Ionizing Plasma (IP) or in Non-Equilibrium Ionization (NEI). The candidate SNe may occur in a thin ISM so that the surface brightness of the SNRs would be below the resolving capability of the GSPC. 

\citet{Koy89} re-examined the thermal plasma in the GRXE using more extended GSPC data set of 27 XB-free fields in the Galactic plane, and analyzed the X-ray spectra. The best-fit temperature and \EWFeK~are $\sim$3\,--\,14\,keV 
and $\sim$0.24\,--\,1.5\,keV, respectively. Thus, the extensive data set of the GSPC provides larger position-dependent variations in the temperatures and \EWFeK~than those of \citet{Ko86a}, and hence the  argument of the point source origin for the GRXE (section 2.1.1) becomes more unlikely. 

\begin{figure}[!h]
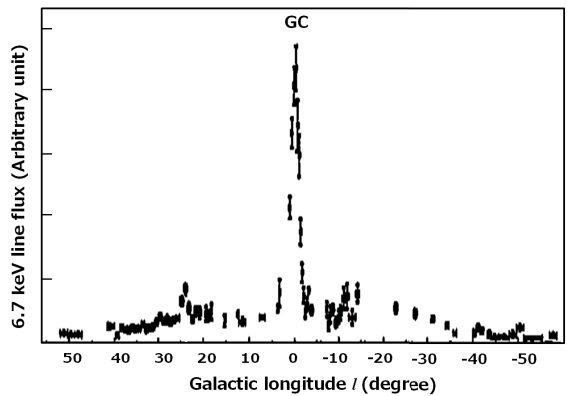
 % figure 1
\begin{center}
\FigureFile(80mm,60mm){eps-figure/Ginga-Plane.eps}% figure 1
\end{center} 
\caption{The longitude distribution of the 6.7\,keV line along the Galactic plane taken from the  Ginga  Galactic plane survey 
(From \cite{Ko89}).} 
\label{fig:Ginga-plane}
\end{figure}

The Large Area proportional Counter (LAC) on board the Ginga satellite surveyed the Galactic plane in the \FeK~line band with the FWHM beam size of
$\timeform{1.1D}\times\timeform{2.0D}$~\citep{Ko89}. 
Since the \EWHea~of XBs, the brightest point sources in the GRXE region, is only $\lesssim$~50\,eV \citep{Hi87}, the \FeK~line  profile is free from possible contamination of bright XBs (figure 1). This is an advantage of the \FeK~line over the continuum X-ray band (e.g. 2\,--\,10 keV band) for the study on the global spatial structure of the GRXE.
The total flux of the \FeK~line is $\sim 10^{37}$\Lu, about 10\,\% of the 2\,--\,10\,keV band flux ($\sim10^{38}$\Lu).  

\citet{Ya95} examined the center energy of the \FeK~line and \EWFeK~ as a function of the plasma temperature. 
Since the center energy of the \FeK~line and the \EWFeK~are systematically lower than those expected from a collisional ionization equilibrium (CIE) plasma of $\sim$5\,--\,10\,keV temperature in one solar Fe abundance, they estimated  the ionization parameter ($n_{\rm e}t$) to be $10^{10}$\,--\,$10^{11}$cm$^{-3}$\,s. This is consistent with the scenario of \citet{Ko86b} that the GRXE is assembly of young-intermediate aged SNRs, in which the plasma is in IP or NEI. However soon after, the energy down-shift of the \FeK~line is found to be a mixture of \Ka~ (6.40\,keV), \Hea~ (6.68\,keV) and \Lya~ (6.97\,keV) lines (see next paragraph), and hence the IP (NEI plasma)  interpretation is questionable. 

The X-ray CCD detectors on board ASCA had a better energy resolution than any other previous instruments. With ASCA, \citet{Ya96} and \citet{Ka97} obtained the X-ray spectra from the Scutum Arm region at $l\sim 28^\circ$.5. They  
resolved the \FeK~line into \Ka~ (6.40\,keV), \Hea~ (6.68\,keV) and 
\Lya~ (6.97\,keV) lines.  
They also detected the bright \SiHea~and \SHea~ lines. 
Therefore, the GRXE spectra are not single-temperature plasmas, but are  well fitted with a two-temperature plasma model, the LTP of $\sim0.8$\,keV temperature for the \SiHea~and \SHea~ lines,  and the HTP of $\sim$7\,keV temperature for the \Hea~and \Lya~lines.  

\citet{Ka97} reported that the surface brightness  of the LTP and HTP at $(l, b)\sim  (28^\circ.5,  0^\circ)$  
are $\sim2\times10^{-6}$\Fu\,str$^{-1}$ and 
$\sim 5\times10^{-7}$\Fu\,str$^{-1}$ (0.5\,--\,10\,keV), respectively. 
The flux of the LTP is extended to $|b|\sim2^\circ$, larger than the HTP of $|b|\sim \timeform{0.5D}$.  
However, taking into account of the differences of the optical depth, they proposed that the real SH of the LTP  may be equal to the HTP of
$\sim 70$\,pc. \citet{Ya96} found position-to-position fluctuations of the surface brightness, and concluded that point sources of the luminosity larger than $\sim 2\times10^{33}$\Lu~are not the major origin  of the GRXE.   

%%No Galactic object, regardless diffuse or point source, has similar spectrum of the GRXE with temperature of $\sim$5\,--\,10 keV and large \EWFeK ~of $\sim$0.5\,--\,1.1 keV \citep{Ya95, Ka97}. One possible exception is non-mCV, or dwarf nova (DM), but the samples were limited \citep{Mu93}.

\subsubsection{Non-Thermal Emission of the GRXE} % section 2.1.3

In the wide band spectra of the LAC on board Ginga,  \citet{Yamas96, Ya97} found a hard X-ray tail over the hot plasma components above 10\,keV  (the non-thermal component) from the Galactic plane in the regions of $l
=\timeform{-20D}$ to $\timeform{40D}$ at $|b| \lesssim\timeform{3D}$.
This non-thermal flux is smoothly extrapolated to the gamma-ray flux in the Galactic plane.

\citet{Va98} made an averaged spectrum from the Galactic plane of $|l|\lesssim\timeform{30D}$ using RXTE.
The averaged spectrum  in the 3\,--\,35 keV band is fitted with a model of thermal plasma with $\sim$2\,--\,3\,keV temperature and a power-law component of the photon index ($\Gamma$) of $\sim$1.8. 
\citet{Va00} re-examined  the ASCA 
data at  $l\sim 28^\circ$.5, and confirmed the presence of the non-thermal emission.
They proposed that the origin of the non-thermal emission is either bremsstrahlung by low energy cosmic-ray (LECR) electrons (LECRe), inverse Compton scattering of ambient microwave, infrared and optical 
photons by the high energy electrons associated to the LECRe, non-thermal emission from SNRs, or discrete X-ray sources.
In the bremsstrahlung origin, the LECRe produce the \Ka~lines at 6.4\,keV (section 5.1), hence the  \FeK~line energy becomes lower than 6.7\,keV due to the mixture of the \Ka~line and the \Hea~line in a hot plasma. This energy down-shift is consistent with the result of Yamauchi and Koyama (1995).

\citet{Va98, Va00} proposed that the continuum shape is the sum of the non-thermal bremsstrahlung and the thermal plasma; the spectrum is a mixture of a high-temperature plasma (HTP) and bremsstrahlung of LECRe. The best-fit temperature of the HTP by this model is reduced to $\sim$2\,--\,3\,keV from the simple model of $\sim$5\,--\,10\,keV temperature with no bremsstrahlung component.
This relaxes the potential difficulty of the production and gravitational confinement of the HTP.  Since the HTP temperature is typical to SNRs, they re-visited the idea of \citet{Ko86b} that the origin of the HTP in the Galactic disk would be multiple SNe. The surface brightness of these SNRs would be too faint to be resolved into individual SNR \citep{Ko86b}.  The SN rate is estimated to be $\lesssim$5\,/\,century, which  is not unreasonably large.
This scenario, however has a serious problem that the $\sim$2\,--\,3\,keV temperature of the HTP is too low to produce strong \Lya~line, detected with ASCA (section 2.1.2).  

\subsection{The Galactic Center X-Ray Emission (GCXE)} % section 2.2

The X-ray observations of the Galactic Center region were started from the Uhuru satellite \citep{Ke71}. 
The early results in 1970's were summarized by \citet{Pr78}.  After 1980's, the Galactic center observations 
have been  made by many instruments:  Einstein \citep{Wa81}, Spacelab-2 \citep{Sk87}, Spartan-1 \citep{Ka88} 
and  ROSAT \citep{Pr94}.  These instruments (authors) found a hint of diffuse extended emission near at the GC, in addition to many point sources.  

As is shown in figure 1, Ginga found  a bright peak at the Galactic 
center (GC) in the \FeK~line  distribution along the Galactic plane \citep{Ko89}. \citet{Ya90} made the  \FeK~line map near at the GC, and found the emission region is an ellipse of
  $\sim\timeform{1.8D}\times\timeform{1.0D}$ size around \SGRA. The \EWFeK~is variable in the range of $\sim$500\,--\,1300\,eV, which would be due to the position variable \EWKa, found later with ASCA.
This is the first concrete result of the presence of diffuse Galactic center emission, and is referred as the 
Galactic Center X-ray Emission (GCXE).  The  surface brightness of the \FeK~line in the GCXE is about 10 times larger than 
that of the  GRXE \citep{Ko89}. The total X-ray luminosity is estimated to be (0.8\,--\,2.3)$\times10^{37}$\Lu.

The X-ray CCD detectors on board ASCA resolved the \FeK~line in the GCXE into \Ka~ (6.40\,keV), \Hea~ (6.68\,keV) and \Lya~ (6.97\,keV) lines, and found  \SiHea, \SiLya, \SHea, \SLya, \ArHea, \ArLya~and \CaHea~lines \citep{Ko96}.  The spectra of the GCXE are fitted with a thermal bremsstrahlung of $\gtrsim10$\, keV temperature plus many Gaussian lines.  These are similar from position to position except the regions of the Sgr\,A and  Sgr\, B MC complexes.  The plasma temperature of $\gtrsim10$\, keV  is unusually high  even for young SNRs. The total luminosity of the GCXE is estimated to be $\sim10^{37}$\Lu. Together with the uniformity over the GCXE,
\citet{Ko96} suggested that the GCXE  is  due to a large scale diffuse plasma with very high temperature. In this case, however, the plasma is very difficult to be confined by the Galactic gravity.  

\citet{Ta00, Tan02} also  examined the ASCA spectrum of the GCXE and those in  the Scutum and Sagittarius ($l\sim$10\arcdeg) regions (GRXE).  
They claimed that the \Hea~and \Lya~lines are significantly broadened to $\sim$80 eV ($1\sigma$), corresponding to a velocity dispersion of a few thousand km s$^{-1}$, higher than the thermal velocity of the $\sim$10 keV temperature plasma. They argued that the boarding is due to charge exchange (CX) of low-energy cosmic-ray irons. The low-energy cosmic-ray origin is consistent with the presence of a non-thermal component in the GRXE (section 2.1.3). They  also obtained the \EWKa, \EWHea~and \EWLya~ to be $\sim$110, $\sim$270 and $\sim$150 eV, respectively. However, these  small \EWKa, \EWHea~and \EWLya~ and the broadenings of the \Hea~and \Lya~lines are later rejected with Suzaku by \citet{Ko07c} (section 3.3).

The \Ka~line is very clumpy with a strong peak at the giant molecular cloud (MC) Sgr\,B2. GRANAT found an elongated hard X-ray emission (8.5\,--\,19\,keV) parallel to the Galactic plane in correlation to the map of MCs \citep{Su93}. 
They suggested that the high energy X-rays come from nearby compact sources (XBs), which are Thomson scattering by a dense molecular gas.  
The scattered X-ray flux is expected to be more than 10\,\% of the observed hard X-ray flux from the GCXE. They proposed that the remaining flux is due to the past X-ray flare of \SGRA. 

The ASCA discovery of the \Ka~ clump from the Sgr\,B2 cloud \citep{Ko96} strongly supports the idea of the past X-ray flare of \SGRA. 
\citet{Mu00} examined the Sgr\,B2 cloud, and found a very peculiar spectrum, a strong emission line at $\sim$6.4 keV, a low-energy cutoff below $\sim$4 keV and a pronounced edge structure at $\sim$7.1 keV. 
The X-ray spectrum and the morphology are well reproduced by a scenario that X-rays from an external source located in the Galactic center direction are scattered by the molecular cloud Sgr B2 and come into our line of sight. They named the \Ka~source at Sgr\,B2 as the X-ray Reflection Nebula (XRN). The 4\,--\,10\,keV band  luminosity of this XRN is $\sim10^{35}$\Lu. 
Soon after, other XRN candidates, \Ka~clumps are found from the Sgr\,A and Sgr\, C MC complexes (section 5.1), and from the other selected regions (section 5.2). 

\subsection {The Galactic Bulge X-Ray Emission (GBXE)} % section 2.3 

In the early studies, \citet{Co69, Pr80, Wa80, Iw82} found an X-ray emission extended to a high Galactic latitude of SH $\gtrsim$~500\,pc\,--\,1.5\,kpc. The emission is also extended to a large Galactic longitude.  

A secure detection of an extended emission near the GCXE with larger SH than those of the GRXE and GCXE was made with Ginga using the \FeK~line distribution.
\citet{Ya93} found largely extended \FeK~lines by $\sim5$\arcdeg~($\sim700$\,pc) above and below the Galactic plane, in addition to a narrow component of the SH$\sim100$\,pc (GRXE).  
The longitude extension is estimated to be $\sim$1.4\,kpc from \SGRA.  This diffuse X-ray emission is named the Galactic Bulge X-ray Emission (GBXE), because of the association to the Galactic bulge region.  This is the third component of the GDXE recognized after the GRXE and GCXE.

Using RXTE, \citet{Va98} examined the flux distribution in the central $l=\pm30^\circ$ of the Galactic  plane in more detail, and  found two components, a thin and broad disks with the e-folding scales  of $\lesssim\timeform{0.5D}$ and $\sim\timeform{4D}$, or the SHs of $\lesssim$70\,pc  and $\sim$500\,pc, respectively. The longitude extension of the later component (SH$\sim$500\,pc) is, however, not constrained. 

\citet{Re03} observed the area of $|l|\lesssim\timeform{10D}$ and $|b|\lesssim\timeform{10D}$ around \SGRA~in the 3\,--\,10\,keV band. They found that the intensity distribution in the  $|b|>\timeform{2D}$ region is well described by an exponential model with the e-folding latitude of $\sim\timeform{3D}$. The e-folding longitude scale is not determined.  
The best-fit spectral parameters of the larger SH component (GBXE) are not significantly different from those of the smaller SH component (GRXE).

\citet{Re06a} examined the longitude and latitude  profiles at $\timeform{3.0D}<|b|<\timeform{3.5D}$, and $\timeform{1D}<|l|< \timeform{4D}$.
The e-folding latitude is $\sim\timeform{2D}$\,--\,$\timeform{3D}$, while the e-folding longitude is $\sim\timeform{8D}$. Thus, they confirmed the two components  proposed by \citet{Ya93}, the bulge/bar (GBXE) and the disk of the Galaxy (GRXE). The GBXE is more largely extended than the GRXE above and below the plane with the total luminosity of $\sim4\times10^{37}$\Lu.
\citet{Re06b} further examined  the two-dimensional distribution in the \FeK~line, from  the central $15\arcdeg\times15\arcdeg$~region around \SGRA (section 2.1.4). 
They found that the linear correlations between near-infrared and the surface brightness of the \FeK~line are similar between the disk and bulge.  Therefore, they  proposed  that the  populations of the  unresolved X-ray point sources in the disk (GRXE) and the bulge (GBXE) are not different with each other.
\citet{Re12} also found two components of SH$\sim$110~pc and $\sim$260~pc.
in the Galactic latitude of $\timeform{-5D} < b <\timeform{0D}$ at $l=\timeform{18.5D}$. 

In summary, Ginga and RXTE discovered the new component GBXE, which has a larger SH than that of the GRXE.  With RXTE, \citet{Re03, Re06a,  Re06b} suggested no difference in the spectra and point source compositions between GRXE and GBXE. However, the limited spectral resolutions prevented reliable study whether or not the spectrum of the GBXE is different from that of GRXE.  Also, the limited spatial resolution prevented to obtain reliable SH of the GBXE (section 2.1.1).  These issues are solved later with Suzaku (section 3). 

\section{Global Structure of the GDXE} %section 3 

This section  overviews the GDXE, the recent results of Chandra, XMM-Newton, Suzaku and partly NuSTAR.  From the spatial distributions of the \FeK, \SHea~and \SLya~lines, the GDXE is clearly decomposed into three separate components, GCXE, GBXE and GRXE (section 3.1). The X-ray luminosity and spectra of these components are given in  section 3.2. The spectral difference among GCXE, GBXE and GRXE verify the decomposition of the GDXE.  
The detailed structure of the central region of the GCXE is discussed in section 3.3.  

\subsection{Decomposition of the GDXE into the GCXE, GBXE and GRXE} 
% section 3.1 
As are given in sections 2.1, 2.2 and 2.3, the GCXE and GBXE are separated from the GDXE. However a clear separation in both the spatial and spectral profiles has to wait the Suzaku satellite.
\citet{Uc13} made Suzaku surveys around the Galactic center regions of $|b|\le\timeform{0.5D}$, $|l|\le\timeform{1D}$, and additional regions of larger $|b|$ and $|l|$ in the Galactic plane.
Since the position of the Galactic center, \SGRA~is at $(l,  b)= (\timeform{-0.056D}, \timeform{-0.046D})$, 
new parameters of the Galactic 
coordinates ($l_*, b_*$) are defined hear and after, shifting $l$ and $b$ by $\timeform{-0.056D}$ and $\timeform{-0.046D}$, respectively.  

They divided the surveyed regions into many rectangles of $\Delta l_* = \timeform{0.1D}$, $\Delta b_* = \timeform {0.2D}$, 
and  analyzed the X-ray spectra from each rectangle. These  spectra are fitted  with a power-law continuum plus many Gaussians 
lines to represent the K$\alpha$, He$\alpha$ and Ly$\alpha$ lines of S, Ar, Ca and Fe.
The best-fit \FeK~(\Ka, \Hea~and \Lya)~fluxes in the positive $l_*$ region near \SGRA~are larger than those in the negative  
$l_*$  region \citep{Uc13}. This asymmetry is mainly due to the bright SNR Sgr\, A East, the Arches cluster  and XRNe
(see sections 4.1 and 5). Therefore, the line and continuum band distributions  of the GDXE are made excluding these bright local spots.
The profiles are shown by the black circles in figure\,\ref{fig:K-line-dis}.

Figure 2 shows the Galactic longitude distributions  of the \FeK, \SHea~and \SLya~line
%%, and the 2\,--\,3\, keV  and 5\,--\,8 keV band 
fluxes have two-exponential components, the small size ($|l_*|\le\timeform{1D}$\,--\,$\timeform{2D}$), and largely extended ($|l_*|\ge\timeform{2D}$\,--\,$\timeform{3D}$) emissions. 
Therefore, \citet{Uc13} fitted the flux distributions by a phenomenological formula, 

%%\scriptsize
\begin{equation}
A_1\exp(-|l_*|/l_1)\exp(-|b_*|/b_1)+A_2\exp(-|l_*|/l_2)\exp(-|b_*|/b_2)
\end{equation}
%%\normalsize
, where the unit of the flux  is
 photons\,s$^{-1}$\,cm$^{-2}$\,arcmin$^{-2}$. 
The best-fit curves of the \FeK, \SHea~and \SLya~lines are given by the dotted lines in figure\,\ref{fig:K-line-dis}, while the best-fit parameters of these lines and those of the 2.3\,--\,5\,keV 
and 5\,--\,8\,keV band fluxes are listed in table\,2. 
The  best-fit e-folding longitude ($l_1$ and $l_2$) clearly indicate that the GDXE has two components:  the small size ($|l_*|\lesssim\timeform{1D}$), and larger size ($|l_*|\ge\timeform{1D}$) emissions. The former and latter are the GCXE and GRXE, respectively. 

\begin{table}[!ht]%Table 2
\label{table:e-fold-l} 
\caption{Fitting results of two-dimension \& exponential function (after \cite{Uc13})$^\ast$.}
\scriptsize
\begin{center}
\begin{tabular}{lcccc}
\hline
Component& $A_1$ & $A_2$ 	& $l_1$ &$l_2$\\
 &\multicolumn{2}{c}{($10^{-7}$photon\,s$^{-1}$\,cm$^{-2}$\,arcmin$^{-2}$)}
 &\multicolumn{2}{c}{(degree)} \\
\hline
\multicolumn{5}{c}{Sulfur(S)}\\
He$\alpha$	&$7.8\pm0.5$	&$0.75\pm0.08$	&$0.58\pm0.05$ 	 &$52\pm13$\\
Ly$\alpha$	&$0.31\pm0.24$	&$0.32\pm0.05$	&0.58$^\dagger$  &52$^\dagger$\\
\hline
\multicolumn{5}{c}{Iron (Fe)}\\
K$\alpha$ 	&$6.2\pm0.6$ 	&$0.21\pm0.09$  &$0.62\pm0.09$   &$57\pm50$\\
He$\alpha$	&$14.6\pm0.5$ 	&$0.91\pm0.08$  &$0.63\pm0.03$   &$45\pm10$\\
Ly$\alpha$	&$6.1\pm0.3$	&$0.18\pm0.04$  &0.63$^\ddagger$ &45$^\ddagger$\\
\hline
\multicolumn{5}{c}{Continuum Band}\\
2.3--5\,keV	&$126\pm15$	&$16\pm1$	&$0.63\pm0.07$	 &$59\pm6$\\
5--8\,keV	&$101\pm11$	&$6.2\pm0.6$ 	&$0.72\pm0.06$ 	 &$52\pm9$\\
\hline
\end{tabular}

 %table 2
\end{center}
\normalsize
{\footnotesize
$\ast$ Errors are 1\,$\sigma$ confidence levels. \\
%$\star$ In unit of $10^{-7}$photons~s$^{-1}$~cm$^{-2}$~arcmin$^{-2}$.\\
%$\ast$  In unit of degree\\
$\dagger$ Fixed to \SHea. \\
$\ddagger$ Fixed to \Hea.\\
% \\
}
\end{table}
\normalsize

As is noted in section 2.3, another component, the GBXE, has been found near the GCXE.
The SH of the GBXE is larger than those of the GCXE and GRXE. Therefore, the e-folding latitudes of the GCXE and GRXE ($b_1$ and $b_2$) in equation (1) would be contaminated by that of the GBXE, while the e-folding longitudes ($l_1$ and $l_2$) are not significantly affected by the GBXE. Therefore, only the parameters of $l_1$ and $l_2$ are listed in table\,2, but $b_1$ and $b_2$ are excluded from the original table given by \citet{Uc13}.

\begin{figure*}[htbp] % Figure 2 Fe, S profile 
\begin{center}
\FigureFile(50mm,30mm){eps-figure/Uc13-Fig2-FeHea.eps} 
\FigureFile(50mm,30mm){eps-figure/Uc13-Fig2-FeHa.eps}
\FigureFile(50mm,30mm){eps-figure/Uc13-Fig3.eps} 
\FigureFile(50mm,30mm){eps-figure/Uc13-Fig2-SHea.eps} 
\FigureFile(50mm,30mm){eps-figure/Uc13-Fig2-SHa.eps} 
\end{center} 
\caption {Upper panel: the longitude  profiles of  \Hea~(left), \Lya~(center), and~\Ka~(right).Lower panel: the longitude profiles  of \SHea~(left) and  \SLya~(right). (From \cite{Uc13})} \label{fig:K-line-dis}  
\end{figure*}

\begin{figure*}[htbp] % Figure 3 Fe b-profile 
\begin{center}
\FigureFile(50mm,30mm){eps-figure/SH67.eps}
\FigureFile(50mm,30mm){eps-figure/SH64.eps}
\FigureFile(50mm,30mm){eps-figure/SH697.eps}
\end{center} 
\caption {The latitude  distributions of  \Hea~(left), \Lya~(center), and  \Ka~(right).(From  \cite{Ya16})} 
\label{fig:K-line-dis-b} \end{figure*}

The coexistence of the \Hea, \Lya, \SHea, \SLya~and \Ka~lines clearly indicates that the GDXE is composed of three components, 
the HTP (for the \Hea~and \Lya~lines), LTP  (for the \SHea~and \SLya~lines) and CG (for the \Ka~line). 
The profiles of figure\,\ref{fig:K-line-dis} (upper panel) shows that 
the flux ratio of \Lya\,/\,\Hea~in the  GCXE ($|l_*|\lesssim1^{\circ}$) is larger than that of the GRXE~($|l_*|\gtrsim1^{\circ}$).
An opposite trend is found in the flux ratio of \SLya\,/\,\SHea~(figure\,\ref{fig:K-line-dis}, lower panel). The intensity profiles of the GCXE and GRXE regions in  the other key elements, Ar and Ca are approximately in between those of S and Fe.

The temperatures of the LTP and HTP are estimated by the intensity ratios of \SLya\,/\,\SHea, and \Lya\,/\,\Hea, respectively. 
Then, the temperature of the LTP in the GCXE is lower than the GRXE, while that of the HTP in the GCXE is higher than the GRXE (see section 3.2).  Thus, the flux distributions of the \Hea, \Lya, \SHea~and \SLya~lines in the GCXE and GRXE given in figure\,\ref{fig:K-line-dis} (upper and lower panels)  clearly demonstrate that the global spectra of the GCXE and GRXE are different with each other.  

\begin{table}[!ht] %table 3
\caption{Parameters of e-folding latitude  of the GCXE, GBXE and GRXE (after \cite{Ya16})$^\ast$}
\label{table:e-fold-b}
\begin{center} 
\begin{tabular}{llcc}
\hline
Region	&Component&Norm($A$)&e-fold $(b_1$, $b_2)^\ddag$ \\
	&&(\dag)&(degree) \\
\hline
GCXE 
&\Ka	&4.1$\pm{0.2}$		&0.22$\pm{0.02}$ 	\\	% 31
&\Hea	&11.9$\pm{0.6}$		&0.26$\pm{0.02}$ 	\\	% 36
&\Lya	&4.9$\pm{0.2}$		&0.24$\pm{0.02}$ 	\\	% 36
&5--8~keV	&77$\pm{4}$		&0.25$\pm{0.02}$ 	\\	% 35
GBXE
&\Ka	&0.31$\pm{0.15}$	&1.15$\pm{0.36}$ 	\\	% 161 	
&\Hea	&1.14$\pm{0.34}$	&2.25$\pm{0.68}$ 	\\	% 316 
&\Lya	&0.40$\pm{0.12}$	&2.13$\pm{0.66}$ 	\\	% 36
&5--8~keV	&12$\pm{2}$		&1.96$\pm{0.25}$ 	\\	% 275 
GRXE$^\star$ 
&\Ka	&0.26$\pm{0.03}$	&0.50$\pm{0.12}$ 		\\	% 70
&\Hea	&0.65$\pm{0.02}$	&1.02$\pm{0.12}$		\\	% 140
&\Lya	&0.09$\pm{0.02}$	&0.71$\pm{0.29}$ 		\\	% 80
&5--8~keV	&5.4$\pm{0.4}$		&1.04$\pm0.20$		\\	% 140
\hline
\end{tabular} 

 %table 3
\end{center} 
{\footnotesize
$\ast$ Errors are 1\,$\sigma$ confidence levels.\\
$\dag$ In unit of $10^{-7}$photons s$^{-1}$cm$^{-2}$~arcmin$^{-2}$.\\
$\ddag$ The parameter $b_1$ is the e-folding latitude of the GCXE, while $b_2$ is that of the GBXE or GRXE.\\
$\star$  Average of the
 $l_*=\timeform{10D}-\timeform{30D}$ and$l_*=\timeform{330D}-\timeform{350D}$ \\
}
\end{table}

\begin{table}[!ht] %table 4
\caption{The properties of  the GCXE, GBXE and GRXE}
\label{table:GCXE-GBXE-GRXE}
\begin{center}
%\footnotesize
\begin{tabular}{lcccc}
\hline
Component &Norm($A$) &\multicolumn{2}{c}{e-folding $^\ddag$} &Luminosity \\
	 &(\dag)   & ($l$\arcdeg) & ($b$\arcdeg)   &(\Lu)\\
\hline
\multicolumn{5}{c}{GCXE} \\ 
\hline
\Ka		&4.1		&0.62 	&0.22 	& $1.5\times10^{34}$\\	
\Hea		&12		&0.63 	&0.26 	& $5.8\times10^{34}$\\	
5--8~keV	&77		&0.72 	&0.25  	& $3.9\times10^{35}$\\
\hline
\multicolumn{5}{c}{GBXE} \\
\hline
\Ka		&0.31		&-	&1.15	&-\\ 	 	
\Hea		&1.1		&10	&2.25	& $7.5\times10^{35}$\\ 	 
5--8~keV	&12		&8	&1.96	& $5.3\times10^{36}$\\	 
\hline
\multicolumn{5}{c}{GRXE} \\ 
\hline
\Ka		&0.36		&57	&0.50 	& $2.9\times10^{35}$\\	
\Hea		&1.0		&45	&1.02	& $1.4\times10^{36}$\\	
5--8~keV	&7.9		&52	&1.04	& $1.2\times10^{37}$\\	
\hline
\end{tabular} 

 %Table 4 
\end{center}
\normalsize
{\footnotesize
$\dag$ The normalization ($A$) of the GCXE and GBXE are taken from \citet{Ya16} (table 3), while  those of the GRXE are converted to the GC position using the e-fold ($l$)  of  \citet{Uc13} (table 2), in unit of $10^{-7}$photons s$^{-1}$cm$^{-2}$~arcmin$^{-2}$.\\
$\ddag$ The e-folding scales ($b$) are  taken from \citet{Ya16} (table 3), while the e-folding longitude ($l$) for the GCXE and GRXE are taken from \citet{Uc13} (table 2), and those for the GBXE are from  Yamauchi and Koyama (1993), and \citet{Re06a}.\\  
%\dag~ In unit of $10^{-7}$photons s$^{-1}$cm$^{-2}$~arcmin$^{-2}$.\\
}
\end{table}

\citet{Ya16} separately estimated the e-folding latitudes ($b_1$ and $b_2$) using all the Suzaku archive data along and near the Galactic inner disk
 ($|b_*|\le\timeform{3D}$, $|l_*|\le\timeform{30D}$). 
To increase statistics, the data are grouped according to the positions of (a):~$|l_*|\le\timeform{0.5D}$, 
(b):~$l_*=\timeform{358.5D}$, 
(c):~$l_*=\timeform{330D}$\,--\,$\timeform{350D}$, and
(d):~$l_*=\timeform{10D}$\,--\,$\timeform{30D}$.
The position (a) includes mainly the GCXE and some fractions of the GBXE, while the main component in the position (b) is the GBXE 
with small fractions of the GCXE. The data in the positions (c) and (d) are from the GRXE. 

The intensity profile perpendicular to the Galactic plane in the 5\,--\,8\,keV band and \FeK~line fluxes are made for the regions of (a)\,--\,(d). The profiles near potitoin (a) and (b) show  two component shape.  As examples, the \FeK~lines profiles at the region of (a) are shown in figure\,\ref{fig:K-line-dis-b}. 
In order to make clear the two-component  structure,
  the profiles in the regions (a) and (b) are simultaneously fitted by a 2-exponatial model of,
\begin{equation}
A_1\exp(-|b_*|/b_1)+A_2\exp(-|b_*|/b_2)
\end{equation}
, where the subscripts 1 and 2 represented the GCXE and GBXE, respectively.
The free parameters of respective normalizations ($A_1$ and $A_2$) and the e-folding latitudes ($b_1$ and $b_2$) for the GCXE and GBXE,
are linked in the (a) and (b) regions.
The best-fit profiles of the \FeK~lines at the position (a) are shown by the two solid lines in figure\,\ref{fig:K-line-dis-b}.

On the other hand, the profiles of the (c) and (d) regions show one-exponential shape , and hence fitted with 1-exponatial model of, 
\begin{equation}
A\exp(-|b_*|/b_2) \end{equation}
, where the e-folding latitudes ($b_2$) in the (c) and (d) regions  are linked. 
The best-fit e-folding latitude of $b_1$ and $b_2$ for the \FeK~and the 5\,--\,8 keV band fluxes in the GCXE, GBXE and GRXE are listed in table\,3.  

The regions and luminosity of the \Hea~and \Ka~lines, and the 5-8 keV band in the GCXE, GBXE and GRXE are determined from the e-folding scales of the \Hea~and \Ka~lines, where the longitude  and latitude  scales are taken from tables 2 and 3, respectively. The longitude scale for the GBXE is unclear, and hence taken from the old data of Yamauchi and Koyama (1993), and \citet{Re06a}. The normalization ($A$) at $b_*$=0  are taken from table 3, where those of the GRXE are calculated from the e-folding longitude ($l_2$) in table 2.  Thus determined  regions and luminosity  of the \Hea~and \Ka~lines, and those of the 5\,--\,8 keV band, in the GCXE, GBXE and GRXE are summarized in table\,\ref{table:GCXE-GBXE-GRXE}.
Hear and after, the regions of the GCXE, GBXE, and GRXE are followed from table\,\ref{table:GCXE-GBXE-GRXE}. The study of the GCXE, GBXE and GRXE spectra are given in the next section (section 3.2).

\subsection{X-Ray Spectra and Luminosity  of  the GCXE, GBXE and GRXE} \label{sec:GCXEspec} % section 3.2

\begin{table*}[!ht]% table 5  
\caption{The best-fit parameters of the GCXE, GBXE and GRXE spectra (after \cite{No16}).$^\ast$}
\label{tab:GDXE}
\begin{center}
\begin{tabular}{lccccccc}
       \hline 
&&\multicolumn{2}{c}{GCXE} &\multicolumn{2}{c}{GBXE}&\multicolumn{2}{c}{GRXE}\\
\hline
\multicolumn{8}{c}{Continuum}\\
\multicolumn{2}{l}{\NH~($10^{22}$~cm$^{-2}$)} &\multicolumn{2}{c}{6 (fix)} &\multicolumn{2}{c}{3 (fix)} &\multicolumn{2}{c}{3 (fix)}\\
\multicolumn{2}{l}{Fe K edge$^\dag$} &\multicolumn{2}{c}{$0.24\pm0.01$} &\multicolumn{2}{c}{0 (fix)} &\multicolumn{2}{c}{0 (fix)}\\
\multicolumn{2}{l}{$kT_{\rm e}$ (keV)}&\multicolumn{2}{c}{$14.9^{+0.5}_{-0.6}$} &\multicolumn{2}{c}{5.1$\pm{0.4}$}&\multicolumn{2}{c}{5.0$\pm{0.4}$}\\
\hline
			\multicolumn{8}{c}{Emission lines}\\
Line$^\ddag$	&CE$^\S$  	& Flux$^\star$ &EW$^\S$			&Flux$^\star$ &EW$^\S$		&Flux$^\star$ &EW$^\S$\\
  	\hline
\Ka	&6400  	&3.54$\pm{0.04}$&175$\pm2$ &0.14$\pm{0.02}$&84$\pm10$ &0.16$\pm{0.01}$&118$\pm9$\\
\Hea	&6680	&9.40$\pm{0.05}$&500$\pm3$ &0.70$\pm{0.02}$&463$\pm13$&0.60$\pm{0.02}$&487$\pm13$\\
\Lya    &6966	&3.45$\pm{0.04}$&198$\pm2$ &0.24$\pm{0.02}$&173$\pm13$&0.10$\pm{0.01}$&96$\pm11$\\
\Kb	&7059       &0.44$^\#$	         &26	  &0.01$^\#$	     &14	   &0.02$^\#$           &19\\
\hline
$\chi ^2$/d.o.f.  &   & \multicolumn{2}{c}{331/265}           & \multicolumn{2}{c}{117/84}     & \multicolumn{2}{c}{107/72} \\
\hline
\end{tabular}

 %table 5
\end{center}
{\footnotesize
$\ast$ Errors are 1\,$\sigma$ confidence levels.\\
$\dag$  Absorption depth at 7.11 keV.\\
$\ddag$ ATOMDB~3.0.2 (http://www.atomdb.org/) and Wargelin et~al. (2005).\\
$\S$  Units of CE (line Center Energy) and EW are electron bolt (eV), respctively \\
$\star$  Flux is in unit of $10^{-7}$~photon~cm$^{-2}$~s$^{-1}$~arcmin$^{-2}$.\\
$\#$ Fixed to 0.125$\times$\Ka.\\
}
\end{table*}

The Suzaku spectra in the selected regions of the GCXE and GRXE have been made by several authors \citep{Ko07c, Eb08, Yu08, Yamau09, He13a}.  \citet{Uc13} made the whole spectra of the GCXE and GRXE.  However the regions of the GRXE are limited, and  would be contaminated by the GBXE. The GCXE , on the other hand, is fur bright, and  hence contamination from the GBXE is ignored. The spectrum of the GCXE is fitted with 2-CIE and one power-law model, which are associated with
the highly ionized  atomic lines and the \Ka~line, respectively. The best fit temperatures are $\sim$0.95\,keV and $\sim$7.5\,keV with the iron abundance of $\sim$1.25 solar, while the best-fit photon index and \EWKa~are $\sim$2.1 and $\sim$0.46\,keV, respectively. 

\citet{No16} made the global spectra of the GCXE, GRXE and GBXE in the 5\,--\,10 keV band from the regions of ($|l_*|<\timeform{0.6D}, |b_*|<\timeform{0.25D}$), 
~($|l_*|<\timeform{0.6D},\timeform{1.0D}<|b_*|<\timeform{3.0D}$), and
~($|l_*|=\timeform{10D}$\,--\,$\timeform{30D}$, $|b_*|<\timeform{1.0D}$),
respectively (following table 4).  For the GBXE spectrum, the overlapping region to the GCXE 
($|b_*|< \timeform{1.0D}$) is excluded, to avoid a large contamination from the GCXE spectrum.  For the GCXE spectrum, the bright spots of the \Ka~and \Hea~lines are excluded (see sections 4 and 5). 
They fitted the GCXE, GBXE and GRXE spectra with a model of a bremsstrahlung plus Gaussian lines at 6.40\,keV, 6.68\,keV and 6.97\,keV (\Ka, \Hea~ and \Lya, respectively), including  the other faint lines.  The best-fit results are listed in table\,\ref{tab:GDXE}.

The flux ratio of \Lya\,/\,\Hea~of the GCXE, GBXE and GRXE are $\sim$0.37,  $\sim$0.34 and  $\sim$0.17, which correspond to the CIE temperatures of $\sim6.8$\,keV,  $\sim6.5$\,keV and $\sim5.0$\,keV, respectively.  Thus the plasma temperatures of the GCXE, GBXE  and GRXE determined by the line flux ratio of \Lya\,/\,\Hea~are not largely difference with each other. However the continuum shape (bremsstrahlung) of the GCXE gives the temperature of $\sim$15\,keV, which is significantly larger than those of the  GBXE and GRXE of $\sim$5\,keV (table\,\ref{tab:GDXE}). 
The reason of this apparent inconsistency in the temperatures is found in the different flux ratio  of  \Ka~/~\Hea. The flux ratio \Ka~/~\Hea~in the GCXE is $\sim$0.38, which is significantly
larger than those of the GBXE and GRXE of $\sim$0.20  and  $\sim$0.27, respectively.  This indicates that the hard X-rays of power-raw spectrum  of the CG occupy larger fraction in the GCXE, and hence gives apparently higher bremsstrahlung temperature than those of the GBXE and GRXE. 

Using the parameters of the GCXE, GBXE and GRXE regions in table\,\ref{table:GCXE-GBXE-GRXE} and table\,\ref{tab:GDXE}, 
the total X-ray fluxes of the GCXE, GBXE  and GRXE  in the 5\,--\,8 keV band are estimated, which are also given in table\,\ref{table:GCXE-GBXE-GRXE}.
The 5\,--\,8 keV band luminosity of $\sim3.9\times10^{35}$\Lu, $\sim5.3\times10^{36}$\Lu~and  $\sim1.2\times10^{37}$\Lu~are converted to the 2\,--\,10 keV band luminosity of $\sim1.2\times10^{36}$\Lu,  $\sim1.6\times10^{37}$\Lu~ and $\sim3.8\times10^{37}$\Lu, for the GCXE, GBXE  and GRXE, respectively. These are smaller than the previous reports, due to the smaller e-folding longitude and latitude scales of the GCXE, GBXE and GRXE. However, the quality of the spectra of table\,\ref{tab:GDXE} is the best, in particular, mutual mixing of the GCXE, GBXE and GRXE spectra are minimized.

\subsection{Iron K-shell Line Property of the Central Region of the  GCXE} % section 3.3

\citet{Ko07c} studied the hard X-ray spectrum in the 5\,--\,10 keV band of the central GCXE region of $|l_*|< \timeform{0.3D}, |b_*|<\timeform{0.15D}$. They made the X-ray spectrum excluding the Sgr\,A East SNR (section 4.1.1), but including the bright XRNe at the northeast of \SGRA~ (section 5.2.2). This spectrum is fitted by the same model of section 3.2, a phenomenological model of a bremsstrahlung continuum plus many Gaussian lines. 
The best-fit spectral parameters are given in table 6.  

The \Hea~line is a blend of the resonance, inter-combination and forbidden lines. 
The mixing ratios of these lines depend on the plasma nature such as charge exchange (CX).  Depending on the plasma temperature, satellite lines from iron of less ionized  than He-like iron, such as the dielectronic recombination lines may be contained \citep{Be92,Be93}. All these processes shift the nominal \Hea~energy of 6.68\,keV to a lower energy; the line center of the \Hea~ produced by the CX process is $6666\pm 5$\,eV \citep{War05}. The observed center energy and width of the \Hea~line are  6680$\pm$1\,eV and  $\sim$40\,eV, respectively, consistent with the proper mixing ratio of the resonance, inter-combination and forbidden lines in the normal CIE plasma of $\sim$6\,--\,7\,keV temperature. 

The flux ratio of \Lya\,/\,Hea~is $\sim$0.33, which corresponds  to the ionization temperature of $\sim$6\,--\,7\,keV. 
The flux ratio of \Heb\,/\,\Hea~is $\sim$0.1, which gives the electron temperature of $\sim$6\,--\,7\,keV. 
Thus, the center energy and width of the \Hea~line, and the flux ratios of 
\Lya~and \Heb~lines relative to the \Hea~line (\Lya\,/\,\Hea~and \Heb\,/\,\Hea) favor a CIE plasma of $\sim$6\,--\,7\,keV temperature for the HTP. 

Then, \citet{Ko07c} fitted the GCXE spectrum with a model of CIE plasma.
The best-fit temperature and iron abundance are consistent with those of \citet{Uc13} and \citet{No16} in the whole GCXE area of $|l_*|<\timeform{0.6D}, |b_*|<\timeform{0.25D}$ (see sections 3.1 and 3.2).  This indicates that the HTP spectrum in the central GCXE region of $|l_*|< \timeform{0.3D}, |b_*|<\timeform{0.15D}$ is nearly the same as the whole GCXE;  no significant variation of the HTP is found over the  whole GCXE region.
 
\begin{table}[!ht] %% Table 6  
\caption{The best-fit parameters of the GCXE spectrum (after \cite{Ko07c})$^\ast$.}
\label{tabl:lines}
\begin{center}
\begin{tabular}{lccc}
\hline
       \multicolumn{4}{c}{Continuum}\\
       \hline
$N_{\rm H}$ (cm$^{-2}$) &\multicolumn{3}{c}{6$\times10^{22}$}\\
$N_{\rm Fe}$ (cm$^{-2}$) &\multicolumn{3}{c}{9.7$^{+0.7}_{-0.4}\times 10^{18}$}\\
$kT_{\rm e}$ (keV)	    &\multicolumn{3}{c}{15$^{+2}_{-1}$}\\
       \hline	
       \multicolumn{4}{c}{Emission lines}\\
       \hline
 Identification	&Center Energy 	& Width  & Intensity \\
       		&(eV) 	&(eV)  &(\ddag)\\
	\hline
Fe\emissiontype{I}$^\dag$~K$\alpha$
     &6409$\pm$1 	&33$^{+2}_{-4}$  &43.2$^{+0.5}_{-0.8}$ \\
Fe\emissiontype{XXV}~He$\alpha$&6680$\pm$1&39$\pm$2    &51.0$^{+0.8}_{-0.6}$\\
Fe\emissiontype{XXVI}~Ly$\alpha$       
   &6969$^{+6}_{-3}$	&15$^{+8}_{-15}$	&16.6$^{+0.9}_{-1.1}$\\
Fe\emissiontype{I}$^\dag$~K$\beta$ &7069$^*$      	& 38$^\S$     	 & 6.91$^{+1.12}_{-0.96}$\\
Ni\emissiontype{I}$^\dag$~K$\alpha$       
     &7490$^{+12}_{-14}$    & 0 ($<$28)    &3.05$^{+0.73}_{-0.57}$\\
 Ni\emissiontype{XXVII}~He$\alpha$       
   &7781$^{+24}_{-31}$      & 39$^|$       & 3.97$^{+1.06}_{-0.65}$\\
Fe\emissiontype{XXV}~He$\beta$ 
     &7891$^\#$            & 30 (fixed)   & 4.69$^{+0.81}_{-0.61}$\\
Fe\emissiontype{XXVI}~Ly$\beta$ &8220$^{+31}_{-22}$    & 30 (fixed)   & 2.29$^{+1.35}_{-1.31}$\\
Fe\emissiontype{XXV}~He$\gamma$      
    &8264$^{**}$           & 30 (fixed)   & 3.08$^{+1.32}_{-1.34}$\\
Fe\emissiontype{XXVI}~Ly$\gamma$ 
    & 8681$^{+33}_{-32}$      & 0 ($<$91)    & 1.77$^{+0.62}_{-0.56}$\\
\hline
\end{tabular}
% table 6
\end{center}
{\footnotesize
$\ast$  Errors are 1$\sigma$ confidence levels.\\
$\dag$  Neutral or low ionization state.\\ 
$\ddag$ In unit of $10^{-5}$~photons~s$^{-1}$~cm$^{-2}$.\\
$\star$ Fixed to 1.103~$\times E$ (Fe\emissiontype{I}~K$\alpha$).\\ 
$\S$ Fixed to 1.103~$\times~\sigma$ (Fe\emissiontype{I}~K$\alpha$).\\
$|$  Fixed to $\sigma$ (Fe\emissiontype{XXV}~He$\alpha$). \\
$\#$  Fixed to $110+ E$ (Ni\emissiontype{XXVII}~He$\alpha$). \\
${**}$  Fixed to $44+E$ (Fe\emissiontype{XXVI}~Ly$\beta$).\\ 
}
\end{table}

\begin{figure}[!h]
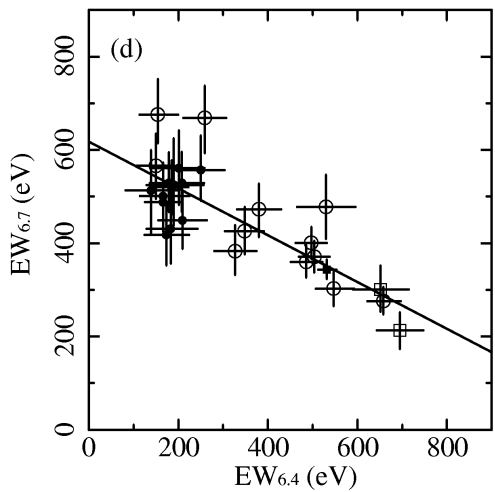
% figure 4
  \begin{center}
  \FigureFile(70mm,70mm){eps-figure/Ko09-Fig2d.eps} 
\end{center}
\caption{The scarred plots of \EWHea~ and \EWKa. The solid line shows the best-fit relation of \EWKa~+2$\times$\EWHea$\simeq$1.2\,keV.
The open and filled circles are data from the Galactic east and west fields, respectively.  
The data from bright \Ka~regions near the Radio arc are plotted by the open squares (From \cite{Ko09}).}
\label{fig:Ko09-Fig2}
\end{figure}

\citet{Ko09} made correlation plot of the \EWHea~and \EWKa~taken from  small areas of $\timeform{4.5'} \times \timeform{4.5'}$.  The correlation plots  are shown in figure\,\ref{fig:Ko09-Fig2}. The solid line shows the correlation function, which is \EWKa\,+\,2$\times$\EWHea$\simeq$1.2\,keV. 
The scattered correlation plots indicate that the mixing ratio of the HTP (\Hea) and CG (\Ka) are different from position to position.  
In the extreme case, where the \EWHea~is 0\,keV (HTP = 0), the \EWKa~is $\sim 1.2$\,keV, or the CG  has  the \EWKa~of $\sim 1.2$\,keV.  
This large \EWKa~ favors that the origin of the \Ka~line is due to the irradiation of external X-ray sources or low energy cosmic ray proton (LECRp) on MCs (see section 5). 
The \EWHea~is $\sim$0.6 keV at \EWKa ~=  0\,keV (CG = 0). This value is consistent with \citet{Ko07c} that the HTP has a CIE spectrum with temperature and  abundances of $\sim$6\,--\,7\,keV and nearly one solar, respectively.  

Figure\,\ref{fig:Ko09-Fig2} shows that the scatter of the \EWKa~ is  larger than that of the \EWHea, which  means that the 
\Ka~flux (CG) is  more clumpy than the \Hea~flux (HTP). 
The \EWKa~and \EWHea~of the west region of \SGRA~(GC west), where free from the bright \Ka~and \Hea~spots,  are given by the filled circles in figure\,\ref{fig:Ko09-Fig2}.
The \EWKa~and \EWHea~ are concentrated in the small parameter space of EW$_{6.7}\sim$510\,eV and EW$_{6.4}\sim180$\,eV, which are consistent with those of the  whole GCXE (table\,\ref{tab:GDXE}), but are significantly larger than the ASCA results of \citep{Tan02}.  Since Suzaku had a larger correcting area and better spectral resolution, the results of Suzaku would be more reliable than ASCA. 
%%In fact, all the Suzaku results of the \EWFeK~are consistent among different authors with different data reduction and analysis (e.g. GC-west:\cite{Ko09}; GC1 and GC2:\cite{Yamau09}; GCXE: \cite{No16}).%%%%%%%

The large and scattered \EWKa~of the GC east (open circles in figure \ref {fig:Ko09-Fig2}) are due to the XRNe (open squares) and\,/\,or the \Ka~line contamination of the surrounding XRNe. 
The XRNe are associated with non-thermal power-law emissions (section 5.2.2).
\citet{Yu08} discovered  a power-law emission from the GCXE with the photon index of $\sim$2. Most of them would be due to the XRNe, because  \citet{Mo15} found 
hard X-ray excesses in the 10\,--\,20\,keV band with NuSATR,  at the positions of the XRNe in the Sgr\,A complex, MC1, MC2, Bridge, and G0.11$-$0.11  and  another \Ka~source,  the Arches cluster (see  tables 9 and 11). The photon index of  MC1 and Bridge are $\sim$2.2 and $\sim$1.8, respectively, similar to the power-law emission of \citet{Yu08}.

 In the close vicinity of \SGRA, the region of  $4\arcmin$\,--\,$13\arcmin$~ from  \SGRA, using XMM-Newton, \citet{He13a} found that the \EWKa, \EWHea, and \EWLya~are $\sim$220\,eV, $\sim$730\,eV, and $\sim$320\,eV, respectively, which are 1.3\,--\,1.5 times larger than those of the whole GCXE region \citep{No16}.
The larger \EWFeK~ would be due to a lower continuum flux than \cite{No16};
\citet{He13a} assumed a 7.5\, keV- plasma for the continuum flux, while
\citet{No16} included  a power-law component associated to the \Ka~line, with the best-fit bremsstrahlung temperature of $\sim$15\, keV, significantly higher than 7.5\,keV.  

\citet{Uc13}  more explicitly fitted the GCXE spectra with a model of 7.3\, keV-plasma and power-law of $\Gamma =1.4$.
The \EWKa~and \EWHea~in the GC west \citep{Ko09} (filled circles in figure 4) are $\sim$180\,eV and  $\sim$510\,eV, respectively. These are  consistent with \citet{No16}.

\citet{Mu04a} made the Chandra X-ray spectrum in the region of  $<9\arcmin$ from \SGRA. They fitted the spectrum with a 2-CIE model, and found that the best-fit temperatures are $\sim$0.81\,keV and $\sim$7.7\,keV. The best-fit Fe abundance is $\sim$0.7 solar. However their model did not include power-law component. % (table 3)
Adding a power-law component of the same flux of \citet{Uc13}, the Fe abundance is revised to be $\sim$1.1 solar, in agreement with \citet{Uc13}.  

In the $3\arcmin$.5\,--\,$5\arcmin$~ring around \SGRA, Uchiyama et al. (2017) analyzed the Suzaku spectrum.  
In order to take into account of overflow flux from the bright SNR Sgr A East, they did simultaneous fit for the GCXE and Sgr A East.  The best-fit \EWKa, \EWHea~and \EWLya~in the GCXE are $\sim$160 eV, $\sim$520 eV and $\sim$190 eV, respectively, in good agreement with \citet{No16} for the whole GCXE spectrum (table\,\ref{tab:GDXE}).

In summary, in spite of significant enhancement of the GCXE fluxes near at \SGRA (section 9.3), the \EWHea~and \EWLya ~are nearly the same in the whole GCXE regions. 
The \EWKa~is highly variable from position to position in the GCXE. 

\section{Local Diffuse Hot Plasma}\label{sec:LHP}  %section 4

This section reviews local hot plasmas in the hard X-ray band, or 
\Hea~emitting plasmas (section 4.1) and in the soft X-ray band, or
\SHea~emitting plasmas (section 4.2).  The relevant region is $|l|\lesssim$1\arcdeg~and $|b|\lesssim$0\arcdeg.5, which includes the full GCXE region (section 3.1). 

\subsection{Hot Plasmas with the \Hea~Line (HTP)} % section 4.1
\label{sec:HP}

This section reports individual local hot plasma with the \Hea~line. The 2\,--\,10\,keV band luminosity of Sgr A East is  $\sim10^{35}$\Lu~, $\lesssim$10\,\% of the GCXE, while that of sum of the other hot diffuse sources  with the \Hea~line is $\lesssim$~$6\times10^{34}$\Lu ~(2\,--\,10\,keV), only $\lesssim$~5\,\% of the GCXE luminosity. 

%%%%%%%%%%%%%%%%%%%%%%%%%%
\subsubsection{Sgr\,A East, the Brightest SNR in the GCXE Region}
 \label{sec:SgrAEast} %section 4.1.1

Sgr\,A East is a  non-thermal radio shell with the size of $\timeform{3.6'} \times \timeform{2.7'}$ \citep{Ek83}. Within 
20\,pc ($\sim \timeform{10'}$) from \SGRA, Sgr\,A East is a unique object surely identified as a young SNR.  
\citet{Ro01} found diffuse hard X-ray emissions from the position of Sgr\,A East with ASCA.
Then, \citet{Sa02} detected K-shell lines from highly ionized Si, S, Ar, Ca and Fe, and established that Sgr\,A East has a thermal plasma. 

\citet{Sa04} made the XMM-Newton spectra from the three annual regions of Sgr\,A East, the regions of radius $r<28\arcsec$ (Center), 
$28\arcsec<r< 60\arcsec$ (Middle), and $r>60\arcsec$ (Outer).
The spectra are fitted with  two thermal plasmas 
of $\sim$0.9\,keV and $\sim$3\,keV (Center), 
$\sim1$\,keV and $\sim$5.5\,keV (Middle), and   
$\sim0.9$\,keV and $\sim$4.4\,keV temperatures (Outer). 
The highly ionized  Fe-rich ejecta shows  significant concentration  towards the center; the Fe abundances are $\sim$4\,solar, $\sim$1.5\,solar  and  $\sim$0.5\,solar in  Center, Middle and Outer, respectively. 
The other elements, S, Ar and Ca are roughly uniformly distributed in the range of $\sim$1\,--\,3\,solar.  They interpreted that Sgr\,A East is a young  SNR of either Type Ia (Ia-SNR) or Core-Collapsed SNR (CC-SNR) of a relatively low mass progenitor star.  

With Chandra, \citet{Ma02} found a  diffuse  plasma
of $\sim$2\,keV temperature and abundances of $\sim$4\,solar inside the radio shell. They proposed that Sgr\,A East is a CC-SNR with the progenitor star mass of $\sim$13\,--\,20$M_\odot$, and that Sgr\,A East is a member of young mixed-morphology (MM-SNR) \citep{Rh98}.
\citet{Pa05} confirmed the center-filled X-ray structure. They made X-ray spectra from the three regions, named  Center, North and Plume. 
The spectra are fitted with two-thermal plasmas of $\sim 1$\,keV and $\sim$5\,keV temperatures in Center, but in North, the temperatures are $\sim 1$\,keV and $\sim$11\,keV.  The Fe abundance shows clear concentration toward the SNR center, from $\sim$1\,solar (Plume) and $\sim$2.5\,solar (North), to $\sim$6\,solar (Center). The iron mass of Sgr\,A East is estimated to be $\lesssim0.27M_\odot$.
The abundances of lighter elements, S, Ar and Ca are roughly 1\,solar, but are higher in  Plume and North than  Center. Therefore, the plasma in North and Plume  would be the shock-heated ISM. 

\citet{Pa05} found a hard point-like source, CXOGC J174545.5$-$285829 (Cannonball) at the northern edge of the SNR. It has a power-law  spectrum of index $\sim$1.6, which is typical to a non-thermal synchrotron emission from a NS magnetosphere.  The absorption ($N_{\rm H}$) is $\sim17\times10^{22}$\,cm$^{-2}$, similar to that of Sgr\,A East of 13\,--\,19$\times10^{22}$\,cm$^{-2}$, and hence  Cannonball is associated to the Sgr\,A East SNR at the same distance. 
Then, the  X-ray luminosity is estimated to be $\sim 3 \times 10^{33}$\Lu~(2\,--\,10 keV), which is  typical to a pulsar and\,/\,or pulsar wind nebula (PWN). From these facts, they suggested that Sgr\,A East is a CC-SNR, and  Cannonball is a high-velocity NS born in the CC-SN.

\citet{Ko07b} obtained a high quality X-ray spectrum of Sgr\,A East with Suzaku. They discovered  many K-shell emission lines 
from highly ionized atoms, which are
\SHea, \SLya,
S\emissiontype{XV}-He$\beta$, Ar\emissiontype{XVII}-He$\alpha$,
Ar\emissiontype{XVIII}-Ly$\alpha$, Ar\emissiontype{XVII}-He$\beta$,
Ca\emissiontype{XIX}-He$\alpha$, 
\Hea, \Lya, \Heb+Ni\emissiontype{XXVII}-He$\alpha$ and \Hec+\Lyb. 
The flux ratios of these lines indicate that Sgr\,A East  has, at least two thermal plasmas. With the 2-temperature CIE model fit, the plasmas temperature  are  found to be $\sim1.2$\,keV and $\sim$6.0\,keV, the mean abundance of Fe is$\sim$ 2.6\,solar, while the other elements are $\sim$1\,solar. The total and iron masses are  $\sim$27$M_\odot$ and $\sim$0.15$M_\odot$, respectively, consistent with a CC-SN origin.  A hint of the Mn\emissiontype{XXIV}-He$\alpha$~line is found at 6.1\,keV, but no hint of the Cr\emissiontype{XXIII}-He$\alpha$~line at 5.6\,keV is found in spite of larger abundance than Mn.

In addition to the two CIE plasmas, a non-thermal component is found, which occupies the major fraction above $\sim$7\,keV. \citet{Pe15} and \citet{Mo15} found strong hard X-rays in the 10\,--\,40\,keV band with NuSTAR from the regions including some fractions of Sgr\,A East. The flux at the \Hea~lines is nearly comparable to the thermal emissions, which roughly agree to the power-law component of \citet{Ko07b}.
\citet{Pe15} proposed that the origin of the power-law component is many faint mCVs in the region of Sgr\,A East. In this case, significant \Ka~line flux should be observed, because mCVs are strong \Ka~emitter (table 14). However no hint of strong \Ka~line is found from Sgr\, A East.  

Another possibility of the power-law component is assembly of non-thermal filaments listed in table 12 \citep{Mu08} plus unresolved non-thermal sources. These would be due to synchrotron emissions by HECRe accelerated by a shock wave of Sgr\,A East. 
The  LECRe and possibly LECRp,  may ionize the Fe\emissiontype{XXV} to higher ions of Fe\emissiontype{XXVI}, hence emit extra \Lya~lines. 
This may lead to the  large flux ratio of \Lya\,/\,\Hea~of $\sim$0.05,
corresponding to the plasma temperature of $\sim$4\,--\,6\,keV, an unusually high among any known SNRs. The center energy of the \Hea~at $\sim$6.65\,keV is also the highest among the normal CC-SNRs.

  \citet{Uc17} re-analyzed the Suzaku spectrum of Sgr A East spectrum  by the simultaneous  fitting with the GCXE spectrum of nearby sky (see section 3.3). They found  a hint of recombining plasma (RP) in Sgr\,A East. These very high temperature plasma of $\sim$4\,--\,6\,keV,  RP and strong power-law component are  unusual  even in the normal CC-SNRs, where presence of circumstellar matter (or MC), a possible origin to make a RP, is more likely than Type Ia SNRs.  These unusual structures  would be related to extreme environments at the Galactic center (GC) region.

%%%%%%%%%%%%%%
Sgr\,A East is an SNR of $\sim10^{3}$ years age, and is located  in the close vicinity of \SGRA. \citet{Ma02} predicted  that the dust\,/\,molecular ridge was compressed by the forward shock of the SN Sgr\,A East. When the blast wave passed over the black hole \SGRA, the compressed dense dust and gas had accreted onto \SGRA, and produced X-ray flares  in $\sim$ a few 100\,years ago (sections 5.2). 

\citet{To06} proposed another idea that the mean accretion rate onto \SGRA~during the past $\sim10^7$\,yr had been extremely higher than the current rate.
The accretion energy in the past was sufficient 
to produce and keep the HTP gas in the GCXE. Also a significant amount of positrons should had been created, which might produce the observed 511 keV annihilation line  from the Galactic bulge.
After the passage of the blast wave of the SN Sgr\,A East in  $\sim$ a few 100\,years ago, the ambient gas had been cleaned-up, leading  \SGRA~ to the  present quiet level (section 7.1). 

Sgr\,A East is the brightest \Hea~source in the GCXE region. The 2\,--\,10 keV band  luminosity is $\sim10^{35}$\Lu~\citep{Ma02, Sa04, Ko07b}, this 
is only  $\sim$8\,\%  of the GCXE.

 %figure 7 {eps-figure/Ko07-SgrA-Fig5.eps} %figure 7
\subsubsection{G0.61+0.01} %section 4.1.2

Although the \Hea~line is smoothly distributed over the Sgr\,B region. \citet{Ko07a} found  a local excess of $\sim5\arcmin\times2\arcmin.4$ size with Suzaku. The position of the  center is  $ (l, b) = (\timeform{0.61D}, \timeform{0.01D}$), and  named Suzaku J1747.0$-$2824.5 (G0.61+0.01). 
The deep XMM-Newton image shows a hint of very weak  enhancement near this position in the low energy band of 2\,--\,4.5\,keV \citep{Po15}.  
This source is, therefore very peculiar dominated mainly in the \Hea~line. 

The X-ray spectrum is fitted with an IP (NEI) plasma model  \citep{Ko07a}. The best-fit plasma temperature, ionization time scale and iron abundance are $\sim 3$\,keV,
$n_{\rm e}t\sim2\times10^{11}$ s~cm$^{-3}$, and
 $5.1^{+1.2}_{-1.1}$\,solar, respectively.
Then the dynamical age and ionization age are estimated to be 
$\sim4\times10^{3}$\,yr and $\sim7\times 10^{3}$\,yr, respectively. 
From these results,  G0.61+0.01 is likely a new  ejecta  dominant type Ia SNR. The absorption column density $N_{\rm H}$ is $\sim1.6\times10^{23}$\,cm$^{-2}$, and hence this SNR would be behind the Sgr\,B MC complex. 
Assuming the distance of 8\,kpc, the plasma mass is estimated to be $\sim1.3M_\odot$, a bit smaller than the typical ejecta mass of Ia SNR. 

Faint emissions are extending around G0.61+0.01.  The south of this emission would be another SNR G0.57$-$0.001 (see section 4.1.3), but the northeast emission would be a part of G0.61$+$0.01. In this case, the whole plasma of G0.61$+$0.01 may be a bit larger than $\sim1.3M_\odot$, consistent with an ejecta of young Ia SNR. The X-ray luminosity of G0.61+0.01, including the faint emission at the northeast, would be at most  $\sim2\times 10^{34}$\Lu~(2\,--\,10\,keV), only $\sim1$\,\% of the GCXE flux 

%figure 8 {eps-figure/Ko07-SgrB-Fig35.eps}  figure 8

\subsubsection{G0.570$-$0.018 and G0.570$-$0.001} %section 4.1.3

\citet{Se02} found a faint X-ray emission in the hard X-ray band images of Chandra and ASCA. The Chandra position is  $(l, b) = (\timeform{0.570D}, \timeform{0.018D}$), hence named  G0.570$-$0.018\,/\, CXO\,J174702.6$-$282733. 
They fit the Chandra spectrum with a phenomenological model of a thermal bremsstrahlung plus a Gaussian line at $\sim$6.5\,keV (\FeK~line), and obtained  the  center energy and \EWFeK~to be $6.5\pm{0.03}$\,keV and $4.1^{+1.4}_{-1.0}$\,keV, respectively. These are consistent with the ASCA results of $6.60^{+0.12}_{-0.09}$\,keV and $3.7^{+3.0}_{-1.2}$\,keV.
Then, they fit the ASCA and Chandra spectra simultaneously with a physical model of IP (NEI) plasma. The best-fit temperature ($kT$), ionization parameter ($n_{\rm e}t$) and iron abundance are $\sim$6\,keV,  $\sim2\times10^{10}$cm$^{-3}$\,s and $\sim$4.5\,solar, respectively. The absorption is $\sim1.4\times10^{23}$\,cm$^{-2}$, and hence this source would be behind the Sgr\,B complex. 
Assuming the distance of 8\,kpc, the X-ray luminosity (2\,--\,10\,keV) is estimated to be $\sim10^{34}$\Lu, which is only $\sim1$\,\% of the GCXE flux.

The Chandra morphology is a ring-like structure  of $\sim10\arcsec$ radius  plus a tail of $\sim20\arcsec$ long. From this small ring and high plasma temperature, \citet{Se02} proposed that G0.570$-$0.018 is a very young SNR of $\sim$100 years old.  The tail would be outflow plasma from this young SNR. However the INTEGRA\,/\,IBIS $\gamma$-ray and VLA radio observations  by \citet{Renaud2006} revealed neither the $^{44}$Ti $\gamma$-ray line nor the radio continuum feature from this very young SNR candidate.

\citet{In09} made the spectra of ASCA (observed in 1994), Chandra (2000), XMM-Newton (2001), XMM-Newton (2004) and Suzkau (2005) from the circle  of the radius of $2\arcmin.5$ around G0.570$-$0.018. They fitted  all the spectra with the same model of power-law plus 3 Gaussian lines for the \FeK~lines  fixing  at 6.40, 6.67 and 6.97\,keV, respectively. 
The best-fit \Ka~line flux (6.40 keV) increased from the ASCA(1994) to Chandra (2000) and XMM(2001) by $\sim$2 times, then turned to decrease to  XMM-Newton (2004) and Suzaku epoch (2005) by factor of $\sim$0.7.
The \Hea~line flux (6.67 keV) was, on the other hand, almost  constant with time. 

With XMM-Newton, \citet{Po15} found an SNR candidate G0.570$-$0.001 in the close vicinity of G0.570$-$0.018. Since the area of \citet{In09} includes both G0.570$-$0.001 and  G0.570$-$0.018, one possibility is that G0.570$-$0.018 is a time variable \Ka~source  (XRN, see table 7), and G0.570$-$0.001 is a \Hea~line emitting SNR candidate. 
However, it is still a puzzle whether G0.570$-$0.001 is a \Hea~line emitting hard X-ray source or soft source with no \Hea~line, because \citet{Po15} detected G0.570$-$0.001 in the soft X-ray band (2\,--\,4.5\,keV) only. 

\subsubsection{G359.942$-$0.03 and J174400$-$2913} %section 4.1.4 

With the Chandra survey observations, many small diffuse spots of  $\lesssim10\arcsec$ size are found (e.g. \cite{Mu08, Lu08, Jo09}, see section 6). Most of them are  feature-less non-thermal emissions, but   a few sources exhibit strong \FeK~lines.  
\citet{Jo09} found a \Hea~line source, named G359.942$-$0.03.  The spectrum is fitted with a thin thermal plasma model with $\sim7$\,keV temperature and absorption of $\sim2\times10^{23}$\,cm$^{-2}$. The \EWHea~is $\sim0.8$\,keV. These are similar to the HTP.  The detection of the \Hea~line from G359.942$-$0.03 is, however, a bit confusing  possibly due to the limited statistics, because \citet{Mu08} reported no significant \FeK~line (\EWFeK$~\lesssim$~180\,eV) from possibly the same source, named G359.941$-$0.029 (see table 12).

\citet{Ya14} found  another \Hea~line source, named Suzaku J174400$-$2913. The spectrum is fitted with a CIE plasma of 
$\sim 4 $\,keV temperature and $\sim$0.6\,solar abundance. The position coincides to a narrow X-ray filament ($\sim10\arcsec$) of G359.55+0.16 \citep{Jo09}. This source would be aligned with the radio non-thermal filament  G359.54+0.18 \citep{Yu97, Wa02a, Lu03}.  The strong \Hea~line  from G359.55+0.16 is a bit confusing, because no \Hea~line has been  reported \citep{Wa02a, Lu03, Jo09}. However the statistic of  \citet{Ya14} is highest, high enough to conclude that the \Hea~line from  Suzaku J174400$-$2913 is a robust result. 

The presences of strong \Hea~lines of these sources are similar to young SNRs. However the filament-like morphologies  with small sizes ($\lesssim10\arcsec$) and  the X-ray luminosity of $\sim2\times10^{33}$\Lu~(2\,--\,10\,keV) is  one or two orders of magnitude 
smaller than the usual young SNRs.  Therefore, these sources would be either  small fragments of young SNR, 
or in other origins such as a filament produced by magnetic field reconnection and confined by the magnetic field, or a ram pressure confined stellar wind bubble generated by a massive star \citep{Jo09}.
Since these X-ray filaments are very faint ($\sim10^{33}$\Lu),  contributions to the HTP flux in the GCXE would be negligible.
	
\subsubsection{Sgr\,B2 Star Cluster} %section 4.1.5

Sagittarius B2 (Sgr\,B2) is a giant MC complex, located at the projected distance of $\sim$100\,pc from the Galactic center 
\SGRA, and is one of the richest star-forming regions (SFRs) in our Galaxy. It contains many  compact H\emissiontype{II} 
regions (e.g. \cite{Be84}). Thus, Sgr\,B2 harbors many clusters of very young high-mass stellar objects (YSOs). 

With Chandra, \citet{Mu01b, Ta02} found about one and half dozens of compact sources in the Sgr\,B2 region of $\sim3\arcmin\times3.5\arcmin$~area. These sources are  likely  YSOs.  The brightest  2 sources  at the compact HII region, Sgr\, B2 Main, are slightly extended with  sizes of 3$\arcsec$ and $5\arcsec$ radii. The  spectra are fitted with a CIE  plasma of $\sim$1 solar abundance. The  best-fit temperature and luminosity are in the range of $\sim$5\,--\,10\,keV and $\sim10^{33}$\Lu ~(2\,--\,10\,keV), respectively. Another bright source at the position of Sgr\, B2 North of  $25\arcsec\times21\arcsec$ in size has the luminosity of  $\sim10^{33}$\Lu.  From the extended natures and large luminosity, these three sources are likely clusters of YSOs.  

The combined spectrum of the other point sources is fitted with  a CIE plasma model. 
The best-fit temperature is $\sim$10\, keV. The individual luminosity is in the range of  $\sim2\times10^{31}$\,--\,$10^{32}$\Lu.
The temperature is higher than, but the  luminosity is typical to the YSOs. Since no IR nor radio counterpart is found, an alternative idea of these thermal X-ray emissions are isolated white dwarfs powered by the Bondi-Hoyle accretion from the dense cloud gas.  In any origin, these are surely \Hea~line emitters, which contribute to the HTP of the GCXE. The total luminosity is $\sim5\times10^{33}$\Lu (2\,--\,10\,keV), which is $\sim$5\,\% of the total X-ray flux of the Sgr\,B complex.  The major X-ray fraction of the Sgr\,B complex is a diffuse non-thermal emission with prominent \Ka~lines (the XRN, section 5.1.1). 

\subsubsection{Arches Star Cluster} %section 4.1.6

The Arches cluster is one of the most massive star clusters near the GC. It has a total mass of $\sim 10^{4}\,M_{\odot}$ within a compact size of $\sim$0.5\,--\,1\,pc diameter. 
\citet{Yu02b, La04} detected  2 compact X-ray  sources (A2, A1\,N/S) in addition to the diffuse emission from the central region using Chandra. 

\citet{Wa06b} reported that the 2 sources are separated to  3 point sources (A1N, A1S and  A2), which exhibited bright \Hea~lines with temperatures of $\sim$1.8\,keV,  $\sim$2.2\,keV and  $\sim$2.5\,keV, respectively. 
Diffuse thermal emission with a strong \Hea~line is also found near the cluster center of $\le15\arcsec$ radius. The luminosity is $\sim4\times10^{33}$\Lu~(2\,--\,8 keV), about 20\,\% of the sum of A1N, A1S and  A2. 
The total luminosity of the thermal plasma of A1N, A1S,  A2  and  central diffuse source is $\sim3\times10^{34}$\Lu~(0.3\,--\,8 keV), $\sim$2\,\% of the GCXE. 

The Suzaku spectrum of the whole cluster region is fitted  with a two-component model, a CIE plasma of $\sim$1\,solar abundance,  and a power-law component with a \Ka~line \citep{Ts07}. The best-fit temperature of the CIE plasma is $\sim2$\, keV, and hence exhibits strong \Hea~lines. The luminosity of the CIE plasma is $\sim10^{34}$\Lu~(3\,--\,10~keV).  
About half of the X-rays are largely extended diffuse emission of power-law spectrum with strong \Ka~lines (section 5.2.1). 

\citet {Ca11a} examined long term X-ray emissions observed with XMM-Newton in 2002\,--\,2009.  They found a clear  flare with the flux increase of $\sim$70\,\% above the quiescent level. The spectrum in the quiescent state  shows both the \Hea~and \Ka~lines, and hence the spectrum is a combination of thermal plasma (\Hea~line) and non-thermal component (\Ka~line). The best-fit temperature of the thermal plasma is $\sim1.7$\, keV, in good agreement with those of Suzaku \citep{Ts07} and Chandra \citep{Wa06b}.  The total luminosity (2\,--\,10\,keV) is $\sim1.5\times10^{34}$\Lu. The luminosity ratio between  the thermal plasma and non-thermal component is 0.85 : 0.15, which seems inconsistent with \citet{Ts07}. However, taking account of the diffuse nature of the \Ka~line and larger correction area of \citet{Ts07}, the flux ratios may be consistent with each other.  
The flare spectrum does not show significant emission of the \Ka~line, in contrast to the quiescent spectrum.  This also supports that the \Ka~line is different origin from the higher ionization \Hea~line.  The flare spectrum is described well by a CIE plasma of $\sim1.8$\,keV temperature. The total luminosity is $\sim3\times10^{33}$\Lu, one of the largest flares of YSOs.

\subsubsection{Other Young Star Clusters} %section 4.1.7

\citet{Ba03} found  central diffuse emission of a size of \timeform{10"} radius in the immediate vicinity of \SGRA, where Central Star Cluster (CSC) is included. The spectrum is fitted with a thermal plasma of $\sim$ 1.3\,--\,1.6\, keV temperature.  The luminosity is $\sim 2\times10^{34}$\Lu~(2\,--\,10\, keV). The spectrum  has a \FeK~line at 6.5$^{+0.1}_{-0.2}$\,keV, the energy between the \Ka~ and \Hea~lines. They interpreted that the lower line energy than 6.7\,keV is due to ionizing hot plasma (IP).  More plausible idea is that the \FeK~line is a mixture of \Ka~(CG)~and \Hea~(HTP).  The luminosity of the hot plasma is significantly higher than that of single YSO, and hence likely origin is a cluster of YSO, including  colliding winds of OBs  and W-R stars.  

The Sgr\,D complex is  composed of Sgr\,D H\emissiontype{II} 
or G1.13$-$0.10, and an SNR, Sgr\,D SNR or G1.0$-$0.1 \citep{Do79, Do66}. \citet{Sa09} found a hard  diffuse X-ray spot (Diffuse Source 2: DS2)  in  the radio shell of Sgr\,D H\emissiontype{II}.  
The spectrum of DS2 is a  high temperature plasma of $\sim$4\,keV, accompanied by a \Hea~line, possibly, a cluster of YSO in the non-thermal radio shell.

 \citet{No17a} found faint hard X-ray emissions from the Sgr\,D SNR region. Remarkable discovery is an extremely high Ni abundance of $\sim$30\,solar from the northeast shell of the SNR.  Such a large Ni abundance has not been predicted by any model of normal SNR. This anomalous structure would be due to some extreme circum stellar conditions in the GCXE region.
One possible scenario is that Sgr\,D SNR is a CC-SN, and the SN explosion was highly asymmetric so that a part of the neutron-rich (Ni) inner core region was ejected perennially to the northeast shell  of the SNR.

Sgr\,C is also a MC complex, and hence could be \Hea~line sources like Sgr\,B, Arches and Sgr\,D.  However, no hot plasma with a \Hea~line is found, except lower temperature plasmas with a \SHea~line (section 4.2.7), and hence  Sgr\,C would  not contribute to the HTP in the GCXE. 

Sgr\,A is another MC complex, which is associated with many XRNe (table 9).  Although no compact point source with a \Hea~line is found from the  Sgr\,A complex due to the very crowded region, some fractions of HTP may come from this region, possible activities of YSOs.

\citet{La04} found 4 X-ray point sources and diffuse emission from the Quintuplet cluster with the  luminosity of 
$\sim10^{33}$\,--\,$10^{34}$\Lu.
\citet{Wa06b} found 8 X-ray sources in the Quintuplet region. Since the X-ray properties are largely different, they combined 3 bright sources with similar properties (QX2, QX3 and QX4), and fit the spectrum with a CIE model.  The best-fit $kT$ and $N_{\rm H}$ are $\sim$8\,keV and $\sim6\times10^{22}$\,cm$^{-2}$, respectively.  The total luminosity is  $\sim8\times10^{32}$\Lu. 
An extended  emission also found with  luminosity and temperature of  $\sim3\times10^{33}$\Lu~and $\sim10$\,keV, respectively. 

In summary, the total luminosity of the young star clusters with the \Hea~lines is larger than that of the other diffuse hot plasma excluding Sgr\, A East. The sum of the luminosity of all the young clusters (both point sources and diffuse plasma) in the GC, would not exceed $\sim$5\,\% of the GCXE, even if contributions of undetected  X-ray faint young star clusters are taken into account.

\subsection{\SHea~Sources (LTP)} %section 4.2

The soft X-ray plasmas with the size  of $\lesssim$10\arcmin~are given in this section. 
Most of them exhibit strong \SiHea~and \SHea~lines with a moderate temperature of $\sim$1\,keV, and hence these are likely intermediate-old SNRs.  However, some of them show unusual structures as SNRs either in morphology or spectrum.  This would be  closely related  to extreme ISM environment near at the GC.
These soft X-ray plasmas may contribute significant fractions to the LTP flux, but the contribution to the HTP would be ignored.   

\subsubsection{G0.42$-$0.04 (G0.40$-$0.02), G1.2$-$0.0 and G0.13$-$0.12} %section 4.2.1

In the soft X-ray band map of Suzaku, \citet{No08} found an excess spot with elliptical shape of $\sim1\arcmin.8\times2\arcmin.4$ at $ (l, b) = (\timeform{0.42D}, \timeform{-0.04D}$), hence named G0.42$-$0.04
(Suzaku J1746.4$-$2835.4). 
The source has a \SHea~line at $\sim$2.45\,keV, a cut-off below $\sim$2\,keV, and a steep slope above $\sim$4\,keV. The spectrum is fitted with a CIE plasma model with the temperature and abundances of $\sim$0.7\,keV, and  $\sim$0.9\,solar, respectively. 
The absorption column of $\sim8\times10^{22}$\,cm$^{-2}$ is consistent with the Galactic center (GC) distance of 8 kpc. 
Then, the physical size of the ellipse is $\sim$5.6\,pc$\times$4.2\,pc. The X-ray luminosity is estimated to be 
$\sim6\times10^{33}$\Lu~(2\,--\,10\,keV). These values are consistent with an intermediate-old aged SNR. 	

\citet{Po15} found a larger ellipse  at $ (l, b) = (\timeform{0.40D}, \timeform{-0.02D}$), named G0.40$-$0.02, with the size of
 $\sim4\arcmin.7\times7\arcmin.4$.
The spectrum is fitted with a CIE plasma with the temperature and absorption column of $\sim$0.55\,keV and $\sim8\times10^{22}$\,cm$^{-2}$, respectively, both are similar to those of G0.42$-$0.04.
Therefore, G0.40$-$0.02 and G0.42$-$0.04 would be the same object with the distance of 8 kpc. Then, the physical size of G0.40$-$0.02 is $\sim$11\,pc$\times$17\,pc. They  estimated that the dynamical age and thermal energy of G0.40$-$0.02 are $\sim$3700\,years and $\sim1.9\times10^{50}$\,erg, respectively.

From the  radio SNR candidate G1.0$-$0.1 in the Sgr\,D complex \citep{Do79}, \citet{Si01} found a faint diffuse soft X-ray with BeppoSAX, but Suzaku found no soft X-ray from G1.0$-$0.1\citep{Sa09}.
Instead, \citet{Sa09} found an elliptical X-ray spot (Diffuse Source 1: DS1) with the size of $\sim4\arcmin\times7$\arcmin~at the northeast of the radio shell Sgr\,D H\emissiontype{II}. From the position, the spot DS1 is named G1.2$-$0.0. 
The spectrum of G1.2$-$0.0 has He$\alpha$ lines of S, Ar and Ca, and hence is fitted  with a CIE plasma of $\sim$0.9~keV temperature. The abundances of S, Ar and Ca are $\sim$1.6, $\sim$1.8 and $\sim$1.8\,solar, respectively.  The X-ray absorption 
is $\sim 8.5 \times 10^{22}$\,cm$^{-2}$, possibly in or behind the Sgr\,D MC complex. Assuming the distance to be 8\,kpc, the size is estimated to be $\sim8$\,pc$\times16$\,pc.
The plasma temperature, abundances and size suggest that G1.2$-$0.0 is an intermediate-old aged SNR. They reported that the unabsorbed luminosity in the 0.7\,--\,8 keV band is 1.4$\times10^{35}$\Lu. However, this luminosity would be a subject of large ambiguity due to the large $N_{\rm H}$ for the soft spectrum. 

In the  XMM-Newton image, \citet{He13b} discovered a diffuse soft X-ray spot of a  circle of $1.5\arcmin$~radius  near the X-ray filament G0.13$-$0.11 (section 6). From the position, this source is named G0.13$-$0.12.
The X-ray spectrum is fitted with a CIE plasma of $\sim$1.1~keV temperature. 
The absorption is $\sim5.6\times10^{22}$\,cm$^{-2}$,  consistent with being a GC source. 
Then, the  X-ray luminosity is $\sim2.2\times10^{34}$\Lu~(2\,--\,10\,keV).
The abundances of Si, S and Ar are  $\sim$1.4, $\sim$2.0 and $\sim$3.4\,solar, respectively. 
These values are consistent that G0.13$-$0.12 is an intermediate-aged SNR.  Since massive MC, G0.13$-$0.13 is present in this region, alternative idea is a hot Inter Stellar Medium (ISM), or Circum Stellar Medium (CSM) plasmas heated by powerful stellar winds. 

\subsubsection{G359.79$-$0.26 and G359.77$-$0.09} %section 4.2.2

\citet{Mo08, Mo09} found two bright diffuse spots, G359.79$-$0.26 and  G359.77$-$0.09 in the Suzaku~\SHea~line 
image near the GC. The sizes are 
$\sim4.0\arcmin\times2.6\arcmin$~and $\sim4.9\arcmin\times2.4\arcmin$, respectively.
They reported that 
the X-ray spectra of G359.77$-$0.09 and G359.79$-$0.26 are fitted with a CIE plasma  model, with the temperatures of $\sim$0.7 and $\sim$1.0\,keV, respectively.  
The diffuse source G359.77$-$0.09 exhibits clear He$\alpha$~lines of Si, S, and Ar
with abundances of $\sim$0.9\,solar, $\sim$0.7\,solar and $\sim$0.9\,solar, respectively. 
The other diffuse source G359.79$-$0.26 exhibits He$\alpha$~lines of Mg, S, S, Ar and Ca with 
abundances of $\sim$1.3, $\sim$1.1, $\sim$1.4, $\sim$1.7 and $\sim$1.4\,solar, respectively. 
The absorption column densities of G359.79$-$0.26 and G359.77$-$0.09  are $\sim 5\times10^{22}$\,cm$^{-2}$ and
 $\sim7\times10^{22}$\,cm$^{-2}$, consistent with GC sources at 8.0 kpc, and hence the sizes of G359.79$-$0.26 and G359.77$-$0.09 are  $\sim$9.3\,pc$\times$6\,pc
and  $\sim$11.4\,pc$\times$5.6\,pc, respectively. The luminosity is in the range of $\sim$4\,--\,6$\times10^{33}$\Lu.
In the XMM-Newton image,  \citet{He13b, Po15} confirmed the soft X-ray sources , G359.79$-$0.26 and G359.77$-$0.09.  The plasma temperatures, absorptions, abundances and luminosity are all consistent with \citet{Mo08, Mo09}.

Although, the two SNR candidates, G359.79$-$0.26 and G359.77$-$0.09 are spatially separated by $\sim0\arcdeg.2$, all the physical parameters are nearly the same. These sources make-up a single elliptical ring with the diameters and width of 
$\sim$$\timeform{20'}\times\timeform{16'}$  and $\sim\timeform{6'}$\,--\,$\timeform{9'}$, respectively.
The  X-ray spectrum from the ring is fitted with  a CIE  plasma model of $\sim0.9$\,keV temperature and the Si, S and Ar abundances of $\sim$1.0, $\sim$1.2 and $\sim$1.4\,solar, respectively.  The thermal energy of the ring is $\sim10^{51}$\,erg.  Therefore, the ring and the two sources G359.79$-$0.26 and  G359.77$-$0.09 would be comprised  a single source, a super bubble (SB) with 
diameter and width of $\sim $40\,--\,50\,pc and $\sim$15\,--\,20\,pc, respectively.  Since the shape of the ring is nearly symmetry with respect to the center point at $(l, b)$ = $(359\arcdeg.83, -0\arcdeg.14)$, an alternative idea would be that the ring is a hyper nova remnant \citep{Mo09, He13b}. 

% figure 9 {eps-figure/Mori2009_fig2_GC-bubble.eps} %figure 9

\subsubsection{Diffuse Soft Sources Near \SGRA~(NW, SE, E)} %section 4.2.3

In the Chandra image near \SGRA,  \citet{Ba03} found diffuse sources, named NW and SE. Since these are bright in the 1.5\,--\,6\, keV band, but weak in the 6\,--\,7\,keV band images, these are soft X-ray sources, unrelated to the Sgr\,A East SNR. Instead, they proposed these are bipolar-flows from \SGRA, although no spectral information was available. 
In the XMM-Newton X-ray image near \SGRA, \citet{He13b}
found three diffuse sources with the size  of
$\sim\timeform{1'} \times \timeform{0.7'}$, 
$\sim\timeform{1.7'} \times \timeform{1'}$ 
and $\sim\timeform{1'} \times \timeform{1'}$,
and named NW, SE and E.  The sources, NW and SE are the same sources of \citet{Ba03}. 

The spectra of NW, SE and E are fitted with a CIE plasma of $\sim0.9$\,keV, $\sim1.1$\,keV and $\sim1.0$\,keV temperatures. The Si, S and Ar abundances are in the range of $\sim$1.0\,--\,0.6, $\sim$0.6\,--\,0.9 and $\sim$0.7\,--\,1.5\,solar, respectively.
The absorptions of NW, SE and E are $\sim8\times10^{22}$,
 $\sim6\times10^{22}$ and $\sim6\times10^{22}$\,cm$^{-2}$, respectively, consistent with the  GC distance of 8\,kpc. 
Then, the sizes of  NW, SE and E are estimate to be $\sim2.3\times1.6$,  $\sim4.0\times2.3$  and $\sim2.3\times2.3$\,pc$^2$,  respectively,
while the luminosity  (2\,--\,10\,keV) is $\sim1.5$\,--\,$2.7\times10^{34}$\Lu. 
The temperatures, abundances and luminosity are similar to, but the sizes are significantly smaller than, those of  the intermediate-aged SNRs.  \citet{He13b} estimated the electron  densities of NW, SE and E to be $\sim$4.6\,--\,9.9 \,cm$^{-3}$, significantly higher than that of ISM or CSM.
The small size of elongated shape pointing to \SGRA, high density and closeness from \SGRA, 
lead to an alternative scenario that NW, SE and E are outflows from \SGRA~(section 7.3).

\subsubsection{G359.1$-$0.5: A Recombining Plasma}%section 4.2.4

G359.1$-$0.5 is a shell-like radio SNR with the size of $\sim10\arcmin$~in radius  \citep{Do79}. Center-filled thermal X-rays are found 
with ASCA \citep{Ba00}, and hence G359.1$-$0.5 is a MM-SNR.  The X-ray spectrum has prominent \SiHea~ and \SLya~lines. This is very peculiar because S is more highly ionized (H-like) than the lighter element Si (He-like).  

\citet{Oh11} observed G359.1$-$0.5 with Suzaku and made a high quality spectrum. The problem of the peculiar spectrum of ASCA is solved by adding radiative recombination continuum (RRC), a saw-teeth continuum shape made by a transition of free electrons to the K-shell of Si\emissiontype{XIII} and S\emissiontype{XV}. 
The strong RRC structures indicate that the plasma is in over-ionization (recombining plasma: RP). In fact, the observed spectrum is well-fitted by a RP model. 

%%%The best-fit electron temperature of $\sim$0.3\,keV is far smaller than the ionization temperature of  $\sim$0.8\,keV.%%% 
The best-fit $N_{\rm H}$ of  $\sim2\times10^{22}$\,cm$^{-2}$ seems smaller than that of the GCXE. However, taking account of the small e-folding latitude  of $\sim0^\circ.25$ of the GCXE (tables 3, 4), this value is consistent with that G359.1$-$0.5 is located near the boundary of the GCXE.

TeV gamma ray emission  (HESS J1745$-$303) and a hint of the \Ka~line are found near this  SNR \citep{Ah08, Ba09}. 
Since the \Ka~line is hardly produced in this low-temperature SNR, this would be due to a bombardment of low energy cosmic rays (LECR) to a surrounding cool MCs. These LECRs would preferentially ionize Si and S, and would make a RP. The other idea to make a RP is cooling of electrons by thermal conduction to adjacent MCs or adiabatic expansion when the shock breaks thorough a surrounding dense MC gas to a tenuous ISM. These MCs should be more numerous in the GC region compared to the other regions of the Galaxy. The ratio of the ionization temperature  ($\sim$0.8\,keV) to the electron temperature  ($\sim$0.3\,keV) of G359.1$-$0.5 is the largest among the known 
$\sim 10$\,RP-SNRs (e.g. \cite{Ko14,Sa14, Wa16}), suggesting some extreme MC, ISM or CSM conditions are presented in the GC, or near around G359.1$-$0.5. 

\subsubsection{G359.41$-$0.12 and Chimney} %section 4.2.5

\citet{Ts09} found an ellipse and a chimney-like structure in the \SHea~line image of Suzaku  near the Sgr\,C complex region.  These are  named G359.41$-$0.12  and  Chimney. The morphology of  Chimney is
similar to a PWN with NS moving to one direction. However no NS candidate, is found at the head region. Furthermore, the spectrum is fitted with a CIE plasma model with $\sim$1.2\,keV temperature.  The spectrum of G359.41$-$0.12 is also fitted with a CIE plasma model of $\sim$0.9\,keV temperature.
The absorption columns of G359.41$-$0.12 and Chimney are very large and are nearly the same of  $\sim1.2\times10^{23}$
and $\sim1.0\times10^{23}$\,cm$^{-2}$, respectively. The abundances of S and Ar are both $\sim$1.7\,solar.
These indicate  that  G359.41$-$0.12 and  Chimney  are in the same origin located at the same distance on the same line-of-sight, possibly just behind a dense MCs in the Sgr\,C complex at 8\,kpc.  The X-ray luminosity (1.5\,--\,8\,keV) of G359.41$-$0.12 and Chimney are, then estimated to be $\sim$2\,--\,$4\times10^{34}$\Lu.

The sum of the thermal energies of G359.41$-$0.12 and Chimney is estimated to be $\sim$1.4$\times$10$^{50}$\,erg. 
The dynamical time scales of G359.41$-$0.12 and Chimney are $\sim$2.5$\times10^4$  and $\sim$4$\times10^4$\,years, respectively.  These values are typical to a single Galactic SNR.
\citet{Ts09} proposed that Chimney is an outflow plasma, extending about 30\,pc from an SNR candidate G359.41$-$0.12. However the highly collimated outflow of $\sim$5\,pc width and $\sim$30\,pc length, emanating from G359.41$-$0.12 is very unusual as the result of a single SN. One possibility is that many MCs in the Sgr\,C complex deformed a spherical SN expansion to the outflow like  expansion. Still to make a highly collimated uni-polar structure would be difficult.  Some other extreme conditions of ISM in the GCXE or near at Sgr\,C complex  would have a responsibility on the outflow like morphology of Chimney.

\subsubsection{Diffuse Plasma Near 1E\,1740.7$-$2942} %section 4.2.6

1E\,1740.7$-$2942 is the brightest XB near the GC region. Since the time variability and 
spectral behavior are similar to those of the low state of Cygnus X-1, the archetypal BH candidate, 
1E\,1740.7$-$2942 would be another BH candidate. It is named the Great Annihilator (GA), because a hint of electron-positron 
annihilation line at 511 keV was found (e.g. \cite{Su91}).  Although, no further evidence for the annihilation line has been found 
so far by other instruments, this source has drawn great attentions, because non-thermal double jet-like structures were  found in the radio band. Thus, the GA is referred  as a micro quasar \citep{Mi92}. 

Two diffuse X-ray sources are found around the GA \citep{Na10}. One is M359.23$-$0.04, a bright \Ka~ spot (see section 5.3.2), and the other is a thermal plasma source G359.12$-$0.05, near the position of the radio SNR  G359.07$-$0.02 \citep{La00}.  The spectrum of G359.12$-$0.05 is fitted with a CIE plasma model of $\sim$0.9\,keV temperature and absorption column of $\sim7\times10^{22}$\,cm$^{-2}$. The abundances of Si, S and Ar are $\sim$1.2, $\sim$1.4 and  $\sim$1.5\,solar, respectively.

The $N_{\rm H}$  is typical column density of the GC region of $\sim6\times 10^{22}$\, cm$^{-2}$, and hence G359.12$-$0.05 is a  
GC source at 8\,kpc.  From the $\sim$0.9\,keV temperature, radius of $\sim\timeform{12'}$  and flux of 
$\sim6.3\times10^{-4}$~photons$^{-1}$~cm$^{-2}$, the luminosity and dynamical age are  estimated to be $\sim5\times10^{33}$\Lu~(1\,--\,10\,keV, absorbed flux)
and $\sim6\times10^4$\,yr, respectively. 
These values are  consistent with an intermediate-old aged SNR, and hence G359.12$-$0.05 comprises a unique system, an SNR associated with a black hole candidate, the GA or micro quasar. 

\subsubsection{Other SNR Candidates with Soft X-Rays} %section 4.2.7

\citet{Po15} found  3 diffuse soft X-ray sources,  named G0.52$-$0.046, G0.570$-$0.001 and G0.224$-$0.032 with  the deep survey of XMM Newton.
The soft source G0.570$-$0.001 is located in the close vicinity of the hard X-ray source G0.570$-$0.018 (section 4.1.3), but these two are separate sources.
The spectra of G0.52$-$0.046, G0.570$-$0.001 and G0.224$-$0.032
are fitted with thermal plasmas of temperatures of $\sim$0.8, 0.6 (fixed) and $\sim$0.5\,keV,  with absorption columns ($N_{\rm H}$) of $\sim$8$\times10^{22}$,
 $\sim$1$\times10^{23}$ and 
$\sim$7$\times10^{22}$\,cm$^{-2}$, respectively, and hence these  are likely located near the GC.  In the GC distance of 8\,kpc, the plasma sizes are $\sim6\times12$\,pc$^2$,  $\sim4\times7$\,pc$^2$ and $\sim5\times11$\,pc$^2$, respectively.  The dynamical ages and the total thermal energies are $\sim$1700, $\sim$1600 and $\sim$1800\,years, and $\sim5\times10^{49}$, $\sim3 \times10^{49}$ and $\sim3\times10^{50}$\,erg, respectively. 
As is noted in section 4.1.3, a possibility that G0.570$-$0.001 is a \Hea~line emitting hard X-ray source, is not fully excluded. 

\section{The \Ka~Clumps} \label{sec:x-echo} %section 5

\begin{figure*}[!htb]
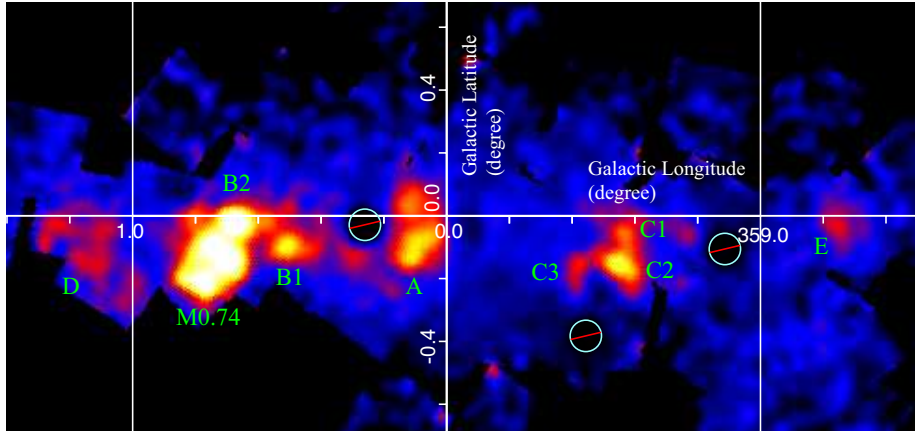
 %figure 5
\begin{center}
\FigureFile(120mm,80mm){eps-figure/64-map.eps}
\end{center}
\caption{The XRN complexes  near the Galactic center: Sgr\,D, B (M0.74, B2, B1), A, C (C3, C2, C1), and E (from the left to the right). 
The white circles are positions of bright X-ray binaries and are excluded from this map (From unpublished Suzaku results).} 
\label{fig:6.4keV-map}
\end{figure*}

The \Ka~line distribution in the GCXE is not uniform, but clumpy as is shown in 
figure\,\ref{fig:6.4keV-map}. The \Ka~flux from these clumps occupy nearly half of that in the GCXE. 
This section reviews the local diffuse sources, which emit strong  \Ka~lines (\Ka~clumps).  The mechanisms of the  \Ka~emission
and results of the \EWKa~values are given in section 5.1. The \Ka~clumps with origins of past big flares of \SGRA~are given in section 5.2, while the \Ka~clumps with the other origins than the \SGRA~flare are in section 5.3.

\subsection{Mechanisms of the \Ka~Emission and \EWKa}% section 5.1

When an X-ray or a low-energy cosmic ray (LECR) hits a neutral Fe atom in a cold cloud, an electron hole is made in the K-shell of the neutral Fe (K-shell ionization). Then another electron in the L-shell falls to the hole,  and  re-emit a characteristic X-ray at 6.4\,keV. This process produces an \Ka~line.
If the K-shell ionization source is an X-ray of a power-law spectrum with the  photon index $\Gamma$, $I(E)\propto E^{-\Gamma}$, the \EWKa~is given by;
\begin{equation}
EW_{6.4}= 
4\left(\frac{1}{\Gamma+2}\right)\left(\frac{6.4}{7.1}\right)^\Gamma\left(\frac{1}{1 + \cos^2\theta}\right)\,{\rm keV} 
\end{equation}
, where the fluorescent yield ($Y_{{\rm K} \alpha}$)
, the density ratio of iron to electron ($n_{\rm Fe}/n_{\rm e}$)
, the differential Thomson scattering cross section ($d\sigma_{\rm T}/d\Omega$)
 , and the photo-absorption cross  section by Fe atoms  ($\sigma_{\rm Fe}$) are
$\sim 0.34$,  $\sim 4\times 10^{-5}$, $\sim 4.0 \times 10^{-26} (1 + \cos^{2}\theta)$\,cm$^{2}$, and  $2 \times 10^{-20} (E/7.1\,{\rm keV})^{-3}\,{\rm cm}^{2}$, respectively.
In a typical value of $\Gamma=1.5$ and scattering angle of $\theta=90^{\circ}$, the \EWKa~is $\sim$1\,keV.
The \Ka~line flux is proportional to $N_{\rm H}$,  when the target cold gas is optically thin, or $N_{\rm H}$ is  $\lesssim10^{24}$\,cm$^{-2}$. 
Therefore, in order to produce detectable \Ka~line flux as is shown in figure\,\ref{fig:6.4keV-map}, the absorption K-edge of the neutral Fe at 7.1 keV should be large enough of $N_{\rm H}\gtrsim10^{23}$\,cm$^{-2}$.

If the K-shell ionization source is a charged particle with the number distribution of power-law function of $N(E)\propto E^{-\Gamma}$, the \EWKa~is given as a function of the spectral index ($\Gamma$) \citep{Do11}. 
In the case of an electron, the \EWKa~is $\sim$250\,--\,400\,eV, almost independent of $\Gamma$.
In the case of protons, the \EWKa~depends largely on $\Gamma$.  In the  normal case that $\Gamma$ is $\gtrsim$1, the \EWKa~is $\gtrsim$1\,keV. 

The largest cross sections of the K-shell ionization of protons and electrons are around the energy of a few $\sim$10\,MeV, and  a few $\sim10$\,keV, respectively \citep{Ta12}. The particle of these energy is called the Low Energy Cosmic Ray proton (LECRp) or the Low Energy Cosmic Ray electron (LECRe).  Both  the  LECRp and LECRe  can penetrate only $N_{\rm H}\lesssim10^{22}$\,cm$^{-2}$. Therefore, if the absorption depth of the iron K-edge at 7.1 keV is shallow of $N_{\rm H}\lesssim10^{22}$\,cm$^{-2}$, the ionization source is likely an electron or proton. 

\subsection{X-ray Reflection Nebula (XRN)}% section 5.2

In figure 5, bright \Ka~ clumps are  found in the MC complexes of Sgr\,A, B, C, D and E in the Central Molecular Zone (CMZ).
 The spectra have a mean \EWKa~of $\sim$1\,keV and a pronounced edge structure at 7.1\,keV of $N_{\rm H}\sim10^{23}$\,--\,$10^{24}$\,cm$^{-2}$. Thus the origin of the bright \Ka~ clumps would be an X-ray irradiation.
The flux and morphology are often variable in the time scale of a few\,--\,10 years.
This section overviews the bright \Ka~line clumps, named  the X Ray Reflection Nebulae (XRNs), and discuss  the nature, structure and origin of the XRNe.

\subsubsection{Sgr\,B: A Prototype of the X-ray Reflection Nebula (XRN)}
\label{sec:sgrb} % section 5.2.1

\begin{table*}[!ht] % table 7
\caption{6.4 keV clumps near the Sgr\,B complex}
\begin{center}
\begin{tabular}{lccccll}
\hline
Name ($l, b$)$\ast$  &$N_{\rm H}$	& $EW_{\rm 6.4}$	&$F_{\rm x}$ (4--10 keV)$^\star$	& Area & Vari$^\#$ &Instrument$^\dag$ \\
&($10^{22}$cm$^{-2}$) &(keV) 	& ($10^{-12}$~erg~s$^{-1}$~cm$^{-2}$)	&(arcmin$^2$) & & Reference$^\ddag$\\
\hline
Sgr B2 ($0.66, -0.05$)	& $83^{+25}_{-20}$&$2.9^{+0.3}_{-0.9}$ &13&28&-& A, a \\
Sgr B2 ($0.66, -0.04$)	& $88^{+20}_{-15}$	& $2.1 \pm{0.2}$&12&11 &- & C, b\\
Sgr B2 ($0.7, -0.04$)	& $40\pm{15}$	& $2.2 \pm{0.1}$&4.5&13&- & B, c\\
Sgr B2 ($0.66, -0.03$)	& $50\pm{13}$ &$1.2^{+0.7}_{-0.3}$ &$1.9\pm{0.2}$ (10 --40\,keV) &7 &Y&N+X, j\\
\hline
Sgr B2 (0.66$-$0.03)& 96$^{+25}_{-8}$ & $1.13^{+0.05}_{-0.02}$ & 11&9&-& S, f\\
M0.74$-$0.09 &40$^{+14}_{-11}$ &$1.55^{+0.37}_{-0.26}$ &3.0&9 \\ 
\hline
Sgr B2 ($0.67, -0.02$) & 84$^{+38}_{-12}$ & -& 4.3~(5--10 keV)$^{||}$ & 9 &Y &S, i\\
Sgr B2 ($0.67, -0.02$)& 88$^{+38}_{-12}$ &- &2.1~(5--10 keV) $^{**}$ &9 &Y	& \\
M0.74$-$0.09 &57$\pm{6}$ &-&1.7~(5--10 keV) $^{||}$ &7&Y &\\ 
M0.74$-$0.09 &65$\pm{18}$ &-&0.9 (5--10 keV) $^{**}$ &7&Y&	\\ 
\hline
G0.66$-$0.13	&$30^{+38}_{-9}$& -& $0.9\pm{0.1}$ (10--40\,keV) & 9&Y & N+X, j\\
G0.570$-$0.018& 13.9$^{+3.3}_{-3.2}$&- &1.2--1.5~(2--10~keV)& 20 &Y &C, d, h \\
Sgr B1 (0.51, $-$0.10)&15$^{+2}_{-1}$ &1.4$\pm{0.3}$ &2.2~(2--10 keV) &22&-&S, g \\
\hline
Sgr B1+G0.570 	&-&0.57$\pm{0.07}$& 9.7~(2--8 keV)& 77	&-&C, e \\ 
Sgr B2+ M0.74+ G0.66 &-& 1.15$\pm{0.15}$ &11~(2--8 keV)& 96&\\
\hline
\end{tabular}
%$1.4\pm{0.2}^\dagger$, $1.1\pm{0.2}^\dagger$ 1.3\pm{0.2}^\dagger$$1.3\pm{0.2}^\dagger$
 %table 7
\end{center}
{\footnotesize
$\ast$ Some numerical values in the columns, Name ($l, b$) and  Area may have errors of $\lesssim0\arcdeg.01$  and $\lesssim10$\,\%, respectively, because these are read from the original figures in the References.\\
$\#$  With multiple observations, flux variability is found (Y) or not(N).\\
$\dag$ Instrument, A: ASCA, C: Chandra, B: BeppoSAX, S: Suzaku, N: NuSTAR,
X: XMM-Newton. \\ 
$\ddag$ Reference, a: \citet{Mu00}, b: \citet{Mu01b}, c: \citet{Si01}, d: \citet{Se02}, e: \citet{Yu07}, f: \citet{Ko07a}, g: \citet{No08}, h:\citet{In09}, 
i:  \citet{No11}, j: \citet{Zh15}\\
$\star$  Unabsorbed flux, but reference (f) is absorbed flux.\\
$||$	 2005 observation.\\
$**$  	2009 observation.\\
}
\end{table*}

As is noted in section 2.2, the XRN scenario of the \Ka~line from the Sgr\,B2 clouds is proposed by \citet{Ko96, Mu00} with the ASCA observation. 
BeppoSAX found hard X-ray emission from the Sgr\,B2 cloud \citep{Si01},
but the line center energy of
$6.5\pm{0.07}$\,keV is higher than those from \citet{Ko96, Mu00}, and any other later observations with better energy resolution instruments (see table 7).

With Chandra, \citet{Mu01b}  found a diffuse emission of nearly one order of magnitude brighter than those of the X-ray emitting young stars in the Sgr\,B regions 
(section 4.1.5). The morphology is a convex shape of $\sim2^\prime \times 4^\prime$ size  facing to \SGRA. The X-ray peak is shifted from the core of the MC toward \SGRA~by
$\sim1^\prime$. 
The X-ray spectrum exhibits pronounced \Ka, \Kb~lines, deep Fe\emissiontype{I} K-edge at 7.1\,keV and large photoelectric absorption at low energy.
The absorption-corrected X-ray luminosity is $\sim10^{35}$\Lu. These are nearly the same as the ASCA results. 
Using the best-fit spectral parameters and the geometry of the MC, 
\citet{Mu01b} simulated the X-ray properties with the XRN scenario in the case that $N_{\rm H}$ is larger than $10^{24}$\,cm$^{-2}$.
Their simulation successfully reproduced the convex  morphology and the peak shift of $\sim1^\prime$ toward the GC or \SGRA.
Thus, the  X-ray morphology and spectrum are all well explained by the XRN scenario, where the irradiation source is located toward the GC, or \SGRA~itself. 

The Suzaku observations by \citet{Ko07a} provided separate maps of the \Ka~and \Hea~lines. 
Although the \Hea~line is smoothly distributed over the Sgr\,B region except a faint clump of the SNR candidate G0.61$-$0.01 (section 4.1.2), the \Ka~line image is more clumpy with local excesses at the positions of the Sgr\,B2 cloud and at $(l, b) = (\timeform{0D.74}, -\timeform{0D.09})$. The latter clump is called   M0.74$-$0.09 (see figure\,\ref{fig:6.4keV-map}).

%%%%%%%%%%%%%%%
\begin{figure}[!htb]
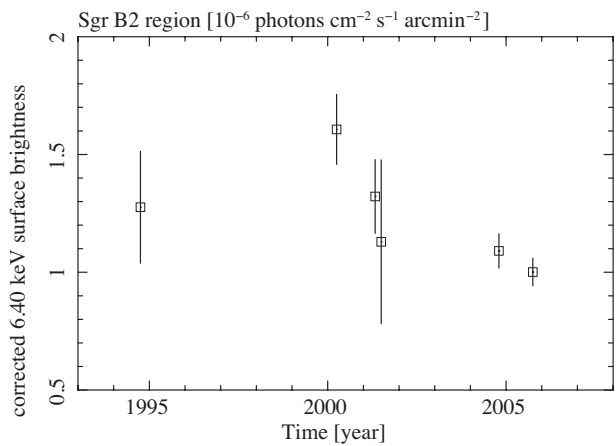
 %figure 5
\begin{center}
\FigureFile(80mm,60mm){eps-figure/In09-B2-lc.eps} %figure 5
\end{center}
\caption{The light curve of the \Ka~ line in the Sgr\,B2 cloud. The data points are ASCA observation in 1994, 
the Chandra and XMM observations in 2000 and 2001, the XMM observation in 2004 and the Suzaku observation in 2005 
(From \cite{In09}).}
\label{6.4-B2-lc}
\end{figure}

The \Hea~line flux is constant with time in all the regions including near the Sgr\,B complex. By contrast, a time variability of the \Ka~line~in the Sgr\,B complex regions is discovered for the first time  by  \citet{Ko08} between the Chandra and Suzaku observations separated by 5 years. The time variability is confirmed by the more extended data set from ASCA, Chandra, 
XMM-Newton and Suzaku  in the time span of more than 10\,years \citep{In09}.
The long term time variability of the \Ka~line in the giant MC Sgr\,B2 is shown in figures\,\ref{6.4-B2-lc}. 
The \Ka~line flux of ASCA and Chandra  in the observations of 1994 and 2000 was 1.4 times stronger than those of XMM-Newton and Suzaku observations of 2004\,--\,2005. 

The fluorescent \Ka~lines should be accompanied by the Compton scattered hard X-rays from the cloud. Indeed, the INTEGRAL satellite found a diffuse hard X-ray source, named  IGR J17475$-$2822 at the position of the Sgr\,B2 cloud \citep{Re04}.
With INTEGRAL of more than 40 Msec observations from 2003 to 2009, \citet{Te10} found that the flux of IGR J17475$-$2822 decreased  with the time scale of 8.2$\pm{1.7}$\,years. 

\citet{No11} made an additional Suzkau observation in 2009. They
compared the two  Sgr\,B2 spectra observed in 2005 and 2009. These  spectra   include all the essences of the XRN scenario: the rapid and correlated time variability of the \Ka~line and the continuum fluxes, the large \EWKa, and the deep K edge absorption.
The X-ray flux of the \Ka~line decreased in correlation with the hard-continuum flux by a factor of 0.4\,--\,0.5 in four years, which is almost equal to the light traveling 
time across the Sgr\,B2 cloud. The same flux decrease in 2005\,--\,2009 is also found from  the other \Ka~line clump in the Sgr\,B complex, M0.74$-$0.009. 

\citet{Zh15} detected hard X-rays up to 20\,keV with NuSTAR from the Sgr\,B region.
The photon index is determined to be $\Gamma\sim$2,
similar to typical AGNs. They also reported  the time variability of the Sgr\,B2 cloud using XMM-Newton; the \Ka~flux showed continuous decease from 2001 to 2012\,--\,2013. This is a smooth extension of the decreasing trend in 2000\,--\,2009.  NuSTAR  shows a hint that the flux became constant after 2013 to 2014 at a level of 20\,\% of that in 2001.

\citet{Zh15} discovered a new \Ka~cloud in the Sgr\,B complex, named  G0.66$-$0.13.  This source  is time variable of 
increasing \Ka~flux during 2001\,--\,2013. In 2013, the flux was brighter than the Sgr\,B2 cloud, but in 2014, it became below the detection limit. 
The 8\,--\,12 keV band flux of NuSATR in the non-thermal continuum emission dropped by 50\,\% in 2013 from that measured with XMM-Newton in 2012.
Due to the faintness, however, no detailed spectral information is available for G0.66$-$0.13.

Another \Ka~clump, named  M0.51$-$0.10 is found at the position of the Sgr\,B1 cloud (\cite{Yu07} and figure\,\ref{fig:6.4keV-map}).
With Suzaku, \citet{No08}  studied in detail,  and found that \EWKa~ and $N_{\rm H}$ are $\sim$1.4\,keV and  $\sim1.5\times10^{23}$\,cm$^{-2}$, respectively. 
 Therefore,  M0.51$-$0.10 (Sgr\,B1) is likely another XRN in the Sgr\,B complex. 

As is noted in section 4.1.3, the young SNR candidate G0.570$-$0.018 is discovered with Chandra, then confirmed with ASCA \citep{Se02}.  \citet{In09}, however found  a  time variability of \Ka~line from  this source in the ASCA, Chandra, XMM-Newton and Suzaku observations. The \Ka~line flux changed by factor of $\sim$0.5, but  \Hea~line was constant with time. Thus, G0.570$-$0.018 is an XRN, but the X-ray would be contaminated by the nearby SNR G0.570$-$0.001, which may emit \Hea~and \Lya~lines (section 4.1.3).  

All the \Ka~clumps (candidates of XRNe) near the Sgr\,B complex are listed in table 7.  In this table, the \EWKa~in Sgr\,B1+G0.57$-$0.018 \citep{Yu07} is exceptionally small compared to those of other regions and authors. Since \citet{Yu07} used larger photon collecting areas than those of the other authors, the value of \EWKa~ would be affected by the local background of the GCXE with small \EWKa, and hence gives the smaller \EWKa~than the other authors of the smaller photon collecting areas. On the other hand, the Sgr\,B2+M0.7$-$0.09+G0.66$-$0.13 clouds \citep{Yu07} have a large XRNe area, and hence observed \EWKa~wolud be normal.  From the Sgr\,B2 cloud,  unusually large \EWKa~of $\sim$2\,--\,3 keV are reported by \citet{Mu00, Mu01b, Si01}.  These would be due to the limited spectral resolution and statistics of ASCA and BeppoSAX compared to the other later observations with Suzaku, Chandra and XMM-Newton. 

The X-ray spectra and time variability in the flux and morphology of the \Ka~clumps in the Sgr\,B complex  strongly support the XRN scenario. The \Ka~line  and associated power-law continuum are due to  "an X-ray echo" , or X-ray fluorescence (\Ka) and Thomson scattering (power-law) of an external X-ray source. 
The required flux of the external X-ray source depends on the distance to the \Ka~clump.  Even in the minimum case that the external X-ray source is inside the \Ka~clump of $\sim$1\,pc size, the flux should be larger than  10$^{37}$\Lu. Furthermore, this high flux should be  an averaged value in more than $\sim$1 year.  Since no such Galactic X-ray source is found near the Sgr\,B complex, a unique candidate for the  irradiating X-ray source is the SMBH, \SGRA.  One plausible scenario is that \SGRA~exhibited large flares with the luminosity of more than $\sim10^{39}$\Lu (averaged in $\gtrsim$1\,year). The fluorescent\,/\,Thomson scattered X-rays by the Sgr\,B clouds  have now arrived at the Earth, after  traveling extra time between the direct pass from \SGRA~and the pass via the Sgr\,B clouds, which is a few hundred light-years.
From all the observational results, the Sgr\,B \Ka~clumps are regarded as  the most secure example of the XRN, hence may be called  a prototype of the XRN. 

\citet{Su98, Od11} proposed  that the morphologies, spectra and the time variations of scattered and fluorescent X-rays are the useful diagnostics for the study of the XRN scenario.  The detailed simulation is made by \citet{Od11}.  In order to compare  the observations to this simulation, very fine and complicated observational results are required, which is a subject of future study.

\subsubsection {The Northeast Region from \SGRA: Sgr\,A XRN
 Complex}\label{sec:Radio-Arc} % section 5.2.2

The \Ka~line clumps in the Sgr\,A  complex located at the northeast of \SGRA~are firstly found with ASCA \citep{Ko96}. However, unlike the Sgr\,B complex, no detailed study was made due to its highly complex structure.  
With Chandra, \citet{Ko04} found \Ka~complex in the MC at $(l, b)\sim(0.12, -0.12)$  \citep{Ts97, Ok01}, and proposed these are XRNe candidates. 
\citet{Pa04} found at least three \Ka~clumps (No1, 2 and 3 in table 9) with large  EW$_{6.4}$ of $\sim1$\,keV from this region. 
\citet{Mu07} listed up 2 \Ka~sources named Feature\,1 and Feature\,2, the same sources as No1 and No2, respectively.
They found that Feature\,2 showed flux (4\,--\,10\,keV) decline by 26$\pm{7}$\,\%  from the 2002 to 2004\,--\,2005 observations.  Although, Feature\,1 showed no flux change, this source showed a hint of a morphology change during this period similar to Feature\,2  

\citet{Lu08} found two \Ka~sources near the radio and X-ray filament F10 (table 12). These sources are named  East and West, the same sources as  No 1 and No 2, respectively. Although they suggested electron origin, the discovery of time variability by \citet{Mu07} strongly supports an X-ray irradiation scenario.

\citet{Mu08} listed small size ($\lesssim0\arcmin.1$) filament-like sources near the GC (section 6.1). Some  of them exhibited large \EWKa.  Therefore these would be bright fragments of normal XRNe in the XRN  complex region.  This unusual morphology, on the other hand, leads to suspect different origins than XRN. One plausible idea is that the filament-like \Ka~line source is due to the bombardment of  LECRp on  low-temperature gas  confined by strong magnetic fields. 

\citet{Po10} made the \Ka~line image using $\sim$1.2\,Msec data of XMM-Newton in the time span of $\sim$8\,years. They divide the image into 4 regions, 
Bridge, G0.11$-$0.11, MC1 and MC2. The region of Bridge is further divided into 7 sub-regions (Bridge 1\,--\,7).  The fluxes of the  \Ka~line from these regions show different time variation with each other.  The \Ka~line flux in MC1, MC2,  Bridge 5\,--\,7 are constant with time, those in  Bridge 1\,--\,4 increase with time,  and the other \Ka~clump G0.11$-$0.11 shows decrease with time.  An apparent superluminal motion of a light front is found in Bridge 1.
%% with the time scale of about 15 light-years.  

\citet{Ca12}, on the other hand, divided the XMM-Newton image of the \Ka~line complex into 9 regions, and found flux increasing clumps, B2, C and  D, and decreasing clumps,  B1 and F.  The fluxes in the clumps  A, E, DS1 and DS2 are constant in time.  

Using the Chandra data, \citet{Cl13} examined the details of the time variability and spatial structure of the \Ka~clumps in the regions of \citet{Po10}. 
They divided  Bridge 1 into Br1a\,--\,Br1e  with each size of $26\arcsec\times61\arcsec$, and selected  the same small bright spot Br2f from Bridge 2. 
They found   short flares of the \Ka~line in the 10 years light curves from  Br1a\,--\, Br2f.  The peak position moved in the order of the look-back time in the light-curve of Br1a\,--\,Br2f, during 2005\,--\,2012.  This confirmed the superluminal motion suggested by \citet{Po10}.
Since the flare peak is $\sim$10 time brighter than that in quiescent, they proposed  that the past flare of \SGRA~is not single but composed of multiple flares with  the peak luminosity of $\sim10^{39}$\Lu. They also divided the clump MC1 into MC1a\,--\,f with each size  of  $26\arcsec\times61\arcsec$, and found increases of the \Ka~flux in  MC1c\,--\,MC1f, and decreases in MC1a and MC1b.
They further selected  13 small bright spots in the 4\,--\,8 keV band from
Bridges 1 and 2, G0.11-0.11, MC1 and MC2, with a $15\arcsec$ square, and found time variability in the light curves of the  4\,--\,8 keV band.  The fluxes of the spots,  Br1p, Br1m, Br2o, G011s and MC1j, are increasing.
From the former 3 spots, short flares of time scale of a few years were observed. The fluxes of the spots, G0.11u, G0.11r, MC1k and MC2n, are decreasing during the time span of 10 years. The fluxes of the other four spots were constant with time.

\citet{No10} made a high quality spectrum with Suzaku from the  brightest K$\alpha$ region at the northeast of \SGRA.  They fitted the spectrum with a combined model, two CIE plasmas and a power-law continuum with many Gaussian lines. The CIE plasmas represented the background spectrum (GCXE) in these clumps. They discovered K$\alpha$ lines from neutral Ar, Ca, chrome (Cr) and manganese (Mn), in addition to the already known K-shell lines from the neutral Fe and Ni.
The best-fit parameters of the Gaussian lines are given in table 8. 
They determined the Fe abundance in the GCXE around these XRNe to be $\sim$1.1\,--\,1.2\,solar.
Assuming the scattering angle of $\theta=\timeform{90D}$, the  \EWKa~of $1150\pm{90}$\,eV in these  XRNe is converted to the Fe abundance 
of $1.2\pm{0.1}$\,solar\footnote{The Fe abundance in the original paper by \cite{No10} is 1.6\,solar, but they used different abundance table.}.
Thus, the Fe abundance in the GC region estimated with \Ka~line in the XRNe and  \Hea~line in the GCXE, are almost the same of $\sim$1.1\,--\,1.2\,solar, consistent with those of \citet{Ko07c, Uc13, Mu04a}\footnote{This Fe abundance is corrected to that in the same model as the other authors, which includes a power-law component.}. 
The EW of the K$\alpha$~lines of Ar\emissiontype{I}, Ca\emissiontype{I}, Cr\emissiontype{I} and Ni\emissiontype{I} are roughly consistent with the solar abundances (table 8). 

\begin{table}[!ht] %table 8
\caption{K-shell lines from heavy elements from the Radio arc region (after \cite{No10}).}
\begin{center}
\begin{tabular}{lccc}
\hline
Line  &Energy 	& Intensity	& EW \\ 
	&(keV)	&(\dag)	&(eV)\\	
\hline
Ar\emissiontype{I}-K$\alpha$&	2.94$\pm{0.02}$&	170$^{+60}_{-40}$&	140$\pm{40}$\\ 
Ca\emissiontype{I}- K$\alpha$&	3.69$\pm{0.02}$&	54$^{+14}_{-9}$&	83$\pm{13}$ \\
Cr\emissiontype{I}- K$\alpha$&	5.41$\pm{0.04}$&	9.5$\pm{2.5}$&		24$\pm{7}$\\ 
Mn\emissiontype{I}- K$\alpha$ &	5.94$\pm{0.03}$&	7.4$\pm{2.2}$&		22$\pm{7}$\\ 
\Ka&					6.404$\pm{0.002}$&	340$\pm{10}$&		1150$\pm{90}$\\ 
\Kb&					7.06(fixed)&		40$\pm{3}$&		160$\pm{20}$\\ 
Ni\emissiontype{I}-K$\alpha$ &		7.48$\pm{0.02}$&	18$\pm{3}$&		83$\pm{13}$\\ 
\hline
\end{tabular}

 % table 8
\end{center}
{\footnotesize
$\dag$ In  unit of $10^{-6}$ photons~s$^{-1}$~cm$^{-2}$.
}
\end{table}

NuSTAR detected non-thermal continuum X-rays  \citep{Mo15} spatially correlated with the \Ka~ fluorescence line  
from the two \Ka~clumps, MC1 and Bridge. 
They made a Monte-Carlo simulation with the XRN  model for the broad-band X-ray spectrum.  Then, they determined that the intrinsic column density is $\sim10^{23}$~cm$^{-2}$, the primary X-ray spectrum  of a power-law has photon index ($\Gamma$) of $\sim2$, and the flare luminosity of \SGRA~is  $\gtrsim10^{38}$\Lu.

\citet{Lu08} found  a faint source at the east of the southernmost extension of the Radio Arc \citep{Jo09}, and named G0.017$-$0.044.  Since G0.017$-$0.044 has a reasonably large \EWKa~of $\sim$0.62\,keV, this may be an XRN, one of the nearest (in projection) XRN from \SGRA. However, no detailed spectral information to judge the reliable origin is available.

G0.174$-$0.233 is discovered with Suzaku near the Radio Arc \citep{Fu09}. The spectrum exhibits bright \Ka, Ca\emissiontype{I}-K$\alpha$ and a hint of Ar\emissiontype{I}-K$\alpha$ lines. The \EWKa~is $\sim$0.95\,keV, typical to XRN.  The detection of the Ca\emissiontype{I}-K$\alpha$ and  Ar\emissiontype{I}-K$\alpha$
lines from G0.174$-$0.233 \citep{Fu09} is the second case after the Sgr\,A XRN \citep{No10}.  

The physical parameters of the XRNe  in the Sgr\,A complex are summarized in table 9.  Since the Sgr\,A  region is very crowded with many  XRNe, some XNRe overlap with those reported by other authors.  
The most important parameter, \EWKa~should be time constant, regardless the time variable  flare fluxes of  \SGRA, and  free from the  observed ambiguity of the $N_{\rm H}$ values.  If the iron abundances and scattering angle $\theta$ are the same among the XRNe, the \EWKa~should be the same in all the XRNe (equation 4).
Nevertheless,  the reported  \EWKa~shows apparent and systematic variations among the authors and instruments, and in each XRNe. 
For example, No3 \citep{Pa04}, G0.11$-$0.11 \citep{Po10} and F \citep{Ca12} are  the same XRN in position with a similar collecting area.  However, the observed \EWKa~are largely different of $\sim$1.3, $\sim$1.0 and $\sim$1.7\,keV, respectively. 

 The value of the \EWKa~depend on the the continuum flux of the XRNe, which is sensitive to the subtraction of local GCXE. The flux of local GCXE is position dependent, because the SH is very small of $\sim0\arcdeg.25$. 
Since the Sgr\,A complex is crowded with many XRNe, the background (GCXE) position  is very limited. This situation causes the non-negligible \EWKa~difference among the authors and XRNe, which may be called  the systematic errors.
An example of this  systematic error is found in the  \EWKa~and $N_{\rm H}$ differences between \citet{Ca12} and \citet{Po10}. 
The total areas of the total XRNe by \citet{Po10} and \citet{Ca12} are almost the same. However, the mean \EWKa~and $N_{\rm H}$ are, respectively $\sim$0.78\,keV and $\sim6.5 \times10^{22}$ in \citet{Po10}, and $\sim$1.2\,keV and $\sim12\times10^{22}$ in \citet{Ca12}. The background of \citet{Po10} is taken from the source free regions in the GC-west, where the GCXE is systematically lower than the GC-east, near the Sgr\,A XRNe complex (see section 9.3). Therefore the continuum emission of the XRNe is under-subtraction, giving systematically smaller \EWKa~and  $N_{\rm H}$~than those of \cite{Ca12} and any other authors.

Accordingly, the values of the  physical parameters in table 9 should be carefully treated. Taking account of the possible errors of the \EWKa~ in table 9, it is still  worth to note that the \EWKa~ in all the XRNe are roughly consistent with $\sim$1\,keV.

\begin{table*}[!ht] %Table 9
\caption{The 6.4 keV clumps near the Radio arc region.} 
%\footnotesize
\begin{center}
\begin{tabular}{lccccll}
\hline\
Name $(l,\ b)^\ast$ &$N_{\rm H}$&$EW_{6.4}$ &$F_{\rm x}$~(2--10~keV)&Var$^{\#}$ & Area& 
Instrument$^\dag$\\
&($10^{22}$cm$^{-2}$) & (keV)& $(10^{-12}$erg~s$^{-1}$cm$^{-2}$)& -&(Arcmin$^2$)& 
Reference$^\ddag$ \\

\hline
No1~(0.023, $-$0.053)& $32.9^{+4.0}_{-4.2}$ & $1.19\pm0.10$ &1.7&-&0.5 &C, a\\
No2~(0.045, $-$0.081)&$36.8^{+9.6}_{-13.8}$ 
& $1.03^{+0.37}_{-0.23}$&3.1&-&1.5 \\
No3~(0.121, $-$0.137) & $15.8^{+4.1}_{-2.6}$& $1.29\pm{0.10}$&2.0&-&1.3 \\

\hline
All~(0.08, $-$0.08)	&-& $0.67 \pm{0.05}$	&56 	&-	&123	&C, b \\
All~(0.08, $-$0.08)&12.0$\pm{1.1}$ &1.15$\pm{0.09}$ &- &- & 45 &S, h
\\

\hline
Feature\,1~(0.023, $-$0.053)
&$36^{+5}_{-3}$  &$1.00^{+0.24}_{-0.09}$ 
&$0.52^{+0.02}_{-0.03}$~(4--8~keV)$^{||}$ &N&0.9&C, c\\
Feature\,1~(0.023, $-$0.053)
 	&$34^{+5}_{-3}$  &$1.01^{+0.19}_{-0.09}$ 
&$0.48^{+0.02}_{-0.03}$~(4--8~keV)$^\S$ &N&0.9& \\
Feature\,2~(0.045, $-$0.081)
 	&$40^{+20}_{-6}$  &$0.93^{+0.16}_{-0.16}$ 
&$0.77^\pm{0.03}$~(4--8~keV)$^{||}$&Y&2\\
Feature\,2~(0.045, $-$0.081)
 	&$75^{+26}_{-18}$  &$0.69^{+0.22}_{-0.22}$ 
&$0.55^\pm{0.03}$~(4--8~keV)$^\S$ &Y&2\\

\hline 
East clump~(0.023, $-$0.053)& $14\pm{0.2}$ 
& $0.75^{+0.08}_{-0.07}$ &1.1&-&0.9& C, d\\
West clump(0.045, $-$0.081)&$23^{+0.6}_{-0.7}$ 
& $1.07^{+0.16}_{-0.13}$ &1.1&-&2& \\

\hline
G0.014$-$0.054	&-	&$0.87^{+0.45}_{-0.38}$	&0.08~(2--8 keV)  &-&0.072&C,  e \\
G0.021$-$0.051	&-	&$1.0^{+0.57}_{-0.47}$ 	&0.07~(2--8 keV)&-&0.052 \\
G0.039$-$0.077	&-	&$0.66^{+0.35}_{-0.27}$ &0.11~(2--8 keV) &-&0.093\\ 
G0.062+0.010	&-&$1.58^{+1.38}_{-1.35}$ &0.06~(2--8keV) &-&0.33\\ 
G0.097$-$0.131	&-	&$1.19^{+0.42}_{-0.35}$ &0.48~(2--8 keV)&-&1.2\\

\hline
G0.017$-$0.044&-&$0.62^{+0.58}_{-0.34}$ &0.04&-&0.01&C, f \\
G0.174$-$0.233&7.5$^{+2.0}_{-1.7}$&0.95$^{+0.18}_{-0.19}$&0.48&-&3& S, g\\

\hline
Bridge$^{**}$~(0.09,$-$0.08)&$4\pm3$ 
&$0.75^{+0.05}_{-0.03}$$^\star$ &-&Y/N&7.4&X, i; C, k\\
G0.11$-$0.11 	& $7\pm4$ 	& $0.96\pm{0.06}$$^*$	&- &Y	&14\\ 
MC1~(0.020, $-$0.052) & $10_{-2}^{+1}$ 	& 
$0.68^{+0.07}_{-0.02}$$^\star$ &-&N	&2.1\\
MC2~(0.035, $-$0.096)  &$5_{-4}^{+5}$ 	
&$0.72^{+0.12}_{-0.07}$$^\star$&-&N	&1.8\\
\hline
A~(0.022, $-$0.052)	&$18.4_{-2.7}^{+1.5}$&0.9$\pm{0.1}$	&-	&N&2.2&X, j \\
B1~(0.028, $-$0.077)	&$10.2_{-2.0}^{+2.3}$&0.9$\pm{0.1}$	&-	&Y&2.6\\
B2~(0.055, $-$0.083)	&$12.3_{-2.7}^{+3.0}$&1.5$^{+0.3}_{-0.2}$	&-&Y&1.9\\
C~(0.048, $-$0.053)	&$5.8_{-1.9}^{+2.3}$&1.0$^{+0.2}_{-0.1}$	&-&Y&2.0\\
D~(0.072, $-$0.072)	&$13.2_{-4.7}^{+5.0}$-&1.3$^{+0.4}_{-0.3}$	&-&Y&1.8\\
E~(0.090, $-$0.087)	&$9.6_{-1.4}^{+1.7}$&1.4$\pm{0.2}$	&-&N&2.9\\
F~(0.125, $-$0.115)	&$9.2_{-2.3}^{+2.7}$&1.7$\pm{0.2}$	&-&Y&15&  \\
DS1~(0.065, $-$0.020)	&$15.5_{-3.3}^{+3.9}$&0.9$\pm{0.2}$	&-	&N&3.7 \\
DS2~(0.025, $-$0.012)	&14.5$\pm{2.3}$	&0.9$\pm{0.1}$	&-  &N&6.5\\
\hline
\end{tabular}
 %table 9
\end{center}
{\footnotesize
$\ast$ Same as table 7, but for the XRNe in the Sgr\,A complex. \\
$\#$  With multiple observations, flux variability is found (Y) or not(N).\\
$\dag$ Instrument, C: Chandra,  S: Suzaku, X: XMM-Newton.\\
$\ddag$ Reference, a: \citet{Pa04}, b: \citet{Yu07}, c:  \citet{Mu07}, d:\citet{Lu08},
e: \citet{Mu08}, f:\citet{Jo09}, g: \citet{Fu09}, h: \citet{No10}, i: \citet{Po10}, j: \citet{Ca12},  k: \citet{Cl13}.\\ 
$||$ Observed in 2002.\\
$\S$  Observed in 2004\,--\,2005.\\
${**}$ This region is separated to subgroups, Bridge 1\,--\,7.\\
$\star$ Errors are estimated from the flux errors of the \Ka~line.\\ 
}
\end{table*}

\subsubsection{Sgr\,C, D, and E }\label{sec:sgr-cde} % section 5.2.3

\begin{table*}[!ht] %table 10
\caption{The 6.4 keV clumps near the Sgr\,C complex. }
\smallskip
\begin{center}
\begin{tabular}{lcccll}
\hline
Name($l, b$)$^\ast$ &$N_{\rm H}$ & $EW_{6.4}$	&$F_{\rm x}$ (3--10 keV)$^\star$ & Area & 
Instrument$^\dag$\\
&($10^{22}$cm$^{-2}$) &(keV)&($10^{-12}$~erg~s$^{-1}$cm$^{-2}$)	&(arcmin$^2$)& Reference$^\ddag$\\
\hline
Sgr C ($359.42, -0.04$)& $12.6^{+3.5}_{-3.3}$    
&0.8$^{+0.4}_{-0.5}$ &6 (4--10 keV) &23	&A, a\\
G359.45+G359.42 & - & 0.47$\pm{0.10}$ 	& 12 (2--8 keV)	   &77	&C, b\\
\hline
M359.43$-$0.076&$9.2^{+4.9}_{-4.4}$&$2.2^{+0.3}_{-0.4}$&0.27&7.1	&S, c\\ 
M359.47$-$0.15&$8.2^{+3.6}_{-1.7}$& $2.0^{+0.2}_{-0.2}$ & 0.41 &6.9\\
\hline
C1 (359.43, $-$0.076)	&11.6$^{+0.6}_{-0.8}$	&1.1--1.5& -&13	&S, d \\ 
C2 (359.47, $-$0.15)	&17.9$^{+1.9}_{-1.6}$	&1.1--1.6& - &13	\\
C3 (359.58, $-$0.13)	&13.7$^{+0.7}_{-0.8}$	&0.7--1.3& - &31	\\
\hline
\end{tabular}
% table10 
\end{center}
{\footnotesize
$\ast$ Same as table 7, but for the XRNe in the Sgr\,C complex.\\ 
$\star$  Unabsorbed flux, but reference (c) is absorbed flux. Two separate observations of C1 made in 2006 and 2010 show a time variability. \\
$\dag$ Instruments. A:ASCA, C:Chandra,  S: Suzaku \\
$\ddag$ Reference, a: \citet{Mu01a}, b: \citet{Yu07}, c: \citet{Na09}, 
d: \citet{Ry13}.\\
}
\end{table*}

The Sgr\,C complex is a unique star-forming region in the west of the CMZ, which is located at the mirror point of the Sgr\,B complex with respect to \SGRA.  This complex is composed of high-mass YSOs,  H\emissiontype{II} regions and radio non-thermal filaments (NTF) (\cite{Ke13} and references therein). 
ASCA found a diffuse hard X-ray emission from the Sgr\,C complex \citep{Mu01a}. 
The X-ray spectrum is characterized by a large  EW$_{6.4}$ of $\sim 
1$\,keV and a large absorption column of $\sim10^{23}$\,cm$^{-2}$, suggesting that the X-rays are due to fluorescence and scattering of external X-rays.  No adequately bright source in the immediate vicinity of the Sgr\,C complex 
to account for the fluorescence flux is found. 
Thus, with the same reason of the Sgr\,B complex, the irradiating X-ray source would be a past bright flare of \SGRA; the Sgr\,C complex is the second XRN complex discovered after the Sgr\,B complex.

With Chandra, \citet{Yu07} found 5 X-ray spots from the Sgr\,C complex in the 2\,--\,6 keV band image. 
%which are G359.46$-$0.15, G359.45$-$0.07, G359.42$-$0.12, G359.40$-$0.07 and G359.32$-$0.16. 
Two of them, G359.45$-$0.07 and G359.42$-$0.12, are diffuse sources with the size of $\lesssim\timeform{4'}$. They made an X-ray spectrum  from the region including these two diffuse sources.  The mean \EWKa~ is $\sim$470\,eV. 

Suzaku found four diffuse clumps near  the Sgr\,C region \citep{Na09},  named M359.47$-$0.15, M359.43$-$0.07, 
M359.43$-$0.12 and M359.38$-$0.00. Two of them, M359.47$-$0.15 and M359.43$-$0.07 are prominent in the \Ka~line image. The EW$_{6.4}$~values are very large  of $\sim$2.0\,--\,2.2\,keV, nearly 2 times larger than those of the normal XRNe in the Sgr B and Sgr A complexes. This is  puzzling, but possibly some systematic errors are involved. The absorptions ($N_{\rm H}$) are $\sim10^{23}$\,cm$^{-2}$, consistent with that the sources are in the MCs of the Sgr\, C  complex.

\citet{Ry13} re-analyzed the same region, and found 3 \Ka~diffuse sources, named  C1, C2 and C3, where the positions of C1 and C2 coincide to those of M359.47$-$0.15 and M359.43$-$0.07, respectively. These sources have reasonable EW$_{6.4}$ of $\sim$1.1\,--\,1.6\,keV.
The \Ka~line of C1 increased by 8\,\%  (2.9$\sigma$ confidence) from the 2006 to the  2010 observations.  The time variability of C1, and the large \EWKa~of 1\,--\,1.6\,keV of C1, C2 and C3 favors the X-ray irradiation scenario (XRNe). 

The summary of the \Ka~sources  in the Sgr\,C complex is given in table 10.  The small \EWKa~value of $\sim$ 470 eV by \citet{Yu07} is marginal whether the origin is LECR electron or X-ray irradiation (XRN). However, the small  \EWKa~value would be due to the larger correcting area of 77 arcmin$^2$ than \citet{Na09} and \citet{Ry13}, and hence the spectrum would be contaminated by the nearby GCXE (see section 5.2.2).  In fact, \citet{Na09} observed the \EWKa~value from nearly the same area of \citet{Yu07} with Suzaku  and found that the \EWKa~is 460\,eV. 

The Suzaku extensive observations near the edge of the GCXE revealed  strong \Ka~lines  from the Sgr\,D and E complexes 
(see figure\,\ref{fig:6.4keV-map}).
Although no detailed follow-up analysis of these complexes have been made, the similarity in spectra and morphologies to those 
of the Sgr\,B,  A and  C  complexes suggests that the Sgr\,D and E complexes  also contain  XRNe. 

\subsection{\Ka~Clumps Other than XRNe} % section 5.3
\label{sec:sgr-cde} 

This section reviews the other \Ka~clumps than those in section 5.2. The origins may not be an X-ray irradiation from \SGRA,  but the irradiation sources would be  either local nearby bright X-ray stars, LECRe or LECRp. 

\subsubsection{Arches Cluster} %section 5.3.1

\begin{table*}[!ht] % table 11
\caption{The 6.4 keV clumps near Arches Star clusters.}
%\smallskip
\begin{center}
\begin{tabular}{lcccll}
\hline
Name ($l, b$)$^\ast$	&$N_{\rm H}$ 	&$EW_{6.4}$ 	&$F_{\rm x}$ (2--8 keV)	&Area
&Instrument$^\dag$\\
	&($10^{22}$cm$^{-2}$) &(keV)	&$(10^{-12}$erg~s$^{-1}$cm$^{-2}$)& (Arcmin$^2$)
&Reference$^\ddag$\\
\hline
SE extension (0.1, 0.02)&$6.2^{+2.7}_{-5.6}$ & $1.4^{+0.9}_{-0.5}$ & 0.54	&0.5	&C, a\\ 
Loop (0.1, 0.05)&$7.2\pm{1.4}$		& $1.00\pm{0.25}$ & -&3.1	&X, b\\ 
Arches  (0.1, 0.03)	&-	&0.81$\pm{0.2}$ 	& 1.2	&1.6	 &C, c\\ 
Arches  (0.1, 0.03) &14$\pm5$&$\sim$1.25--1.42& 1.1 (3--10 keV)&6.2	&C, S, d\\ 
\hline
N (0.13, 0.02)	&9.5$\pm{1.5}$		&1.0$\pm{0.4}$	&-&0.5&X, e	\\
S (0.12, 0.01)	&10.1$\pm{0.7}$	 	&0.9$\pm{0.2}$	&-&0.5&  \\
SN (0.10, 0.01)	&8.5$^{+4.0}_{-3.4}$ 	&1.1$\pm{0.4}$	&-&0.9&\\
DX (0.10, 0.05)	&6(fixed)		&2.6$^{+2.1}_{-1.1}$	&-&0.8&\\
\hline
Cloud region (0.12, 0.02) &$11.3^{+1.9}_{-1.3}$ &1.2$\pm 0.2$&- &1.1&X, f \\
Cloud region (0.12, 0.02) &9.5(fixed) &1.1$^{+0.7}_{-0.5}$   &1.5 (3--20 keV) &1.1&N, g \\
Cloud region (0.12, 0.02) &6.0$\pm{0.3}$ &0.9$\pm{0.1}$&-&1.1&X, h\\
\hline
\end{tabular}

 % Table 11
\end{center}
{\footnotesize
$\ast$ Same as table 7, but for the \Ka~clumps in the Arches cluster. \\
$\dag$ Instruments. C:Chandra,  N: NuSATR, X: XMM-Newton\\
$\ddag$ Reference, a: \citet{Wa06b}, b: \citet{Sa06}, c: \citet{Yu07}, d: \citet{Ts07}, e: \citet{Ca11b}, f: \citet{Ta12}, g: \citet{Kr14}, h:\citet{Cl14}.\\
}
\end{table*}

In addition to the thermal hot plasma (section 4.1.6) in the central region, \citet{Yu02b, La04} found extended emission, named  A3 at the southeast from the core of the Arches cluster with Chanda.  The source A3 has a power-law spectrum with possibly \Ka~lines. However no detail of the  spectrum was reported.  Then after, \citet{Yu07} reported  the \Ka~line emission with \EWKa~of $\sim$0.8\,keV.
A long exposure Chandra observation by \citet{Wa06b} found that the thermal plasma mainly comes from the center region of the cluster, but a strong  \Ka~line  with    power-law continuum emission is coming from  the southeast region of the star cluster, named SE extension. Since the \EWKa~is $\sim$1.4\,keV, most likely origin is bombardment of molecular gas by X-ray photons or low energy protons.

\citet{Ts07} made the Suzaku spectrum from the whole region of the Arches cluster, because Suzaku had no good spatial resolution. The spectrum is fitted with a model of a CIE plasma plus power-law continuum with two Gaussian lines  at 6.4\,keV (\Ka) and 7.05\, keV (\Kb).
Then, the power-law spectrum has a photon index of $\sim$0.7, but no pronounced iron K-edge feature at 7.1\,keV is found. 
Since the narrow band image of 7.5\,--\,10.0\,keV  shows a similar distribution to that of the \Ka~line flux, the \Ka~and \Kb~lines are associated to the  power-law component with the \EWKa~of $\sim$1.42\,keV.
They also examined the Chandra spectra, and found that the power-law index and \EWKa ~to be $\sim$1.2 and $\sim$1.25\, keV, respectively.

XMM-Newton found a big loop-like annular structure in the \Ka~band  of $\sim\timeform{3'}$\,diameter and $\sim\timeform{1'}$\,width around the star cluster, named  Loop \citep{Sa06}. 
They  made an X-ray spectrum from the brightest part of Loop at the southeast of the cluster center, and fitted with a model of a power-law plus a Gaussian line. The best-fit absorption column and photon index are $(7.2\pm{1.4})\times10^{22}$\,cm$^{-2}$ and $1.4\pm{0.6}$, respectively.  The  \EWKa ~is $1.0\pm{0.25}$\,keV.
\citet{Ca11b} found  three  \Ka~ clumps, N, S and  SN. These  comprise a part of the loop-like structure of \citet{Sa06}. The \EWKa~ are all within the range of 0.9\,--\,1.1\, keV. They interpreted that N, S and  SN  are explained by the MC bombardment of LECRs from the Arches cluster stars. 
\citet{Ta12} found a strong \Ka~line from nearly the same regions of N and S, with the \EWKa~of $\sim$1.2\,keV. Since no time variability of the \Ka~line during 2000-2010 is found, they claimed that the origin would be  LECRp. 
 Possible supersonic collision between the stellar wind from star clusters and MCs would make strong shock and hence would become efficient particle acceleration, which makes enough LECR to produce the \Ka~line and power-law flux. 
Although it is not clear whether the origin is  X-rays or protons, the candidate source would be related to high-mass stars in the Arches cluster. 

\citet{Kr14} detected diffuse X-rays up to $\sim$30\,keV with NuSTAR. 
The emitting region is an ellipse of northwest-southeast major axis,
nearly the same regions of N and S of \citet{Ca11b}. They determined the \EWKa~to be $\sim$1.1\, keV. The wide band X-ray spectrum is in broad agreement with the LECRp origin.
Using the XMM-Newton data from 2000 to 2013, \citet{Cl14} examined a long-term time variability of the power-law  emission in the same  eclipse of \citet{Kr14}.  The \EWKa~is $\sim$0.9\,keV.  They found a  flux drop by 30\,\% in 2012, and hence constant flux hypothesis is rejected with more than 4\,$\sigma$ confidence.  From this time variability, they interpreted  that  the power-law emission is due to the reflection of an X-ray transient source in the Arches cluster. 

Most of the authors suggested that the irradiation source for the \Ka~emission is active stars in the core of the Arches cluster, because of extreme activity of the embedded stars than the other star clusters.  If the irradiation sources are the cluster stars in the core, with the mean distance of $\sim$1\,pc from the \Ka~diffuse sources, the X-ray luminosity  should be $>$100 times brighter than any of the observed results of $\sim4\times 10^{33}$\Lu~\citep{Ca11a}.
Therefore, with the same reason as the other XRN candidates (section 5.1), it is also possible that the \Ka~emission from the Arches cluster is  due to the  past flare of  \SGRA.  The luminosity and epoch of the flare are not largely different from those estimated from the XRNe in the Sgr\,A complex, because their positions, \Ka~fluxes and luminosity are not largely different from those of the Sgr\,A  XRNe (table 9).
Thus, the true origin of \Ka~line in this star cluster is a puzzle, whether the  irradiation source is cluster stars or \SGRA, whether it is X-ray or LECRp.

In the XMM-Newton image, \citet{Ca11b} found the other \Ka~ clump, DX  at about 16~pc  west of the cluster center.  The \EWKa~of  DX is very large of $\sim$2.6\,keV. A short timescale variability is found from DX. These are unusual compared to the other \Ka~clumps in the Arches cluster.  Together with the large distance from the cluster, DX would be unrelated object to the Arches cluster.
 
The summary of the \Ka~ line emission from the Arches cluster are listed in table 11. All the \EWKa~ are around $\sim$1\,keV. The clump DX has an exceptionally large \EWKa~of $\sim$2.6\,keV, although the error is large.  As is noted, DX would be unrelated object to the Arches cluster.    

\subsubsection{\Ka~Clump Near the Great Annihilator}
\label{sec:Great-Ani} % section 5.3.2

Suzaku found two diffuse X-ray sources with strong K-shell lines near 1E\,1740.7$-$2942, the Great Annihilator by \citet{Na10}.  One is an SNR candidate G359.12$-$0.05 with a strong \SHea~line (section 4.2.6). The other source, M359.23$-$0.04 has a prominent \Ka~line and locates at the northeast of 1E\,1740.7$-$2942.
The \EWKa~is as large as $\sim1.2$\,keV, and is associated with a MC in the radio CS (J$=1-0$) map \citep{Ts99}. The $N_{\rm H}$ is $\sim3\times10^{23}$ cm$^{-2}$.  Thus, the \Ka~line from M359.23$-$0.04 is likely due to X-ray fluorescence irradiated by an external X-ray source. 
One possible candidate of the external source is  1E\,1740.7$-$2942 itself. Assuming the photon index of 1.4, \citet{Na10} estimated that the required luminosity of  1E\,1740.7$-$294 is $\sim$4\,--\,5$\times10^{36}$\Lu. This is consistent with the observed  1E\,1740.7$-$2942 luminosity of $2.6\times10^{36}$\Lu, if possible systematic errors and time variability are  taken into account. The combination of M359.23$-$0.04 and 1E\,1740.7$-$2942 is a rare case of the association of diffuse \Ka~line and bright binary X-ray source.

\subsubsection{G0.162$-$0.217} % section 5.3.3
From the Radio Arc region, a small spot of the \Ka~line is found at $(l, b)= (\timeform{0.162D}$, $\timeform{-0.217D}$) with Suzaku, 
named  G0.162$-$0.217 \citep{Fu09}. 
This source is located adjacent to the south end of the Radio Arc \citep{La00, Yu04}. The Radio Arc is a site of relativistic electrons, 
which may include LECRe along the magnetic field line of the Radio Arc. Thus, G0.162$-$0.217 would be made by the LECRe. 
The observed \EWKa~is $\sim0.2$\,keV, consistent with the LECRe bombarding scenario.  This type of the faint \Ka~line emitter would be 
more numerous in the GCXE (section 6), and may contribute significant fraction of \Ka~flux in the GCXE (section 9.3). However limited statistics of the current instruments does not allow further search of these potential LECRe sources of the \Ka~lines.

\section{Small Size X-ray Sources or Non-Thermal X-Ray Filaments} 
\label{sec:non-thermal} % section 6

This section reviews small scale diffuse plasmas discovered mainly
with Chandra in the central region (section 6.1) and the outer region (section 6.2) of the GCXE.  The shapes of the diffuse plasmas are mostly either filamentary or cometary  with  non-thermal X-ray spectra. In this section, these sources are called  the X-ray filament. 
The summed luminosity (2\,--\,10\,keV) of the X-ray filaments is $\sim5\times10^{34}$\Lu, only $\sim$4\,\% of the 2\,--\,10\,keV band luminosity of the GCXE ($\sim1.2\times10^{36}$\Lu).

\begin{table*}[!ht] % table 12
\caption{Non-thermal X-ray filament within the 17 arcmin Region.$^\ast$} 
\begin{center}
%\footnotesize 
\begin{tabular}{lcccll}
\hline
Name& Size & EW$_{\rm Fe}$ & $L_{\rm X}$ (2$-$8~keV) &Comment& Reference$^\ddag$\\
$(l, b)$ & (arcsec$^2$) &(eV) & ($10^{32}$\Lu)\\
\hline
G359.945$-$0.044	& 22 & $\le$60	&66& in the Cavity, PWN ?& a, b, c, d, n\\ 
G359.942$-$0.045	& 28 & $\le$220	&16& in the Cavity & a, c \\ 
G359.944$-$0.052	& 20 &$\le$470	&2& Jet, in the Spiral & a, c, j\\ 
G359.950$-$0.043	& 55 &$\le$140	&4& in the Spiral  & a\\ 
G359.933$-$0.039	& 15 &-			&-&F1 & a, b, c \\
G359.956$-$0.052	& 10 &-			&1& in Sgr A East& a\\
G359.933$-$0.037	& 13 &$\le$170	&3& & a, b\\ 
G359.941$-$0.029	& 17	&$\le$180	& 2&F2 Stellar Wind ?& a, c \\ 
G359.925$-$0.051	& 19 &$\le$2060	&3&& a \\
G359.964$-$0.053	& 76 &$\le$70&17& in Sgr A East, PWN ?, F3 & a, b, c, e \\ 
G359.965$-$0.056	& 29 &-	&3& in Sgr A East, F4& a, b \\
G359.921$-$0.052	& 12 &-&-&& a \\ 
G359.962$-$0.062	& 26 &-&2& in Sgr A East	& a \\ 
G359.959$-$0.027	& 34 &$\le$75		&6& knot-1, F5 & a, b, c, f  \\ 
G359.971$-$0.038	&148&$\le$130		&10& 	PWN ?, F6& a, b, c  \\ 
G359.969$-$0.033	& 17 &-			&- & PWN ?& a \\ 
G359.921$-$0.030	& 30 &$\le$1300		&3	&F7 	& a, b \\
G359.915$-$0.061	& 22 &-			&-& & a\\ 
G359.983$-$0.040& -&-		&-&  & a\\
G359.904$-$0.047	& 32 &-			&1& &a \\ 
G359.977$-$0.076	& 26 &-			&-& &a \\ 
%G359.95$-$0.04	&16&	&13&  PWN ? & b, c, d\\
G359.970$-$0.008	& 30 &$\le$110	&6	& knot-2  PWN ?, F8& a, b, c, f, m \\
G359.899$-$0.065	& 30 &-			&3& & a \\
G359.897$-$0.023	& 42 &-			&2&& a \\ 
G359.889$-$0.081	&432&$\le$30		&70& Radio, Sgr A-E, PWN ? & a, b, c, g, h, l, n, o \\
%G0.014$-$0.054	&260&866$_{-252}^{+300}$ & 8 &XRN$^\ddagger$& a\\
G0.008$-$0.015 	&51& - 	&-&&a, c \\ 
%G0.021$-$0.051	&190& $998^{+378}_{-312}$& 7 &XRN& a, c\\ 	
G0.032$-$0.056 	&429&$\le$110	&10&& a, b, c \\ 
G0.029$-$0.080 	&838& - 	&-&& a \\ 
%G0.039$-$0.077	&333&$661^{+230}_{-183}$ &13 &XRN & a \\ 	
%G0.062+0.010	&1187&$1576^{+921}_{-895}$&9 &XRN& a \\ 
%G0.097$-$0.131	&4181&$1193^{+275}_{-231}$ &100 &XRN & a \\
G0.116$-$0.111 	&2257& - 	&-&& a\\ 
\hline
G0.029$-$0.06 	&$6\times47$& - 	&11&Radio, PWN?, F10& b \\
%%G359.959$-$0.028& 40 &-		&6.5 &&b, c \\ %6
%%G359.970$-$0.009& 40 &-		&7.8&&b, c \\  %10
G359.974$-$0.00&$4\times7$ &-		& 4 	&knot-3, F9 & b, f\\ %11
G359.983$-$0.046& -&-		&31	&Canonball, PWN ? & b f\\ 
%%G359.89$-$0.08& 240 &- 		& 210 &  &b \\ %12
G359.90$-$0.06&$3\times6$&-		& 108 	&Radio, Sgr A-F&b, o\\ %14
\hline
\multicolumn{3}{l}{Total luminosity (2--10\,keV)}
&3.8$\times10^{34}$\Lu& & \\
\hline
\end{tabular}

  % table 12
\end{center}
%\normalsize
{\footnotesize 
$\ast$ Sources and parameters in upper rows  are taken from \citet{Mu08} in the order of angular distance from \SGRA, while those of rower rows  are from  \citet{Lu08}. The overlapping sources from both  \citet{Mu08} and \citet{Lu08} are combined in the upper rows.  The sources observed by other authors with the separation angle of $\lesssim0.3\arcmin$~are treated as the same objects.\\ 
%%$\dag$ Instrument, A: ASCA, C: Chandra, B: BeppoSAX, S: Suzaku, N: NuSTAR, X: XMM-Newton.\\
$\ddag$ Reference, a: \citet{Mu08}, b: \citet{Lu08}, c: \citet{Jo09}, d: \citet{Wa06a}, e: \citet{Ba03}, f: \citet{Ko04}, g: \citet{Sa03}, h: \citet{Lu03}, i: \citet{Pa05}, j: \citet{Li13}, k: \citet{Ny13}, l:  \citet{Ny14}, m: \citet{Ny15}, n: \citet{Mo15}, o: \citet{Yu05}.\\
}
\end{table*}

\begin{table*}[!ht]  % table 13
\caption{Non-thermal X-ray filament in the $\timeform{0.5D} \times \timeform{1D}$ region, excluding the inner $17'$ region.$^\ast$}
\label{tab:Joh} \begin{center}
\begin{tabular}{lccll}
\hline
Source name &Size& $L_{\rm X}$ (2$-$10~keV)&Comment &Reference$^\ddag$\\
&(arcsec$^2$) &($10^{32}$\Lu) \\
\hline
G0.223$-$0.012	& 6$\times$80	&15& &a\\
G0.13$-$0.11	& 7$\times$33	&40& Radio, PWN? & a, b, c, e\\
G359.55$+$0.16	& 4$\times$21 	&20&Radio, X-ray thread & a, b, f, g, h \\ %%6.7 keV line
G359.43$-$0.14	& 4$\times$21 	&5&&a\\	
G359.40$-$0.08	& 5$\times$27 	&24&&a\\
G0.17$-$0.42 &180$\times$18 &-& near the Radio Arc & d\\
\hline
\multicolumn{2}{l}{Total luminosity (2--10\,keV) }
&1.0$\times10^{34}$\Lu & & \\
\hline
\end{tabular}

%\hline
%G359.15$-$0.2 &312$\times$54 &&Snake\\
%G359.54+0.18 &56$\times$8 &&Ripple filament\\
%G359.79+0.1 &300$\times$70 && Cresent, Curved filament\\
%G359.85+0.47 &300$\times$54 && Pelican\\
%G359.87+0.44 &420$\times$50 &&with G359.85+0.39, Cane\\
%%G359.91$-$1.03 &136$\times$36 \\
%%%G359.23$-$0.82 &156$\times$108 &&The Mouse\\
%G359.96+0.09 &500$\times$400 &&southern thread\\ 
%G0.16$-$0.15 &1690$\times$145 && Radio Arc\\
%G0.15$-$0.07&---&&Radio Arc\\
% G0.17$-$0.42 &180$\times$18 &&XMM J0173$-$0.413,S5\\
%G0.9+0.1&---&&PWN & d, e\\

 \end{center}%table 13
{\footnotesize
$\ast$ Same as table 12, but listed sources are in the large area of $\timeform{0.5D} \times \timeform{1D}$ region around \SGRA, excluding those of the central region in table 12.\\
$\ddag$ Reference, a: \citet{Jo09}, b: \citet{Wa02b}, c: \citet{Yu02a, Yu02b}, d: \citet{Po15}, e: \citet{Mo15} f: \citet{Wa02a}, g: \citet{Lu03}, h: \citet{Ya14}.\\ 
}
\end{table*}

\subsection{The Central Region of the GCXE} %section 6.1 

\begin{figure}[!h]
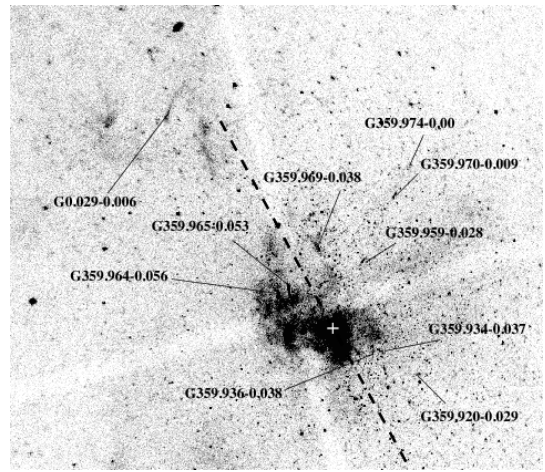
 \begin{center} % figure 7
\FigureFile(70mm,70mm){eps-figure/Filament-Lu08.eps} %figure 7
\end{center} 
\caption{X-ray filaments near \SGRA~in the central region of $6\arcmin.5\times6\arcmin.5$.
The dashed line indicates the Galactic plane, 
and the plus sign denotes the position of \SGRA. The brightest diffuse emission around  \SGRA~is central star cluster, and  the  extended emission at the northeast of \SGRA is  the Sgr\,A East SNR (From \cite{Lu08}).}
\label{fig:Lu2008} \end{figure}

\citet{Mu08, Lu08} made the Chandra map in the close vicinity of \SGRA, the $\sim \timeform{10'} \times \timeform{10'}$ region (within $\sim$20\,pc from \SGRA). They found many small scale diffuse sources, most of them have a filamentary or cometary shape (see figure 7).  
Table 12 is the summary of the X-ray filaments within $\sim$20 pc from \SGRA, where strong \FeK~line filaments in Sgr A East and  near the Sgr\,A complex and the Arches cluster, are excluded, because these are separately discussed in sections 4 and 5.
%%%\citet{Po15} surveyed nearly the same region with XMM-Newton, but the detected diffuse sources are almost overlapped to those of the Chandra observations \citep{Mu08, Lu08}.%%%% 
All the diffuse sources listed in upper rows of table 12, the positions (source name), sizes and  luminosity are  taken from \citet{Mu08}, while those in lower rows  are from  \citet{Lu08}. 

Diffuse sources have been also found or studied by many other authors. Since the typical morphology is filamentary or cometary with the length and width of  $\sim20\arcsec$ and $\sim4\arcsec$ (the mean values of \cite{Mu08, Lu08}), the cited positions may have typical uncertainly of a few $10\arcsec$. Therefore, if the source positions are within $\lesssim20\arcsec$ from \citet{Mu08} or \citet{Lu08}, these sources are regarded as the same objects of \citet{Mu08} or \citet{Lu08}, and the authors names are listed in the column of Reference.

Most of the sources have power-law (non-thermal) spectra with  the photon index of $\Gamma \sim$1\,--\,2.  However, only small fractions of the X-ray filaments are associated to the radio NTF. 
Four X-ray filaments, 359.945$-$0.044, G359.942-0.045, G359.944$-$0.052 
and  G359.950$-$0.043, are located very close to \SGRA~in the Sgr\,A West Cavity or on the Mini Spiral, 
and hence possibly associated to the Central Star Cluster (CSC), or related to the \SGRA~activity.
Other four X-ray filaments, G359.956$-$0.052, G359.962$-$0.062, G359.964$-$0.053 and G359.965$-$0.056  are in the Sgr\,A East SNR. Since the \EWFeK~of these X-ray filaments  are less than $\sim$70\,eV \citep{Mu08}, these are not parts of the hot plasma of Sgr\,A East,  but are more likely non-thermal filaments due to accelerated particles in the SNR Sgr\,A East.

The X-ray filament G359.945$-$0.044 (or G359.95$-$0.04 in \cite{Wa06a}, \cite{Lu08}
and  \cite{Jo09}) is one of the brightest filament separated only 0$\arcmin$.1 from Sgr\,A*.  It has a bright head and faint tail structure. The spectrum  shows softening from the head to tail of photon index $\Gamma$ from $\sim$1.3 to $\sim$3.1 \citep{Wa06a}.
The total X-ray luminosity is $\sim10^{34}$\Lu, with an averaged  photon index of $\sim$1.7\,--\,1.9 \citep{Mu08, Lu08, Jo09}. These values are consistent with a PWN. 
However, no X-ray pulsar is found from the head. Thus, an alternative scenario is a ram pressured magnetic tube, which traps TeV electrons accelerated by \SGRA~(e.g. \cite{Wa06a}).  Hard X-rays above 10\,keV are detected with NuSTAR from this source \citep{Mo15}. The 20\,--\,40\,keV luminosity is $\sim7\times10^{33}$\Lu.

Hard X-rays above 10\,keV are also detected with NuSTAR from the other bright X-ray filaments, G359.983$-$0.046 (J174545.5$-$285829: the Cannonball), 
G359.89-0.08 (Sgr\,A-E) and G359.97-0.038, \citep{Ny13, Ny15, Zh14, Mo15}. With the wide band spectra, X-ray luminosity and photon indexes are determined to be
$\sim10^{33}$\,--\,$10^{34}$\Lu, and $\sim$1.3\,--\,2.6, respectively.  

From the star formation rate at the Galactic center, \citet{Mu08} estimated that  $\sim$20 PWNe are expected in the Galactic center. 
However, only small fractions of the X-ray filaments are suspected to be  PWNe \citep{Mu08}. Therefore, most of the PWNe would have the X-ray luminosity of
$\lesssim10^{31}$\Lu, bellow the detection limit of the sources in table 12.  

The X-ray filaments, G359.959$-$0.027, G359.970$-$0.008  and  G359.974$-$0.00  have  narrow features associated to the radio filaments, and are roughly aligned on a slightly curved line crossing over  \SGRA.  These are named knot-1, 2 and 3 by \citet{Ko04}. The other narrow filament, G359.944$-$0.052, named Jet by \citet{Li13}, has well collimated structure pointing to \SGRA located the other side of the Galactic plane. The detailed discussion on these filaments related to the past high activity of the SMBH, \SGRA~ are given in section 7.3. 

In table 12, the X-ray luminosity is available for 2 thirds of the X-ray filaments, with the luminosity of 
$\sim10^{32}$\,--\,$10^{34}$ \Lu~(2\,--\,8\,keV). The summed  luminosity (2\,--\,10\,keV) is $\sim3.8\times10^{34}$\Lu.
The luminosity of the remaining 1 third  is unavailable due to the faintness.  Assuming the luminosity of the remaining sources is near the lower limit of the observable luminosity of $\sim10^{32}$\Lu, the summed luminosity of all the X-ray filaments is estimated to be $\sim4\times10^{34}$\Lu, only 3\,\% of the GCXE.

\subsection{The Outer Region of the GCXE} % section 6.2
% of $\timeform{1.1D} \times \timeform{0.57D}$} 

The high spatial resolution survey of the whole GCXE of $\sim\timeform{2D}\times\timeform{0.8D}$ around \SGRA~is first made
by \citet{Wa02a} with Chandra. Although 8 bright radio non-thermal-filaments are included in this region, only one source, G359.54$+$0.18 \citep{Yu97} is found as an X-ray filament, named  the X-ray thread. 
With the deeper Chandra survey of $\timeform{1.1D}\times \timeform{0.57D}$ around \SGRA, \citet{Jo09} found 17 X-ray filaments, a dozen of them are located in the inner GCXE region of $\sim\timeform{10'} \times \timeform{10'}$ and are listed in table 12. The source list in the whole GCXE of $\timeform{1.1D} \times \timeform{0.57D}$, excluding the inner region are given in table 13.

\citet{Po15} surveyed nearly the same region with XMM-Newton.  Since most of them except G0.17$-$0.42,  are overlapped with \citet{Jo09}, the sources listed  in tables 13 are mainly due to Chandra observations.  The References in table 13 indicate the other authors who observed the relevant sources.

The X-ray filament G0.13$-$0.11 is the brightest, and is the most studied X-ray filament in this region. This name and its features are confusing, because   other nearby objects with similar names but different natures are presented.
The radio MC, G0.13$-$0.13 is found to be a \Ka~line source \citep{Yu02a, Yu02b}, and is  claimed that the \Ka~line is due to LECRe.  However the selected area is rather large ($\timeform{4'} \times \timeform{3'}$), and hence is contaminated by a nearby XRN, G0.11$-$0.11 (table 9).  The diffuse soft X-ray source, a candidate of an intermediate-aged SNR, G0.13$-$0.12 is near at the filament G0.13$-$0.11 (section 4.2.1).

\citet{Wa02b} found that the X-ray spectrum of G0.13$-$0.11 is a simple power-law  with the luminosity of $\sim3\times10^{33}$\Lu.  From the head of G0.13$-$0.11, a point source CXOGCS\,J174621.5$-$285256 is found  with a power-law  photon index and the 2\,--\,10\,keV band luminosity of $0.9^{+0.9}_{-0.7}$, and $\sim8\times10^{32}$\Lu, respectively.  This luminosity is $\sim$30\,\% of the whole G0.13$-$0.11. 
The morphology, spectrum and luminosity indicate that G0.13$-$0.11 
is the leading edge of a PWN, produced by a pulsar CXOGCS\,J174621.5$-$285256 moving in a strong magnetic field environment. 
The main body of this PWN is likely traced by a bow-shaped radio feature, which is apparently bordered by G0.13$-$0.11, and is possibly associated 
with the prominent non-thermal radio filaments. The origins may be due to synchrotron radiation, or inverse Compton scattering of far-infrared photons from dust by the relativistic electrons responsible to the radio synchrotron emission. The magnetic field strength is estimated to be 0.08\,mG within the radio NTF \citep{Yu05}. 

The source density in the outer region (table 13) is far smaller than that of the inner region (table 12).  This is, at least partly, due to the higher detection threshold luminosity in the larger area of the outer region than that of the inner region. In fact, most of the X-ray luminosity (2\,--\,10 keV) of the table 12 sources is  $\lesssim10^{33}$\Lu, while those in table 13 are  $\gtrsim10^{33}$\Lu. The source sizes of  the inner region are also smaller than those of the outer region, except some distance X-ray filaments from \SGRA.  

The summed  luminosity (2\,--\,10\,keV) of all the resolved X-ray filaments is  $\sim1.0\times10^{34}$\Lu.  Assuming that G0.17$-$0.42  has the luminosity of $\sim5\times10^{32}$\Lu, the lower limit of the detection threshold, the total luminosity is estimated to be $\sim1.1\times10^{34}$\Lu,  $\sim$1\,\% of the GCXE and one quarter of all the point sources in the inner GCXE region (section 6.1).

\section {Past X-Ray Flares of \SGRA} % section 7

\SGRA~is the brightest radio point source located at the dynamical center of Our Galaxy.  The observations of accurate  motions of the IR stars in the close vicinity of \SGRA~revealed that the mass of \SGRA~is $\sim4\times10^{6}$\,$M_\odot$ (e.g. \cite{Ge10}), and hence  established that \SGRA~is a SMBH. 
The fine X-ray image with Chandra resolved \SGRA~from nearby X-ray sources for the first time \citep{Ba03}. The resultant X-ray flux is 
very low of $\sim10^{33}$ \Lu. This quiescent flux of the SMBH would be  due to a small mass flow rate within the Bondi radius. The accretion flow structure of $\sim\timeform{0.6}$\,pc size would be marginally resolved \citep{Ba03}. 

Many X-ray outbursts, possibly due to fluctuations of mass 
accretion rate of a time scale $\sim$ minutes\,--\,hours, have been observed with the rate of $\sim$1\,flare\,/\,day.
The photon index in the flare spectrum is $\Gamma\sim$2, similar to AGNs. The maximum peak luminosity is $10^{35}$\,--\,$4\times10^{35}$\Lu, $\sim$ a few 100 times of quiescent flux of $\sim10^{33}$\Lu~\citep{Po03, Po08}.
Still, this maximum flux is extremely lower than any of the AGNs.  Has \SGRA~ always been quiet in the past ?  This section reviews  possible relics of big flares or high activities of \SGRA~in the past: the
X-ray echo from \SGRA~(XRNe) (section 7.1), recombining plasma near at \SGRA~(section 7.2), and jets and outflow structures from \SGRA~(section 7.3). 

\subsection{X-ray Echo as a Relic of the Past Activities of \SGRA} 
% section 7.1
\begin{figure*}[htbp]
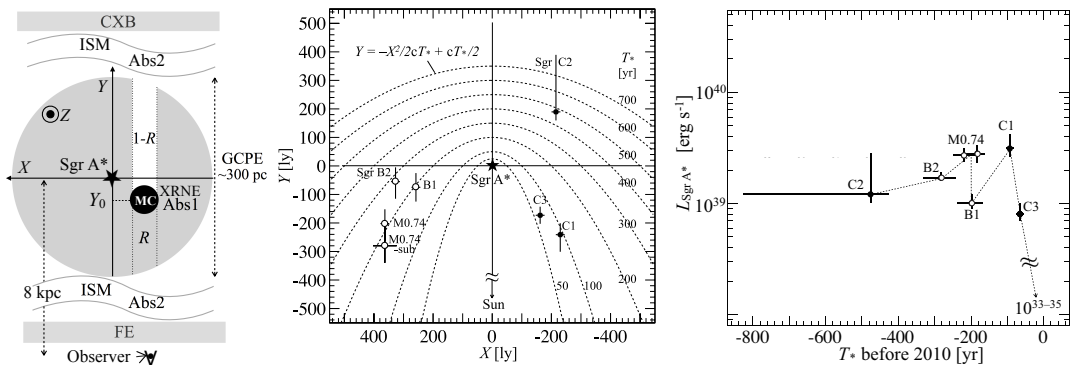
 \begin{center} %figure 6
\FigureFile(35mm,35mm){eps-figure/X-tomo-Ryu13.eps} 
\FigureFile(50mm,50mm){eps-figure/Ryu13-L.eps}
\FigureFile(55mm,55mm){eps-figure/Ryu11_pastlc_SgrAstar.eps}
\end{center}  \caption{
Left: Schematic view of the X-ray tomography to get the line of sight position of the XRNe. The notations of FE and Z are foreground emission and Z-axis of the coordinate (X-Y-Z), respectively.  Center: Face-on view of the positions of XRNe in the Sgr\,C complex  (filled circles) and those in the Sgr\,B complex(open circles).The parabolas (dashed lines) represent the equal-time delay ($T_{*}$) contours for the X-ray echoes from \SGRA. Right: X-ray light curve of \SGRA~in the past $\sim$500 years. Data points of Sgr\,C and Sgr\,B are shown by 
filled and open circles, respectively (From \cite{Ry09} and \cite{Ry13}). 
The present luminosity of \SGRA~is quoted from \citet{Ba03, Po03}.}
\label{fig:X-tomo-Ryu} 
\end{figure*}

%%%%%%%%%\begin{figure*}[hbtp]  % figure 22\begin{center}
%\FigureFile(50mm,30mm){eps-figure/GCXE_ModelB.eps} % figure 22
%\FigureFile(50mm,30mm){eps-figure/GBXE_ModelB.eps}
%\FigureFile(50mm,30mm){eps-figure/GRXE_ModelB.eps}
%\end{center}
%\caption{The spectra and best-fit curves  of the GCXE (left), GBXE (center) and GRXE (right)with a combination of the mCV (red), non-mCV (blue) and AB (orange) spectra.The black line is CXB (From \cite{No16}) }\label{fig:FIT-GDXE-PS} \end{figure*}

The XRNe are dense MCs of large $N_{\rm H}$ (see section 5). 
Thus, the observed GCXE spectrum  behind the XRN (here, GCXE1) is largely absorbed by the MC, while that in front of the XRN (here, GCXE2) is not absorbed by the MC.   The observed spectrum  at the XRN position is the sum of the GCXE1, GCXE2 and XRN spectra. 

\citet{Ry09} fitted the Suzaku X-ray spectra from Sgr\, B2 XRNe by the combined models of GCXE1 plus XRN, and GCXE2, where the former component has a large absorptions of $N_{\rm H}(Abs1)$ due to the absorption of the MC, while the  later has a small $N_{\rm H}(Abs2)$.  The parameters of  $Abs1$ and $Abs2$ are absorptions by the MC+ISM,  and by the ISM only, respectively. The schematic view is given in figure 8 (left).
The best-fit $N_{\rm H} (Abs1)$ is $>10^{23}$\,cm$^{-2}$, consistent with the absorption of the dense MC+ISM (table 7), while the best-fit $N_{\rm H} (Abs2)$ is $\sim6\times10^{22}$\,cm$^{-2}$, consistent with that of the ISM.  
Assuming that the GCXE is spherically extended around \SGRA~with a uniform flux density, the line-of-sight position of the XRN is approximately estimated by the best-fit flux ratio of the GCXE1 and GCXE2 spectra.
The projected position (2-diemnssional) and the line-of-sight position provides the 3-dimensional position of the XRN. The positions in the face-on-view of Sgr\,B XRNe are given in figure 8 (center).
Here and after, this method of 3-dimensional position determination is called, the X-ray tomography technique.

The Sgr\,C complex is composed of  three  XRNe (C1, C2 and C3: table 10). These are located nearby in projection with each other.  \citet{Ry13} applied the X-ray tomography method to the XRNe, C1, C2 and C3, and found that these XRNe are largely separated in the line-of-sight positions; C1 and C3 are near side of the Galactic plane, while C2 is far side. These are separately associated to the  MCs in different velocity ranges of the radio observations. The face-on-view of the positions of  C1, C2 and C3 are shown in figure 8 (center).  

Using the 3-dimensional positions, the distances of the XRNe from \SGRA~are determined. Then, using the best-fit  fluxes and MC absorptions ($N_{\rm H}$) of $N_{\rm H}(Abs1)-N_{\rm H}(Abs2)$, the past luminosity of \SGRA~is estimated. The age of the past is  light-traveling time difference between the direct pass to the Earth from \SGRA, and the pass from  \SGRA~via the XRNe and to the Earth. 
Thus, the X-ray tomography analyses of many XRNe make the X-ray activity history of \SGRA~as is shown in figure 8 (right).  

The X-ray luminosity of \SGRA~has  been  continuously high level of $L_{\rm X}\sim$\,(1\,--\,3)$\times10^{39}$\Lu~in the past of $\sim$70\,--\,500 years ago. Then, about 70 years ago, the luminosity  dropped down to the current low level. 
The averaged past luminosity is $\sim$4\,--\,6 orders of magnitude higher than  the present luminosity. In addition, at least two short-term flares with the  timescale of a few years are found. Thus, the high luminosity level of the $\sim$70\,--\,500\,years ago is not due to a single flare of long duration, but would be due to multiple, overlapping short flares.

To fill the blank ages of $<70$\,years ago in the past  \SGRA~activity history, 
similar tomography method should be applied to the Sgr\,A XRNe complex, because these XRNe are the nearest sample ($\sim30$\,--\,80 light-years in projection)  to \SGRA.  However, the tomography method requires very accurate spectra with Suzaku. Unfortunately, the XRNe density in the Sgr\,A MC complex is too high (see table 9) to be separately observed with the limited spatial resolution of Suzaku. 

Using  XMM-Newton, \citet{Po10} determined the line-of-sight positions of XRNe in the Sgr\,A MC complex.  They assumed a long flare ($\sim$ 100\,year) of constant flux of $\sim10^{39}$\Lu, and that all the XRNe in the Sgr\,A complex are behind  the projected  Galactic plane.  From the  observed  fluxes and $N_{\rm H}$ of the XRNe, they determined the distance from \SGRA, and hence predicted the 3-dimensional positions of the XRNe,  Bridge, M1, M2 and G0.11-0.11.  This method of \citet{Po10}, however gives  no information on the light curve of \SGRA, because the flux is a priori assumed to be constant. 

\citet{Ca12} also determined the line-of-sight positions of XRNe in the Sgr\,A MC complex using equation (4); the XRNe of the minimum \EWKa~have the scattering angle of $\theta=90\arcdeg$, on the line of the projected Galactic plane. The other XRNe with larger \EWKa~would be located  either in front or behind the line of $\theta=90\arcdeg$ (equation 4).  To resolve this degeneracy of in front or behind, they grouped the 9 XRNe in table 9 into 3 groups, according to their time variability profiles.  Group 1 is decreasing flux (B1 and F), Group 2 is increasing flux  (B2, C and D), and Group 3 is constant flux (A, E, DS1 and DS2) (see section 5.2.2). 
They assumed that the XRNe in the same group were irradiated by the same flare of a few ten years duration (observed time scale of the variability). Then, using the observed \EWKa, they determined the 3-dimensional positions of the XRNe.  From the observed fluxes and the column density $N_{\rm H}$ of the XRNe, they made the X-ray light curve of \SGRA~in the past of $\sim$70\,--\,130 years ago, with the average luminosity of $\sim10^{37}$\,--\,10$^{38}$\Lu.

As is noted in section 5.2.2, \citet{Cl13} found many bright \Ka~flares  from small spots of  $26\arcsec\times61\arcsec$ ($\sim$1\,pc$\times2.4$\,pc) 
or $15\arcsec$ ($\sim$0.6\,pc) square.  
Since the flare peaks are  $\sim$10 time brighter than those  in the quiescent level, they  proposed  that the past flare of \SGRA~are composed of multiple short flares of $\sim$a few years duration with the peak luminosity of $\sim10^{39}$\Lu,  overlaid on the lower level and  longer duration ($\sim$10 year) flares.

The key factor to determine the 3-dimensional positions of the XRNe by \citet{Ca12} is reliability of the absolute \EWKa.  However the observed \EWKa~ has large systematic errors as is noted in sections 5.2.2. The \EWKa~of the XRNe depends on the subtraction of a local GCXE background. Since \citet{Ca12} used a common GCXE background at $b = -0\arcdeg.12$ for all the XRNe, 
but each XRN is located at different position of $b$, the \EWKa~should be $b-0\arcdeg.12$-dependent. In fact, $b-0\arcdeg.12$ and \EWKa~are not randomly distributed, but show a clear anti-correlation as expected from the local GCXE subtraction effect. 
Accordingly, the line-of-sight positions, hence the predicted light curve has significant systematic errors. 
The results of \citet{Po10} of the 3-dimensinal positions are  inconsistent with those of \citet{Ca12}, not only due to the different assumption of a single constant flux flare \citep{Po10} of $\sim10^{39}$\Lu, or multiple short flares of various luminosity of \SGRA \citep{Ca12}, but also due to the systematic errors of local GCXE subtraction.

Ignoring these uncertainty, the predicted X-ray light curve of \SGRA~by \citet{Ca12} in the recent past of $\sim$70\,--\,130 years ago comes to the decaying phase in the past $\sim$70\,--\,500\,years light curve in figure 8 (right). However the flux is $\sim$10 \,--\,100 times lower than those of C1 and C3 in the same epoch.  If all the XRNe are located in front the line of $\theta=90\arcdeg$,  in contrast to the assumption of  \citet{Ca12}, the light curve  would be systematically sifted toward recent ages, at least, less than $\sim$30\,--\,80 years. 
In the close vicinity of \SGRA, there exist two giant MCs, the 50 km~s$^{-1}$ and  20 km~s$^{-1}$  clouds.  However, no \Ka~ line is found from these clouds \citep{Pa05, Po10}. This would be due to the largely declined flare flux of \SGRA~in the recent decades. 

\subsection{Recombining Plasma: Another Relic of \SGRA~Activity}\label{sec:GCflare} %section 7.2

 \SGRA~ is located in the  Sgr\,A East SNR. If \SGRA~had been very bright of   $\sim10^{39-40}$\Lu~ during $\sim$70\,--\,500 years ago (see section 7.1), He-like iron  in the hot plasma  would be partially  photo-ionized to H-like iron, then the plasma  would emit the radiative recombination continuum (RRC) at $\sim$8.7 keV  \citep{Oz09}. The RRC structure is recently found in the Suzaku spectrum of Sgr\,A East \citep{Uc17}.  Therefore, the  RRC  in the  Sgr\,A East spectrum  can be regarded as another relic of the past flares of \SGRA in $\sim$70\,--\,500 years ago.  

\citet{Na13} found a possible relic of more energetic flares in the far past.
They found a peculiar X-ray plasma named GC-South at ($l$, $b$)= ($\timeform{0D}$, $\timeform{-1.5D}$). The emission region is an ellipse with  $\sim21^\prime \times 8^\prime$ in the major and minor radius. The jet-like structure is elongated toward \SGRA.  

The X-ray spectrum of the GC-South plasma exhibits emission lines from highly ionized Si and S. Although the X-ray spectrum of the GBXE around GC-South is  well fitted  with a CIE plasma (section 3.2), that of GC-South cannot be fitted with a CIE plasma, leaving saw-teeth shape residuals at $\sim$2.5\,keV and  $\sim$3.5\,keV, which are attributable to the RRCs of He-like Si and S, respectively \citep{Ya09}. In fact, the GC-South spectrum is well fitted with a RP model.  The electron temperature is $\sim$0.46\,keV, while 
the ionization temperature was $\sim$1.6\,keV  in the 
initial epoch, and the plasma is now in a recombining phase after the relaxation time scale $n_{\rm e}t$ (electron density $\times$ elapsed time) of $\sim5.3\times10^{11}$cm$^{-3}$\,s. 

The absorption column density of the GC-South plasma is consistent with that of the Galactic Bulge (GB). Thus, the GC-South plasma is likely located in the GB region (at 8\,kpc distance). Then, the full size of the plasma, the mean electron density ($n_{\rm e}$), and the thermal energy are estimated to be $\sim 97\times37$\,pc$^2$, $\sim 0.16$\,cm$^{-3}$  and $\sim1.6\times10^{51}$\, erg, 
respectively \citep{Na13}. Then, the RP plasma age $t~(=\,n_{\rm e}t/n_{\rm e}$), is $\sim10^{5}$ years.
%%The size and luminosity are too large as a single SNR.  However, the jet-like morphology and off-plane position do not favor the multiple SNe (or super bubble) scenario.%%%%%%

Possible scenario is that the almost fully ionized  (at least, for  Si and S) plasma is made by a bright flare X-rays of \SGRA~of 
$\sim 10^5$ years ago, and the plasma is now in recombining phase (RP). Using this scenario, \citet{Na13} argued that the past flare 
luminosity of $\sim10^{5}$ years ago is near the Eddington limit of $\sim10^{44}$\Lu, more energetic than those of recent flares of $\sim$70\,--\,500\,years ago.

\subsection{The Other Possible Relic of \SGRA~Activity}\label{sec:GCflare} % section 7.3

As is noted in section 6.1, Chandra found 3 filaments (knot-1, knot-2 and knot-3) near \SGRA. These are aligned on almost a straight line, but is slightly curved pointing to \SGRA.  With a power-law model fit, the $N_{\rm H}$ are found to be $\sim($10\,--\,16$)\times10^{22}$\,cm$^{-2}$, consistent with the GC distance.  Then, the size and luminosity of knot-1, knot-2 and knot-3 are nearly the same of $\timeform{10''}\times\timeform{4''}$, and $\sim($2\,--\,6$)\times10^{32}$\Lu, respectively \citep{Ko04, Mu08}.  The power-law photon indexes are flat of $\lesssim$1.3.  From these facts, \citet{Ko04} suggested that the three filaments have the same origin; knot-1, knot-2 and knot-3  would be due to   sequential plasma ejections from a single source, \SGRA.

The other jet-like structure G359.944$-$0.052 (Jet, table 12) has the size of $\sim2\arcsec\times19\arcsec$,  located  at the close vicinity in the southeast from \SGRA with the major axis pointing to \SGRA.  \citet{Li13} found that the spectrum of Jet is a power-law  with the photon index, absorption column and luminosity (2\,--\,10\,keV) of  $\sim$1.8, $\sim12\times10^{22}$ cm$^{-2}$  and  $\sim2.4\times10^{32}$\Lu, respectively. The large absorption column suggests  that Jet is located at the GC distance. The photon index and luminosity are typical to a jet of synchrotron emission. 
The position and major axis of Jet aligns with the curved line  connecting \SGRA, knot-1, knot-2 and knot-3.  Thus, it may be conceivable that Jet is 
 a counter jet of  knot-1, knot-2 and knot-3, or these are highly collimated  magnetized outflows of relativistic particles emanating from \SGRA \citep{Li13}.

The ejected epochs of these jets (outflows)  can be determined from the 3-dimensional ejection angle and the velocity of the jets, which are all unknown. However the projected distances of the jets are small ($\sim$0.7\,--\,8\,pc), and hence the jet ejections would be recent events.
If the past flares of \SGRA triggered the jet ejections, the flare energies would be significantly large in order to produce such  prominent jets.

As noted in section 4.2.3, \citet{He13b} found the diffuse thermal sources NW, SE and E in the GC region from the XMM-Newton image. They suggested these 3 sources are  young-intermediated aged SNRs. However, some aspects are unusual; the sizes are smaller than typical SNRs of $\sim$1\,keV temperatures, the plasma density is very high of $\sim$4.6\,--\,9.9 \,cm$^{-3}$, and the morphology shows bipolar-flow structures  emanated from \SGRA~or Sgr\,A East SNR with angles nearly perpendicular to the Galactic plane. The locations are only $\sim$5\,--\,10 pc away from \SGRA. Therefore, they proposed an alternative scenario that these thermal plasmas are outflows driven by intermittent outbursts of \SGRA.  Assuming the velocity is 1000~km~s$^{-1}$, high speed stellar wind of massive star, the timescale for the plasma to reach the 6$\arcmin$~(14 pc) distance, the most remote position  of these plasmas is $\sim10^{4}$\,--\,$10^{5}$\,years. This is the same age of GC-South (section 7.2).

The Fermi Bubbles are largely extended GeV gamma ray sources of $\sim 50^{\circ}$  above and below the GC \citep{Do10, Su10}. These would be due to 
starbursts or a nuclear outburst which happened near the Galactic center in $\sim$10\,Myr ago. The same idea is firstly proposed by \citet{So00} to account the North Polar Spur (NPS). 
The morphology is spatially correlated with the WMAP haze, and the edges of the bubbles also line-up with NPS in the ROSAT X-ray maps. 
Suzaku revealed  a large amount of neutral matter absorbing the X-ray emission towards the bubble direction 
as well as the existence of the $\sim 0.3$\,keV temperature plasma.
These  are naturally interpreted 
as shock-heated Galactic halo during the bubble expansion \citep{Ka13}.
The 511 keV line emission by INTEGRAL \citep{Je06, We08} would be another hint of the past activity near the GC, or \SGRA.

\section {Methodology to the Origin of the GDXE} % section 8

The long standing questions regarding the origin of the GDXE are; how much fractions of the GDXE are resolved into point sources, and what are the populations of the point sources. The candidate point sources should have similar spectra (plasma temperatures of $>$ a few keV) to the GDXE and reasonably bright in the 2\,--\,10\,keV band to explain the GDXE flux. These Galactic point sources
are, hear and after,  defined as the X-ray Active Stars (XAS). The majority of the XASs are the mCVs, non-mCVs and ABs.

In the previous sections, many arguments for the origin of the GDXE are given along these questions.  However the predictions are often quantitatively inconsistent from author to author or instrument to instrument.
The reasons of these inconsistency are mainly due to large errors of the observed physical parameters (both statistical and systematical), and partly due to confusing definition of XASs ( mCVs, non-mCVs and ABs), differences in the energy band of XLF and that of the analysis method of \EWFeK~and so on. 

Thus this section discuss in detail on these issues, adopting two methodology to the GDXE origin.  One is direct resolution of the GDXE into the XASs. In this section, this approach is referred to as the Flux Integration Method (FIM) of the XASs (section 8.1). The other is a quantitative estimation whether or not, and how much the GDXE spectrum is reproduced by the spectra of the XASs, which is referred as the Spectrum Accumulation Method (SAM) of the XASs (section 8.2). 

Non-negligible systematic errors for the origin of the GDEX, regardless in the FIM or SAM approaches, are found in the spectra (e.g. \EWFeK) and fluxes of the GDXE and the XASs.  The next subsections give interpretations and discussions with critical comments on these systematic errors. 

\subsection {Flux Integration Method (FIM)} %section 8.1

The FIM approach is the development of the previous study given in section 2.1.1. 
If the instrument has enough power to resolve the XASs down to the luminosity limit of $\gtrsim10^{27}$~\Lu~(2\,--\,10\,keV), the lowest luminosity of the XASs  with the plasma temperature of $\sim$ a few keV, the FIM is very simple and straightforward approach. 
However, even with the highest spatial resolution and deepest exposure observations of Chandra, the resolved XASs are limited in the high luminosity range of $\gtrsim4\times10^{29}$\Lu, which is achieved only for the GBXE \citep{Re09, Ho12}.
Therefore, the XAS fraction must be estimated by the extrapolation of the observed fraction from the high luminosity band to the lowest luminosity limit of $\sim10^{27}$\,--\,$10^{28}$~\Lu, using the empirically made XLF, the cumulative X-ray luminosity as a function of the luminosity of the resolved XASs.  

A problem of the FIM is that actually resolved XAS fraction and its XLF have significant uncertainties, variations from  author to author, namely the systematic errors. The systematic errors would come mostly from the  Non X-ray Background (NXB) \footnote{Non X-ray Background (NXB) is cosmic ray induced background and exhibits some K$_\alpha$ lines of various elements.} subtraction, which is serious in the low luminosity band of $\lesssim10^{31}$\Lu. 
These systematic errors are separately discussed in the next sections  9.1, 9.2 and 9.3 in detail in the cases of the GBXE, GRXE and GBXE, respectively.

As for the XLF, \citet{Saz06} constructed an XLF (2\,--\,10\,keV) in the luminosity band of $10^{30}$\,--\,$10^{34}$\Lu~using the XASs
in the solar neighborhood observed with the RXTE and ROSAT.  They claimed that the XLF is  mainly composed of CVs and ABs with the flux per $M_\odot$ of $\sim1.1\times10^{27}$\Lu$M_\odot^{-1}$, 
and  $\sim2.0\times10^{27}$\Lu$M_\odot^{-1}$, respectively. The composition ratio of the CV $vs.$ ABs is 1:2.
However, a large error exists in the conversion process of the ROSAT luminosity band of 0.1\,--\,2.4\, keV to the RXE  luminosity band of 2\,--\,10\,keV. 
In fact, they estimated that the systematic error in the 
conversion process is $\ge$\,50\,\%.  Possibly, the XLF in the lowest luminosity range of $\sim5\times10^{27}$\,--\,$10^{30}$~\Lu~has even larger systematic error. 

\citet{Wa14b} constructed another XLF (2\,--\,10\,keV) in the luminosity band of $10^{28}$\,--\,$10^{34}$\Lu, using  the Galactic ridge survey data of XMM-Newton. They claimed that the XLF is composed of CVs and ABs with the fluxes per $M_\odot$ of 
$\sim2.5\times10^{27}$\Lu$M_\odot^{-1}$ 
and  $\sim1.1\times10^{28}$\Lu$M_\odot^{-1}$, respectively. The composition ratio  of the CVs $vs.$ ABs is 1:4. %%%
These are largely different from that of \citet{Saz06}. 

\citet{Re09, Ho12} made the other XLF (6.5\,--\,7.1\,keV) in the luminosity band of $4\times10^{29}$\,--\,$10^{33}$\Lu, using the Chandra data in the GBXE field. The shape of these XLFs are quite different. From the shape of their XLF, \citet{Re09} claimed
that the main component are mCVs (high luminosity band) and ABs (low luminosity band). 
On the other hand, \citet{Ho12} claimed that the composition is mainly mCVs, quite different from \citet{Re09, Saz06, Wa14b} of the ABs dominant compositions.  These apparent inconsistency in the XLF composition among the authors would come partly from the energy band difference. The ABs spectra become much softer toward the lower luminosity limit, and hence the  contribution in the high energy band  (e.g. 6.5\,--\,7.1\,keV) become smaller than those of canonical energy band of 2\,--\,10 keV.  The other possibility would be  confusion in the definition of the CV and AB; whether the non-mCV (dwarf nova) is included to the CV, included to the AB, or independently treated.

Accordingly, the FIM should be applied  separately for the mCVs, non-mCVs and ABs with unified energy band.  
The sum of these separate FIM estimations is the final solution of the XAS fraction in the GDXE.  

\subsection {Spectrum Accumulation Method (SAM)} % section 8.2

The SAM approach is the development of the early studies given in section 2.1.2.  Using Suzaku, \citet{Yu12} predicted that most of the flux of the GDXE are due to the mCVs. \citet{Ho12, He13a} predicted the same conclusion using
Chandra and  XMM-Newton spectra, respectively. 
These scenarios of the mCV dominant origins, however, have a serious problem that the \EWFeK~of the mCVs are far smaller than those of the GDXE.  
On the other hand, \citet{Xu16} found that the integrated spectra of the non-mCVs (DNe) in the Suzaku archive, has comparable \EWFeK ~to that of the GDXE, and argued that the GRXE is mainly composed of the non-mCVs, as was previously proposed by \citet{Mu93}.

The SAM approaches of  \citet{Xu16, He13a, Yu12, Ho12} to predict the XAS origin for the GDXE have all common problems.
They applied selected candidates of XASs, only the mCVs and ABs (except non-mCVs dominant scenario by  \cite{Xu16, Mu93} of non-mCVs-dominant scenario). They used the limited information of the \EWFeK~for all the relevant objects, and did not separately examine the origin of the GCXE, GBXE and GRXE involving all the possible candidate XASs.
One important note, related to the \EWFeK~is, that the Fe abundances in the observed thermal plasmas of the mCVs, non-mCVs and ABs are different to be $\sim$0.3, $\sim$0.6, and $\sim$0.2\,solar, respectively \citep{No16}. In particular, the observed Fe abundances in the mCV plasma of far less than 1-solar are often ignored in most of the mCV dominant scenario. The reasons of the low Fe abundance and the difference among the XASs would be due to the different production and emission mechanisms of the plasmas among the XASs. The real physical process is unsolved problem, but out of the scope of the review.  

As for the \EWFeK, one technical note is that the default choice of the XSPEC package, $eqwith$ uses the continuum flux in the energy range of $\pm{0.3}$\,keV of the center energy of the relevant iron K-shell line.  Thus, depending  on the data analysis process, the value of \EWHea, for example, would be underestimated by the extra flux of the adjacent \Ka~and \Lya~lines. In the case of two-component spectra of the GDXE , a thermal plasma plus a power-law emission, the \EWKa~may be confused, whether it is estimated under the continuum shape of the thermal plasma (\EWHea~and \EWLya), the power-law continuum (\EWKa), or the sum of the both components. 

 To avoid the source-to-source and author-to-author mismatches in the estimation of \EWFeK, and utilize the proper comparison
between those of the GDXE and XASs, unified data process and analysis for all the GDXE and XASs by the same author are preferred.
\citet{No16}  determined the \EWFeK~of the GDXE and XASs, using all the Suzaku archive in unified analysis. The results of the GDXE are given in table\,\ref{tab:GDXE}.  

The \EWFeK~of the XASs have been measured by many authors, but the qualities and samples  were limited (e.g. \cite{Ya16, No16} and references therein). \citet{No16} found in the best quality Suzaku spectra that the \EWHea~and \EWLya~in the XAS spectra are well explained by a CIE plasma with the free parameters of temperature and Fe abundance.  
The \Ka~line flux would be due to the surrounding cloud, and hence \EWKa~is estimated by the parameters of the covering solid angle ($\Omega$), the absorption column  ($N_{\rm H}$), and the flux above 7.1 keV.  
They found a good correlation between  \EWKa~and the temperature  of the  mCVs, non-mCVs and ABs with each different free parameters of $\Omega\times N_{\rm H}$. Thus the \EWKa~ are quantitatively included with the parameter $\Omega\times N_{\rm H}$  for each mCVs, non-mCVs and ABs into the CIE model. Then, all the observed \EWFeK~are well explained by the CIE model, here and after, the 1-T model.

\citet{No16} found that the X-ray luminosity of XASs are well correlated to the temperature of the 1-T  model.  In the XLF, the relevant luminosity ranges are  $\sim10^{31}$\,--\,$10^{34}$\Lu,
$\sim10^{29}$\,--\,$10^{32}$\Lu~and $\sim10^{27}$\,--\,$10^{30.5}$\Lu, corresponding to the temperature ranges of $\sim$10\,--\,20\,keV, $\sim$3\,--\,10\,keV and $\sim$1\,--\,3\,keV, for the mCVs, non-mCVs and ABs, respectively (table 14).  Then, they constructed a two-temperature CIE model (2-T model), as good approximated spectra in the relevant luminosity bands of the XASs \footnote{It is better to refer the original paper of \citet{No16}, because the process to construct the 2-T models is very complicated.}.

In principle, a multi-T model would be more appropriate than the 2-T model to incorporate the temperature dependent XLF. 
However, in reality, the 2-T and multi-T models show no large difference beyond the observed statistical errors in the XAS spectra.  
Accordingly, the GCXE, GBXE and GRXE spectra should be compared with a combination of the 2-T models of the XASs \citep{No16}. The best-fit results are given in table\,\ref{table:fraction}.
Since the \EWFeK~of the GDXE are more similar to  the non-mCVs than any other XASs, it would be reasonable that the non-mCVs occupies the largest fraction in the best-fit results (table 15). 

The best-fit $\chi^2$\,/\,d.o.f for the GCXE, GBXE and GRXE spectra are 2637\,/\,276, 148\,/\,95 and 282\,/\,91, respectively. Thus the SAM predicts that a combination of the XASs can explain the spectra of the GBXE, but not for the GRXE and GCXE spectra. Detailed discussions are separately given in sections 9.1, 9.2 and 9.3 for the GBXE, GRXE and GCXE, respectively.

The SAM approach using the 2-T  model is less sensitive to the assumed XLF than the FIM.  This is a large advantage of the SAM over the FIM.  Possible systematic errors due to NXB subtraction would be also less sensitive than the FIM.
In fact, the \EWFeK~ in the Suzaku GCXE spectra were almost the same among the authors  (see section 3.3), although the data reductions and analysis methods were independent. 
The 2-T model of \citet{No16} smeared-out position-to-position variations in the GDXE, because they used the larger area for the GDXE spectra than any of the other authors or instruments.

 The realistic error of the most important parameter, the mean \EWFeK~of non-mCV, is statistically estimated using the standard deviation of the 13 non-mCVs sample \citep{No16}, and resultant 1-$\sigma$ error is $\sim10$\,\%.  As the results, the best-fit composition ratios of the mCVs, non-mCVs and ABs (table 15) would have a systematic error of $\sim\pm{0.1}$.  
This, however, has no serious impact on the discussion for the origin of the GDXE given in section 9. 

\begin{table}[!ht] %table 14
\caption{Physical Parameters of the  mCV, non-mCV and AB.$^\ast$}
\begin{center}
\begin{tabular}{lccl}
\hline
& mCV	&non-mCV	&AB\\
\hline
$kT_{\rm e}$ (keV)	&$23.3^{+5.1}_{-3.7}$&$10.7\pm{1.7}$&$4.25\pm{0.18}$\\
Lx (\Lu)  &$10^{31}$--$10^{34}$&$10^{29}$--$10^{32}$&$10^{27}$--$10^{30.5}$\\
SH (pc)$^\dag$	&130--160	&130--160	&150--300 \\
\EWKa (eV)  	&169$\pm{5}$	&82$\pm{7}$ 	&28$\pm{5}$\\
\EWHea (eV)		&118$\pm{7}$	&451$\pm{10}$	&327$\pm{8}$\\  
\EWLya (eV)		&60$\pm{7}$	&167$\pm{9}$	&45$\pm{6}$\\ 
\hline
\end{tabular} 

 %table 14
\end{center} 
{\footnotesize
$\ast$ Errors are 1\,$\sigma$ confidence level.\\
$\dag$  \citet{Ak08}, \citet{Pa84} and \citet{St93}.\\
}
\end{table}

\begin{table}[!ht] % Parameter of combined fit Table 15
\caption{The best-fit fraction of the mCV, non-mCV and AB spectra (after \cite{No16}).}
\label{table:fraction} \small \begin{center}
 \begin{tabular}{lccc}
 \hline
	& GCXE		& GBXE 		& GRXE \\
\hline
Source &\multicolumn{3}{c}{Fraction$^\dag$}\\
mCV	&0.04($<0.01$) 	 & 0.03($<0.09$)	&0.10$\pm{0.05}$ \\
non-mCV	&0.96$\pm{0.01}$&0.67$\pm{0.06}$ 	&0.51$\pm{0.06}$\\
AB	&0.00 ($<0.01$)	 &0.30$\pm{0.03}$ 	&0.39$\pm{0.02}$ \\
\hline
$\chi^2$/d.o.f.& 2637/276(9.55)	&148/95(1.56)	 & 282/91(3.10) \\
\hline
\end{tabular}

 %Table 15
\end{center} 
{\footnotesize
\dag Fraction of the surface brightness of mCV, non-mCVs and ABs.
After \citet{No16}.
}
\end{table}

\subsection {Combined Approach of the FIM and SAM} %section 8.3

 In the FIM approach, the essential points are to obtain a reliable fraction of the resolved  XASs, and a reliable XLF down to the limiting luminosity of $\sim10^{27}$\Lu. The best instrument for the FIM is Chandra, because of the best spatial resolution of $\sim1\arcsec$, two orders of magnitude better than Suzaku. The weakest point is its large NXB, about 10 times larger than Suzaku. Therefore, possible flux errors due to the NXB subtraction are not negligible for the low surface brightness sources, the GBXE and GRXE, and faint XASs.

In the SAM approach, the essential points are to obtain reliable spectra of the GDXE (GBXE, GRXE and GCXE), and the XASs (mCVs, non-mCVs and ABs).  The key parameters are the values of \EWFeK, which are  sensitive to the continuum levels, or the NXB subtractions.  Suzaku is the  best instrument for the SAM approach, because of the reasonably large  effective area  and good spectral resolution.  The NXB is about 10 times lower than Chandra, and the stability and reproducibility of the NXB are  fur better than Chandra and XMM-Newton. 

Whichever the FIM and SAM, to minimize possible systematic errors are to  utilize  the best instruments for FIM and SAM, and carry out unified analysis for the GDXE and XASs. In order to minimize the author-to-author systematic errors, simultaneous and unified study with Chandra (FIM) and Suzaku (SAM) by the same researchers is important. 
Currently, no such unified work has been available.  Therefore, independent approaches by the FIM and SAM should be complementally applied to the origins of the GDXE. If the fluxes are not explained by the FIM, and\,/\,or if the spectra are not well explained by the SAM, new  sources other than the known XASs must be involved, regardless point-like or diffuse.  

So far, the point source origin of the GDXE led by the FIM, is more widely accepted, because the FIM is a simple approach, and has no risk to involve any new Galactic sources, or new physical processes other than the emissions of the known
XASs. 
On the other hand, although the SAM did not exclude the contribution of the known  XASs in some fractions, the SAM did not elude a risk to involve new objects or new concepts of uncommon physical processes.  Due to this risk, the SAM approach has been less accepted.  In the next section, the origin and structure of the GBXE (section 9.1), GRXE(section 9.2) and GCXE (section 9.3) are discussed separately, applying equally both the FIM and SAM. 

\section {Origins of the GCXE, GBXE and GRXE} % section 9

The true origins of the GDXE are not conclusive, due to non-negligible errors of the observed results. The important fact is that the \EWFeK~and \SHFeK ~ are all different among the GCXE, GBXE and GRXE. Therefore, the origin of the GBXE, GRXE and  GCXE should be separately discussed, which are given in subsectios 9.1, 9.2 and 9.3, respectively.

\subsection{Galactic Bulge X-Ray Emission (GBXE)}% section 9.1

\citet{Re09} conducted deep observations ($\sim$1~Msec) in the region of $(l, b)=(\timeform{0.1D},-\timeform{1.4D})$, named the Chandra Bulge Field (CBF).  Although the CBF is near the GC, \citet{Ya16} found that the flux ratio of the GBXE component to that of the GCXE at $(l, b)=(\timeform{0.1D},-\timeform{1.4D})$  
is more than $\sim$10 (see figure\,\ref{fig:K-line-dis-b}).  
Thus, the CBF is not in the GCXE region but is almost in the pure GBXE region. 

In the central region of the CBF, \citet{Re09} applied the FIM. 
The XLF in the 6.0\,--\,7.1\, keV band shows a slow increase in the luminosity range of $\sim10^{30}$\,--\,$10^{32}$\Lu~(2\,--\,10\,keV). The resolved XAS fraction at $10^{30}$\Lu~ (2\,--\,10\,keV) is $\sim50$\,\%, then turns to a rapid increase in the luminosity range of $\sim4\times10^{29}$\,--\,$10^{30}$\Lu, and finally the resolved XAS fraction becomes $\sim80$\,\% of the GBXE at the lowest luminosity limit of $\sim4\times10^{29}$\Lu~(2\,--\,10\,keV).
They predicted that the rapid increase of the XLF in the low luminosity band is due to the increasing contribution of ABs, and hence contribution of ABs to the GBXE is very large at the low luminosity band.  

 In the same  central region of the CBF, \citet{Ho12}  made another XLF using the same data set and  energy band of \citet{Re09}.
His XLF shows a monotonous increase toward the low luminosity.
The resolved XAS fraction at $10^{30}$\Lu~is already $\sim$60\,--\,70\,\%, then becomes slow increase to $\sim$70\,\% at $\sim4\times10^{29}$\Lu~(2\,--\,10\,keV). 
He predicted that the smooth XLF is due to a single class of the XASs, namely mCVs, and suspected that more than $\sim$70\,\% of the GBXE is resolved into mCVs, in contrast to the prediction of \citet{Re09}.

\citet{Mo13} reported that the \EWFeK~of the resolved point sources in the CBF is $\sim$100~eV in the luminosity range of
$\gtrsim10^{32}$~erg~s$^{-1}$ (2\,--\,8\,keV), 
where they regarded the candidate point sources are the mCVs and AGNs (see also \cite{Ho09}). The \EWFeK~increases by a constant rate  in the range of $7\times10^{30}$\,--\,7$\times10^{31}$~erg~s$^{-1}$ (2\,--\,8\,keV) (see figure 13 of \cite{Mo13}), possibly due to an increasing contribution of the non-mCVs and\,/\,or ABs.
In the range of $\lesssim7\times10^{30}$~erg~s$^{-1}$ (2\,--\,8\,keV), the \EWFeK~ become nearly equal to $\sim$300~eV, where main contributors would be  non-mCVs and\,/\,or ABs.  This  trend, at least semi-quantitatively, is consistent with \citet{Re09}, but is against the mCV dominant scenario of \citet{Ho12}. 

The causes of the significant difference of the XLF profiles between \citet{Re09} and \citet{Ho12} would be found in the difference of the resolved XASs in the low luminosity band of $\lesssim 10^{30}$~erg~s$^{-1}$ (2\,--\,10\,keV). Most of the resolved XASs in this faintest luminosity range are uncommon between \citet{Re09} and \citet{Ho12}.  These inconsistency may come from the difference of the NXB estimation and related analysis, because the surface brightness of the NXB  at $\sim$6\,--\,7\,keV is $\sim$10 times larger than the X-ray flux in the CBF \citep{Ho12}. In fact, \citet{Ho12} re-analyzed the same data of \citet{Re09}, and found  that the XLF inconsistency is disappeared. 

Another problems in these authors is  their flat spectra  of the CBF X-rays and those of the resolved XASs. 
These are found in figure 3 of \citet{Re09} and figure 6b of \citet{Ho12}. 
Comparing with the GBXE spectrum of \citet{No16}, the flat spectra of \citet{Re09} and  \citet{Ho12} would be due to  under-subtraction of the NXB \footnote{The spectral comparison between  \citet{Re09}, \citet{Ho12}, and \citet{No16} is not straightforward, due to different unit of the vertical axis (intensity) of keV~s$^{-1}$~keV$^{-1}$,
keV$^2$~s$^{-1}$~keV$^{-1}$, and photons~s$^{-1}$~keV$^{-1}$, respectively}.

 \citet{Mo13} reported that the \EWFeK~in the CBF X-rays may be  $\sim$580\,eV \footnote{This result is confusing because \citet{Mo13}  stated  this value in the text, however $\sim$300\,eV of the resolved XASs at the luminosity of $\lesssim10^{30}$\Lu~is
found in their figure 13, and they argued most of the CBF flux is resolved to XASs.}
, which are significantly smaller than that of the whole GBXE by \citet{No16} of $\sim$720\,eV (table\,\ref{tab:GDXE}). This discrepancy would be either the difference of the selected regions, or more likely due to incomplete NXB subtraction of Chandra.

\citet{Yu12} observed  the GBXE and GRXE regions with Suzaku. The spectra are fitted with a model spectrum of the mCVs. The essence of this fit is to reproduce the  \EWHea~and \EWLya~values, by the two free parameters of the mCV mass (white dwarf mass) and Fe abundances; the  free parameter of mCV mass tunes the ratio of \EWHea\,/\,\EWLya~(plasma temperature), while the Fe abundance tunes  the absolute  \EWHea~(and \EWLya) value.  The other important value of the \EWKa, is independent free  parameter. Therefore, their mCV-dominant model should obviously gives a nice fit to the spectra of the GBXE and GRXE, in particular in the most important energy band of around \FeK~lines. 
The best-fit mCV mass and Fe abundance for the GBXE are $\sim0.7M_\odot$ and $\sim$0.8\,solar, while those for the GRXE are $\sim0.7M_\odot$ and $\sim$0.9\,solar, respectively. The 10\,\% smaller abundance of the GBXE than the GRXE is consistent with that the \EWHea~and \EWLya~of the GBXE are about 10\,\% smaller than those of the GRXE (table\,\ref{tab:GDXE}).  

Since the Fe abundance in the hot plasma of the mCVs in the solar neighborhood is $\sim0.3$ solar (e.g. \cite{Ya16}, \cite{No16}, and references), the hot plasma in the mCV-dominant model by \citet{Yu12} should have $\sim$3 times larger Fe abundance than those of the solar neighborhood.
However, the flux profiles along the large range of $|l|$= 1\arcdeg\,--\,100\arcdeg~in the continuum and in the \FeK~line bands are globally very similar with each other \citep{Re06a, Re06b} (see sections 2.1 and 2.3).  Therefore, the Fe abundance should be  nearly constant in the wide range of the GRXE and GBXE regions.  As is noted in section 5.2.2, the Fe abundance over the whole GCXE is nearly the  same of $\sim$1.1\,--\,1.2 solar.
The infrared star observations also show a global uniformity of Fe abundances  in the wide range from the Galactic ridge to the Galactic center (e.g. \cite{Cu07}, and references therein). Thus the Fe abundance is almost the same in the whole GRXE, GBXE and GCXE regions, in conflict to the mCV dominant scenario for the GBXE  spectrum  by \citet{Yu12, Ho12}.

The \SHHea~and \SHLya~of $\sim$310~pc ($2\arcdeg.2$) and the \SHKa~of $\sim$160~pc ($1\arcdeg.1$) in the GBXE (\cite{Ya16}, and see table\,\ref{table:GCXE-GBXE-GRXE}) are globally consistent with those of the XASs of $\sim$130\,--\,300\,pc (see table 14). 
Therefore, \citet{No16} tried  the SAM approach with the 2-T models of the  XASs spectra (see 8.2).
Unlike \citet{Yu12},~the 2-T models of \citet{No16} are based on the real observed values  of the temperatures and Fe abundances to predict all the \EWFeK~ for all the XASs.  

\citet{No16} obtained a reasonable fit with $\chi^2$/d.o.f=148\,/\,95 by the combined 2-T models (table\,\ref{table:fraction}). Then, they concluded that the major fraction of the GBXE is due to the non-mCVs and ABs.  
The predicted ABs ratio of 30\,\%  is far smaller than the AB-dominant scenario by \citet{Re09} and far larger than the mCV-dominant scenarios by \citet{Yu12, Ho12}. 
Although  the flux of the non-mCVs is about 10 times lower, the space density is 10 times larger than those of the mCVs \citep{Pa84}. Therefore, it may not be surprising  that the non-mCVs is the  main contributor to the GBXE, in contrast to the many previous predictions .

In summary, ignoring  possible systematic errors in the FIMs, a common consensus is that $\sim$70\,--\,80\,\% of the GBXE flux is explained by either the mCVs, non-mCVs,  ABs, or some mixture of these sources \citep{Re09, Ho12}.  This prediction is consistent with the SAM prediction \citep{No16}.  However, the composition ratios are different between the FIM and SAM and even among the FIMs.  Furthermore, this prediction comes from the works of limited area of the GBXE, the CBF of ~$|l|=\timeform{0.0D},~ |b|=\timeform{-1.4D}$ , in the FIM,  and the off-plane field of
 $|l|<\timeform{0.6D},~  \timeform{1.0D}<|b|<\timeform{3.0D}$ in the SAM. 
Therefore,  unsolved  questions are still remained;  which is the major contributor, the mCVs, non-mCVs or ABs ?  How much is the mixing ratio of these sources in all the GBXE region ?  

\subsection {Galactic Ridge X-Ray Emission (GRXE)} % section 9.2

The deepest  point source survey in the GRXE was made with Chandra  by \citet{Eb01, Eb05} near at $(l, b)=(\timeform{28.5D}, 
\timeform{-0.2D})$, named the Galactic Ridge Field (GRF). \citet{Eb05} resolved $\lesssim$10\,\% of the GRF flux into XASs at the detection threshold luminosity of $\sim2\times10^{31}$\Lu~(2\,--\,10~keV). 
In the same region (GRF), \citet {Re07a} reported that the resolved XAS fraction is $\sim$19\,\% in the luminosity range
of $\sim10^{30}$\,--\,$10^{32}$\Lu~(1\,--\,7~keV). This ratio is converted to $\sim$10\,\%  at the detection threshold luminosity of $\sim2\times 10^{31}$\Lu, using the the mean XLF \citep{Saz06, Ho12, Wa14b}.

\citet{Wa14b} applied the FIM to the origin of the GRXE. They assumed  that the resolved XAS  fraction is the same as
\citet{Eb05, Re07a}. Then, they extrapolate this fraction to the low  luminosity limit of $\sim10^{28}$\Lu~ along to their XLF, which is made by the XMM-Newton archives. 
They argued  that more than 90\,\%  of the GRXE flux in the GRF is resolved into the XASs (2\,--\,10\,keV) with the composition  ratio of the CVs:ABs of about 1:4 (section 8.1).
This FIM  results however, may have a large uncertainty due to 
the ambiguity of the XLF profiles sensitive to the ABs temperature near at the low luminosity limits of less reliable data.  
 Furthermore, the observed surface brightness of the GRXE obtained with Chandra 
\citep{Eb01, Eb05, Re07a} and XMM-Newton \citep{Ha04}  are  about 1.3 times larger than those of Suzaku \citep{Eb08, No16}.
These differences may be also due to some systematic errors of the NXB subtraction in the Chandra and XMM-Newton data.

For the SAM approach, \citet{Wa14a} analyzed the XMM-Newton slew survey on the Galactic plane data, and
reported  that the spectrum of the resolved XASs in the luminosity band
of  $\gtrsim8\times10^{32}$\Lu~(2\,--\,10\,keV) has the \EWKa, \EWHea, and \EWLya~of $\sim$90, $\sim$170, and $\sim$80\,eV, respectively, and hence they claimed that the origin of the GRXE is mCVs in the luminosity band of  $\gtrsim8\times10^{32}$\Lu~(2\,--\,10\,keV) .  
This conclusion may be correct, because the \EWFeK~values in this high luminosity band are very close to those of the mCVs \citep{No16}.

\citet{Xu16} found that Suzaku spectrum  of the non-mCV is similar to the GRXE.  Since the observed luminosity of the non-mCV did not cover the low luminosity band of the XLF,  they  made a model spectrum of the non-mCVs to include full luminosity range of the XLF.  The model spectrum of the non-mCVs is also very similar to the GRXE. Therefore, they predicted that the majority of the GRXE is due to unresolved non-mCVs. However their model of non-mCVs is taken into account of only the  \EWHea~values. 

The spectra of the GRXE have been observed with Suzaku, in the regions of sub-degree to a few degree size at $|l|$ $\sim$8$\arcdeg$, 15$\arcdeg$ and 28$\arcdeg$ on the Galactic plane by \citet{Eb08, Yamau09}. Due to the low surface brightness and limited exposure time of $\sim$50\,--\,100 ksec, all the \EWFeK~are not determined except the \EWHea. The \EWHea~ are variable  from position to position in the range of $\sim$350\,--\,640\,eV, larger than the statistical errors of typically $\sim$100\,eV.  These large position-to-position variations suggest that the origin of the GRXE is not fully due to the assembly of numerous XBSs. 

The large statistical errors and large position-to-position variations of the \EWFeK~ do not allow any quantitative study of the SAM.  To increase statistics and to smear-out the position-to-position variations, \citet{No16} made the GRXE spectrum using all the Suzaku archives. The total exposure time is 3 Msec, two orders of magnitude larger than those of the individual positions, and hence the statistical error of the \EWHea~is reduced to be $\sim$10\,eV (table\,\ref{tab:GDXE}). 
They fitted  the GRXE spectrum  with  a combination of the 2-T models of the XASs (section 8.2).  The fit is rejected with $\chi^2$\,/\,d.o.f=282\,/\,91 (table 15).

Although statistically rejected, the best-fit composition ratio of the mCV, non-mCV and ABs is 1:\,5:\,4, similar to the GBXE of 0.3:\,6.7:\,3 (table 15). Therefore, the shapes of the XLF of the GBXE and GRXE would be similar with each other.
The XAS fraction of the GBXE at the luminosity limit of  $\sim 2\times 10^{31}$\Lu \citep{Re09, Ho12} is $\sim$3 times larger than the GRXE.  If the same XLF of the GBXE is applied to the GRXE, the resolved XAS fraction of the GRXE \citep{Eb05, Re07a} is estimated to be $\lesssim$30\,\% of the GRXE at the same luminosity limit of $\sim 2\times 10^{31}$\Lu.
Thus, observational facts of both the FIM and SAM suggest that the GRXE is composed of not only the XASs, but has an additional component, either new class of point or diffuse sources.

Excluding the poor statistic band of~$\gtrsim$\,7.5\,keV, the largest residual from the combined 2-T model is the excess flux of  \Ka~line. 
\citet{Ya16} found that the \SHKa~is $\sim$70\,pc, smaller than any of the XASs (table 14), but is rather similar to that of the MCs. 
\citet{No15} found the \EWKa~excess, at the east of the Galactic plane of $l_* = 2$\arcdeg\,--\,4\arcdeg, $b_*$=0\arcdeg~compared to the west of the same, but negative longitude.
In the  X-ray spectra from both the east and west regions, the \Hea~and \Lya~ fluxes are almost the same between the east and west, while the \Ka~flux shows nearly 2 times larger  excess in the east compared to the west.

Then, they subtracted the west spectrum from that of the east, and made the X-ray spectrum of the east excess. They fitted the spectrum with a model of a power-law continuum and a Gaussian line for the \Ka~line at 6.4 keV. 
The best-fit photon index and \EWKa~are 3$\pm{1}$ and 1.3$\pm{0.4}$\,keV, respectively.
These values are well explained by a scenario of LECRp bombardment. Thus, a significant contribution of the \Ka~line in the GRXE would come from a bombardment to the MCs by LECRp \citep{No15, Ya16, No16}, although possibility of X-ray irradiation by bright XBs \citep{Su93} or LECRe bombardment \citep{Va00} would be partly possible. 

  \citet{Sa14, Sa16, No17b} found the excess of \Ka~line from regions of seven intermediate-old  aged  SNRs in the region of $6\arcdeg \lesssim l \lesssim 35\arcdeg$ and $|b|\lesssim0\arcdeg.5$. 
 Since the diffuse hard X-ray flux (e.g. 5\,--\,10 keV band) is negligibly small, these SNRs are not identified as point sources, then the excess of the \Ka~flux is regarded to be position-to-position fluctuations in the GRXE.  
Even X-ray undetected intermediate-old aged SNRs would make LECRp, and produce the \Ka~line by the bombardment on the MCs.
 
In summary, the FIM did not explain the full flux of the GRXE by XASs. The SAM  predicts the regions of the excess of \EWKa~over the assembly 
of the mean spectra of the XASs. Thus, an additional component is required, which is spatially clumpy and emits strong \Ka~lines, possibly MCs by the bombardment of LECRp.

\subsection {Galactic Center X-Ray Emission (GCXE)} % section 9.3

\begin{figure*}[!ht]
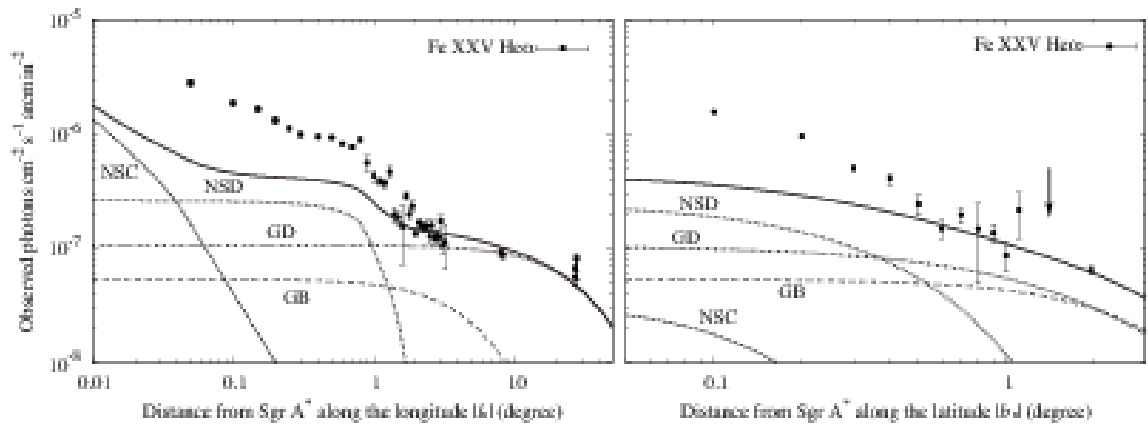
 \begin{center} % 9
\FigureFile(150mm,80mm){eps-figure/Uc11-Fig4.eps}
\end{center} 
\caption{Left: Flux distribution of the  \Hea~ along the Galactic plane in comparison with the SMD. The thick dotted lines are  the  nuclear  stellar cluster (NSC), the thin dotted lines are the central nuclear stellar cluster (NSD).  The dot-dashed lines represent the Galactic dick (GD), while the long-dashed line is the Galactic bulge.   The sum of these components is presented by the solid lines. The GCXE, GBXE and GRXE in X-rays correspond to NSD, GB and GD in infrared (SMD), respectively. Right: same as the left but perpendicular to the Galactic plane. The arrow shows the position of the CBF (section 9.1) (From \cite{Uc11}).}
\label{fig: Uc11-Fig2.ps} \end{figure*}

In the deep observation with Chandra of $\sim$600\,ksec exposure  of $17^\prime \times 17^\prime$  (40~pc$\times$40~pc) field around 
\SGRA,  \citet{Mu03, Mu04a} resolved $\sim$10\,\% and $\lesssim$20\,\% ( 2\,--\,8 keV) into XASs, respectively.
With the $\sim$800\,ksec observation in the fan-shaped region in the GC west of $2^\prime$\,--\, $4^\prime$ from \SGRA, \citet{Re07b} resolved $\sim$40\,\% (4\,--\,8 keV band) into XASs at the same threshold luminosity of $\sim10^{31}$\Lu.
Thus, even if taking account of the energy band difference, there is an extremely large difference by $\sim$2\,--\,4 
between these two authors with the mean (averaged) XAS fraction of $\sim$25\%. 
This mean value is about 2 times smaller than that of the GBXE by \citet{Ho12}. 
Therefore, the XAS fraction of the GCXE would be about half of the GBXE or $\sim$40\% of the GCXE.

With Suzaku, \citet{Uc11} made spatial profiles of the \FeK~lines  in the  GDXE with the resolution of $0\arcdeg1$. Then, they compared  the flux distribution of the \Hea~line, the brightest iron K-shell line,  with the SMD model, where  the SMD flux is normalized to the X-ray flux in the GRXE.  
The results are  shown in figure \ref{fig: Uc11-Fig2.ps}.  
The \Hea~flux in the GCXE region of $|l_*|=0\arcdeg.1$\,--\,0\arcdeg.6 is $\sim$2\,--\,4 times larger than the prediction of the SMD model (the solid lines in figure \ref{fig: Uc11-Fig2.ps}).  
The same result, the excess of \Hea~line  above the infrared flux in the GCXE region is found in the assembly of the infrared stars  obtained by the SIRIUS observations of \citet{Yasui15}.

  \citet{He13a} made  the \FeK~line and the 7.2\,--\,10 keV band profile along the $b_*=0\arcdeg$~and $l_*=0\arcdeg$ lines in the central GCXE region (4$\arcmin$\,--\,13$\arcmin$ from \SGRA).
The flux of the central GCXE region is enhanced  with a sharp peak near at \SGRA.
They predicted that the 2\,--\,10\,keV fluxes per 1$M_\odot$  at $20\arcmin$~ from \SGRA~is $\sim5\times10^{27}$\Lu$M_\odot^{-1}$, while   at $2\arcmin$, it is $\sim1\times10^{28}$\Lu$M_\odot^{-1}$. 
Taking into account of the energy band differences, these are $\sim$1.5 and $\sim$3 times larger than those of the GRXE of 
($3.5\pm{0.5})\times 10^{27}$\Lu$M_\odot^{-1}$ (3\,--\,20 keV)~\citep{Re06a} and the solar neighborhood  of
($3.1\pm{1.1})\times10^{27}$\Lu$M_\odot^{-1}$ (2\,--\,10 keV) \citep{Saz06}, respectively.  

 The systematic enhancement of the fluxes per 1$M_\odot$ toward \SGRA~makes the  FIM approaches for the origin of the whole GCXE to be complicated. On the other hand, \EWFeK~are almost constant in all the GCXE region, except \EWKa~in the XRNe (section 3.3), and hence the SAM approaches would 
be more straightforward for the origin of the whole GCXE.

Since the surface brightness of the GCXE in the iron K-shell band (6.3\,--\,7.1 keV) is $\sim$10 times larger than those of the GBXE and GRXE, reliable fluxes and spectra for the GCXE  may be possible with XMM-Newton and even with Chandra. Therefore the SAM is reliably applied to the GCXE for the  XMM-Newton and Chandra dada in addition to the Suzaku data. 
 \citet{He13a} fitted the XMM-Newton spectra of the central GCXE region with
mCV spectra of unrealistic Fe abundance (\cite{Yu12}, see section 9.1).
Within $\timeform{9'}$ from \SGRA, \citet{Mu04b} made the Chandra spectrum of the resolved point sources in the luminosity range of  $\sim10^{31}$\,--\, $10^{33}$\Lu.  They proposed that the major component in this luminosity band is mCV.
However their estimated \EWKa, \EWHea~and \EWLya ~of
 $\sim140$\,eV$\sim400$\,eV, and $\sim230$\,eV, respectively, % (table 4)
are largely different from the mCV , but is rather similar to the non-mCV (table 14).
 
In order to examine the differences between the global GCXE spectrum  and those of the XASs, \citet{No16} fitted the Suzaku GCXE spectrum with the combination of their 2-T models, which are made from the real observed values  of the temperature,  \EWFeK~for each mCVs, non-mCVs and ABs (section 8.2). 
The combined 2-T model fit is completely rejected with $\chi^2$\,/\,d.o.f\,=\,2637\,/\,276 (table 15), simply because  the  \EWFeK~of the GCXE (table 5) are far larger than any of the XASs (table 14). 

Large excesses  in the GCXE spectrum from the combined 2-T models are found  in the \Ka~and \Lya~lines. This indicates  that the \EWKa~and \EWLya~excess over the 2-T model in the GCXE is larger than those in the GRXE (section 9.2).
 An important note is that the excesses of \EWKa~and \EWLya~in the GCXE are not due to the 1.5\,--\,3 times enhancement of the X-ray luminosity per $M_\odot$ in the GCXE relative to the GBXE and GRXE, but needs new components which exhibit larger \EWKa~and \EWLya~ than any of the XASs.

The SAM results which reject the combined 2-T model fit are consistent with that the GCXE has smaller 
\SHFeK~$\sim$(31\,--\,36)~pc  (calculated from the e-folding $b$ in table 4) than those of the mCVs, non-mCVs~$\sim$(130\,--\,160)~pc and ABs $\sim$(150\,--\,300)~pc, respectively (table 14).  The \SHFeK~of the GCXE are similar to the SH of CMZ \citep{Ts99, We15}, and hence the new components of the GCXE may be closely related to the CMZ, regardless diffuse or point sources. 

The excess of the \EWKa~should be associated with the additional non-thermal X-ray continuum.  As is noted in section 3.3, \citet{Yu08} found  a  power-law emission from the GCXE, most are due to the XRNe. Still some fractions are remained in the non-XRN regions of the GC west (e.g. \cite{Ko09, Uc13}), which would be due to the LECR. If significant fraction of the LECR is LECRp, excess of the \EWKa~ would be obtained. Thus the \EWKa~excess in the GCXE is an enhanced version of the  the GRXE (section 8.3).

The excess of the  \Lya~lines requires another component,  which emit stronger  \Lya~lines than any of the XASs.  In the CMZ region of $|l|\lesssim\timeform{0.3D}$, the longitude profiles of the \Hea~line in the east (positive $l^*$) shows a significant excess over the west, even  excluding the bright SNR Sgr\,A East \citep{Ko07b, He13a}. This excess would be due to larger populations of high-mass stars in the east than the west \citep{Pa04, Mu04a, Ko07b}.
 
In the close vicinity of \SGRA, the \Hea~and \Lya~fluxes relative to the SMD, seems to be larger than $\sim$3 times of the GRXE \citep{He13a}. 
This region corresponds the Nuclear Star Cluster in the CMZ, which is a site of large population of high-mass stars. The high-mass stars may contribute to the GCXE by the putative star-burst activity and\,/\,or frequent SN explosion in the GMZ. The  reconnection  of  strong
magnetic fields, or big outbursts of \SGRA~\citep{In09, Te10, Po10, Ca12, Ry09, Ry13} (section 7)  may also produce hot plasmas, responsible to the strong \Lya~line.
The CMZ and the close vicinity of \SGRA~are unique regions  with the most extreme  physical conditions than any other regions of the Galaxy. Therefore, other physical processes, beyond our current views of the quiet Galactic regions, may be concealable. 

In summary, the FIMs did not explain the full flux of the GCXE by XASs. The SAM result leave significant excess of \EWKa~and \EWLya~ over any combination  of  the XAS spectra. Thus, either diffuse or new type point source are required in the GCXE.  These should have larger \EWKa~and  \EWLya~and smaller \SHFeK
~than any of the XASs. 

\begin{ack}
The author expresses his sincere thanks to all the members of $Kyoto School$ 
for their efficient works, valuable comments and many helps on the study of the Galactic diffuse X-rays.  Particular thanks are due to M. Nobukawa and S. Yamauchi for their great efforts for preparing this draft. This work  is supported by JSPS and MEXT KAKENHI Grant Number 24540229. % 24740123 (M.N.),and 24540232 (S.Y.).
\end{ack}

\end{document}